%% file: TDRCalo.tex
\documentclass[12pt,preprint,aps,showpacs]{revtex4}  
\usepackage{graphicx}

\usepackage{epstopdf} 
\usepackage{graphics}
\usepackage{amssymb}
\usepackage{amsmath}

\usepackage{rotating}  
\usepackage{subfigure}
\usepackage{lineno}

\begin{document}



\title{Electromagnetic Calorimeter for HADES}
\author{W. Czy\v{z}ycki$^1$, E.~Epple$^2$, L.~Fabbietti$^2$, M.~Golubeva$^3$, F.~Guber$^3$, A.~Ivashkin$^3$, M.~Kajetanowicz$^4$, A.~Kr\'{a}sa$^5$, F.~K\v{r}\'{i}\v{z}ek$^5$, A.~Kugler$^5$, K.~Lapidus$^{2,3}$, E.~Lisowski$^1$, J.~Pietraszko$^6$, A.~Reshetin$^3$, P.~Salabura$^4$, Y.~Sobolev$^5$, J.~Stanislav$^5$, P.~Tlust\'y$^5$, T.~Torrieri$^5$, M.~Traxler$^6$\\
\vspace{1cm}
{\it
\mbox{$^1$ Crakow Institute of Technology, Al. Jana Pawla II 37,31-864 Krak\'{o}w, Poland}\\
\mbox{$^2$ Excellence Cluster Origin "Universe", TU-Munich, Boltzmannstr. 2 85478 Garching, Germany}
\mbox{$^3$Institute for Nuclear Research, Russian Academy of Science, 117312~Moscow, Russia}
\mbox{$^4$ Smoluchowski Institute of Physics, Jagiellonian University of Krak\'{o}w, 30-059~Krak\'{o}w, Poland}
\mbox{$^5$ Nuclear Physics Institute, Academy of Sciences of Czech Republic, 25068~Rez, Czech Republic}
\mbox{$^6$ GSI Helmholtzzentrum f\"{u}r Schwerionenforschung, 64291~Darmstadt, Germany}
}
}

\date{\today}
\pacs{29.30, 29.40.Ka, 29.40.Vj}

\maketitle
\tableofcontents
\input{Intro/Intro}
\newpage
\input{Mechanics/Mechanics}

\newpage
\input{Crystals/Crystals}

\newpage
\input{Electronics/FE_Electronics}

\newpage
\input{Aging/Aging}
\newpage
\input{Simulation/Simulation}

\newpage
\input{TestResults/TestResults}
\newpage
\input{Conclusion/Conclusion}

\newpage
\input{Timeplan/Timeplan}

\newpage
\input{./Ref/References}
\end{document}

%% file: Intro/Intro.tex
\section{Introduction}

Dilepton measurements in the energy domain of SIS100 and SIS300 are mandatory to establish
a complete excitation function of the virtual photon radiation from dense nuclear 
matter over the whole energy region available at the current HI machines.  
Dielectron results obtained by the CERES collaboration at the lowest SPS energies of $40$ AGeV
display a large excess in the intermediate ($0.14 < M < 0.6$~GeV$/c^{2}$) mass region of dilepton pairs
pointing to a strong source generated from the high density zone of HI collisions.
A precise determination of this excess, however, depends on a precise knowledge
of the hadronic cocktail, which for 2--40 AGeV, is dominated 
by the $\eta$ Dalitz decay. Furthermore, a convenient
normalization of the dielectron spectra is naturally given by the
$\pi^0$ yield dominating the low mass region ($ M<0.14$~GeV$/c^2$).  These two arguments 
require data on the inclusive cross sections $\pi^0$ and $\eta$ mesons. 

For the SIS18 energy range, production of neutral
mesons has been studied extensively in HI collisions by the TAPS collaboration via
photon calorimetry and in $N-N$ collisions by various detectors at Uppsala, Saclay and COSY. 
However, no respective data are presently available for the energy range 8--40 AGeV, with the consequence that any interpretation of
future dielectron data would have to depend solely on theoretical
models, e.g. transport calculations or appropriate hydrodynamical
models. In order to improve this situation we propose to measure the
respective $\pi^0$ and $\eta$ meson yields together with the
dielectron data in $p-p$ and HI collisions. Furthermore, photon measurement 
would be of large interest for the HADES strangeness program which 
address also spectroscopy of neutral $\Lambda(1405)$ and $\Sigma(1385)$ 
resonances in elementary and HI reactions.

These goals can be achieved by replacing the HADES Pre-Shower detector, located at forward angles
($18^{\circ}<\theta<45^{\circ}$), with an electromagnetic
calorimeter based on lead-glass modules obtained on loan from the OPAL
detector. An additional advantage of such a device would be the improvement of the electron/pion separation at large momenta ($p > 400$~MeV/c) as compared to the
present situation. Note that at lower momenta the electron/hadron
identification is provided by the RICH, RPC and TOF detectors already 
available in HADES.

The total area of the proposed HADES calorimeter amounts to about
8\,m$^{2}$ and covers polar angles between $12^{\circ}$ and $45^{\circ}$ 
with almost full azimuthal coverage. The photon and electron energy 
resolution achieved in 
test experiments  amounts to $\approx$ 6\%/$\sqrt{E}$ which is sufficient 
for the $\eta$ meson
reconstruction with S/B ratio of $\sim 0.4\%$ in Ni+Ni collisions at 8~AGeV. 
 
Below we present details of the detector layout, the support structure, 
the electronic readout and its performance 
studied via Monte Carlo Simulations and series of dedicated test experiments.


%% file: Mechanics/Mechanics.tex
\section{Mechanical Structure}

\begin{figure}[ht]
\begin{minipage}[b]{0.42\linewidth}
\centering
\includegraphics[scale=0.4]{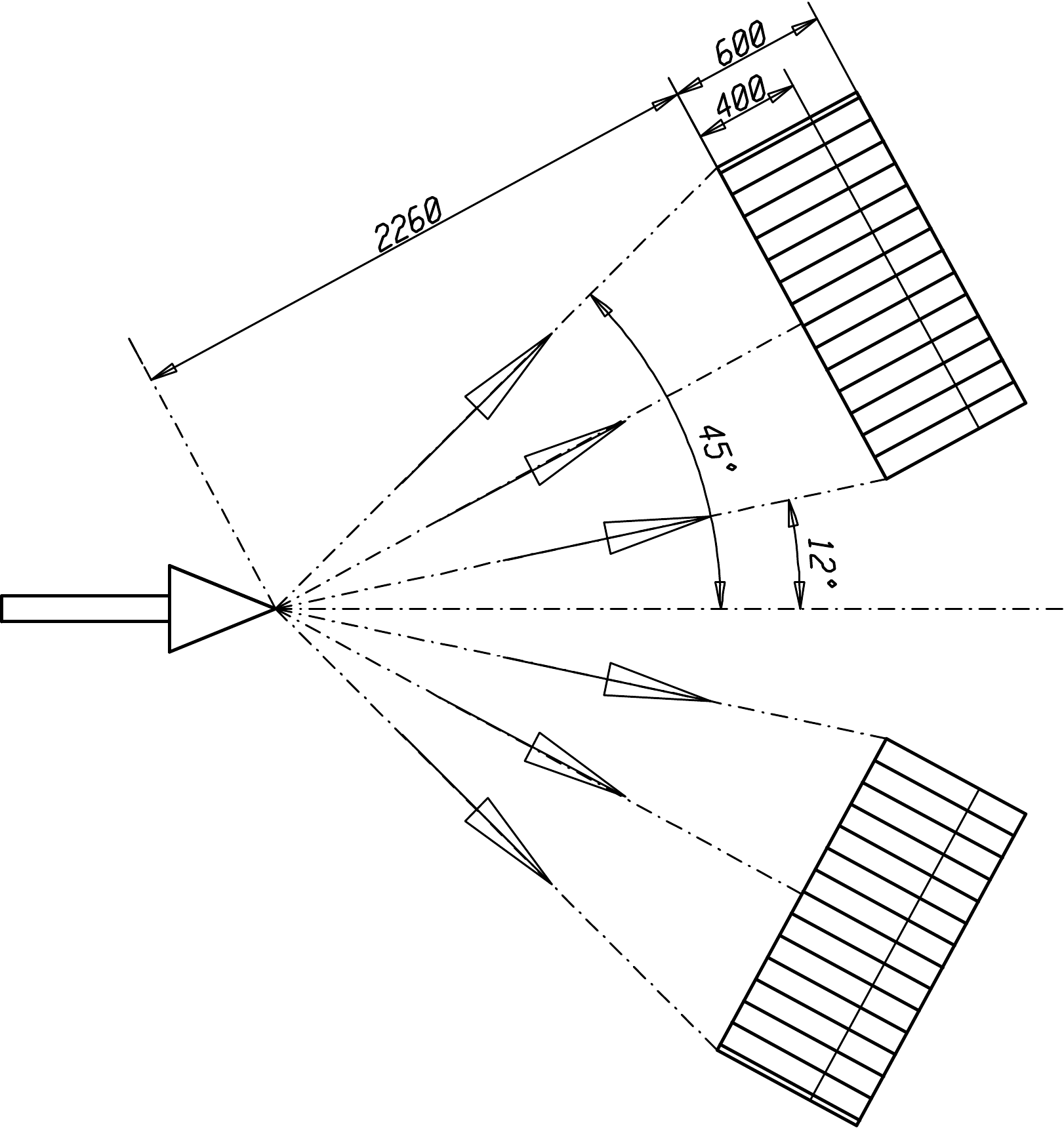}
\caption{Modules layout in vertical plane along the beam axis.}
\label{mech1}
\end{minipage}
\hspace{0.4cm}
\begin{minipage}[b]{0.42\linewidth}
\centering
\includegraphics[scale=0.3]{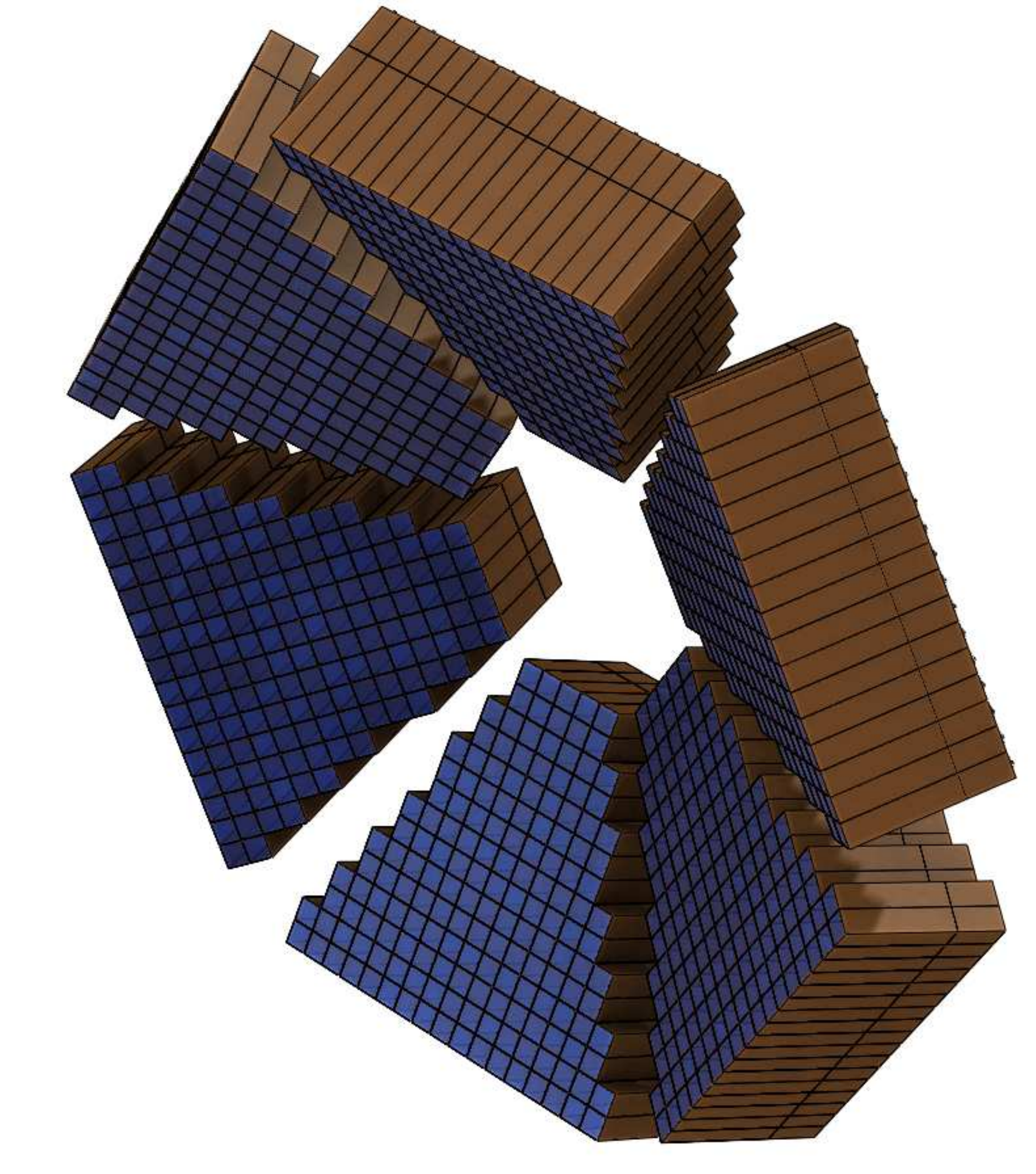}
\caption{3Dl arrangement of the calorimeter modules.}
\label{mech2}
\end{minipage}
\end{figure}

\begin{figure}[!htb]
\begin{center}
\includegraphics[angle=0,scale=0.31]{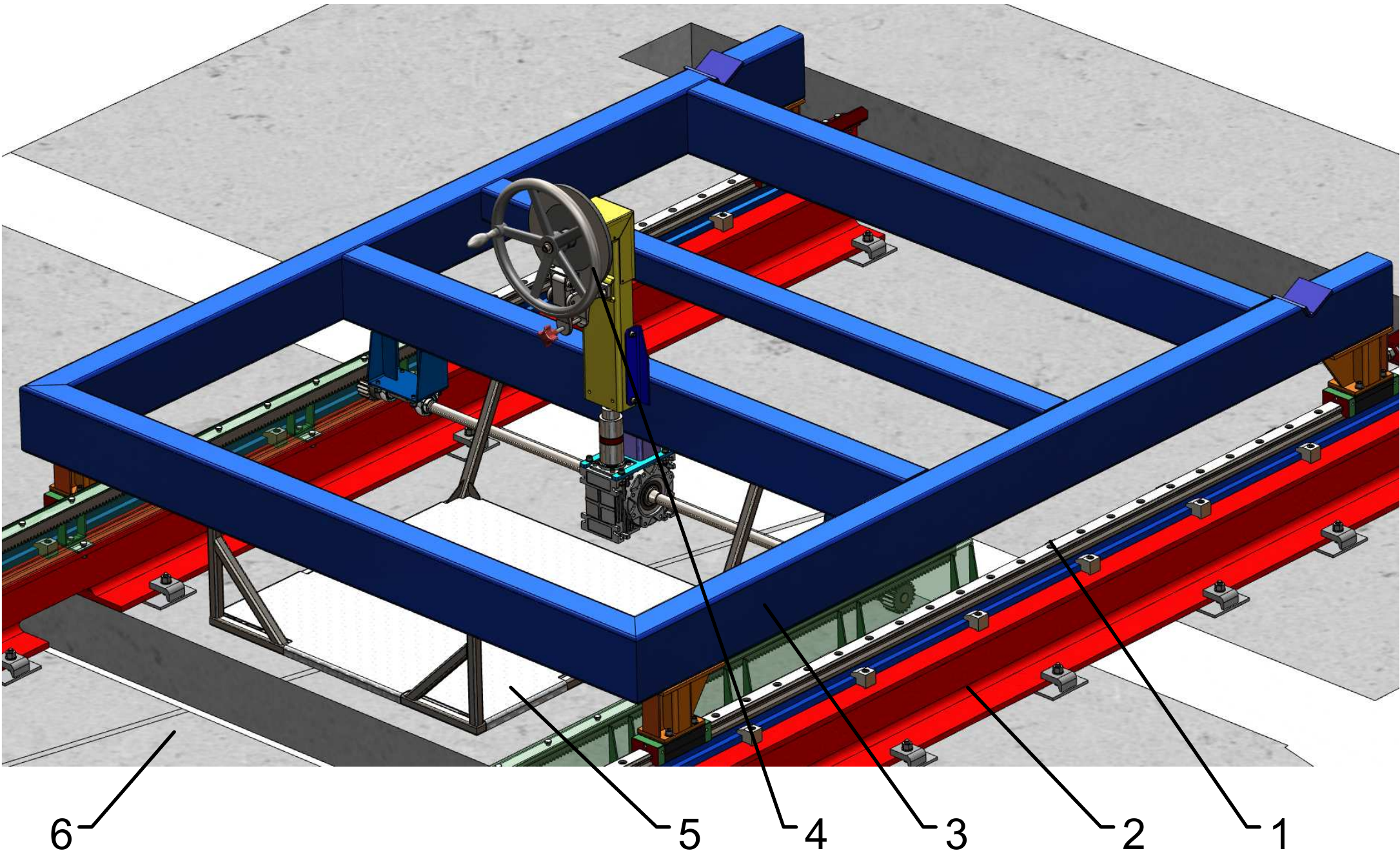}
\caption[]{View of the platform model: 1Ð linear bearing, 2 Ð fixed structure for reinforcing floor, 
3 Ð main frame of the platform, 4 Ð drive mechanism with break system, 5 Ð small platform for operator, 
6 Ð floor. 
\label{mech3} }
\end{center}
\end{figure}

\begin{figure}[ht]
\begin{minipage}[b]{0.35\linewidth}
\centering
\includegraphics[scale=0.22]{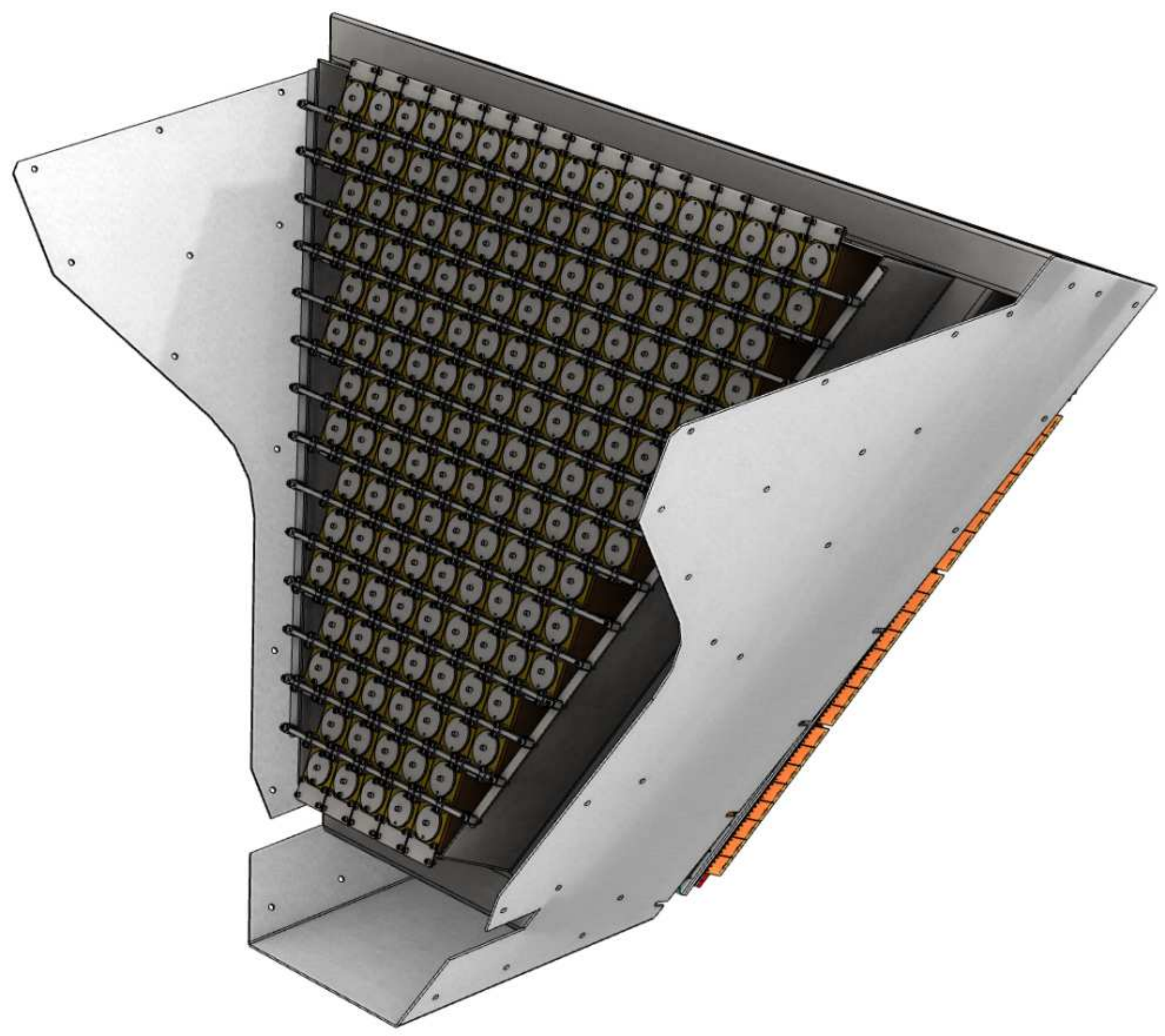}
\caption{Single container with EMC modules.}
\label{mech4}
\end{minipage}
\hspace{0.4cm}
\begin{minipage}[b]{0.55\linewidth}
\centering
\includegraphics[scale=0.28]{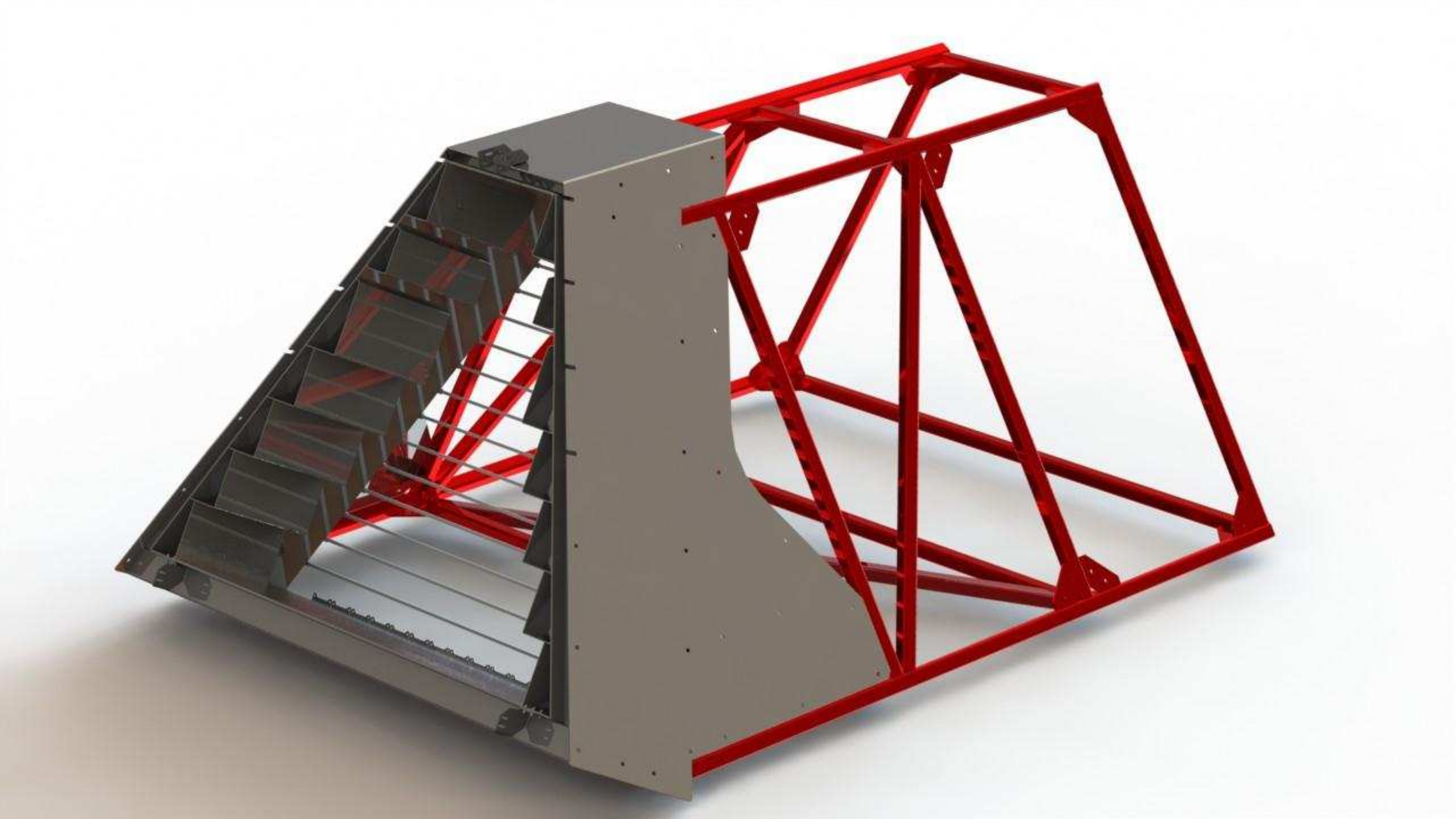}
\caption{Single container segment mounted to construction segment.}
\label{mech5}
\end{minipage}
\end{figure}

\begin{figure}[ht]
\begin{minipage}[b]{0.45\linewidth}
\centering
\includegraphics[scale=0.3]{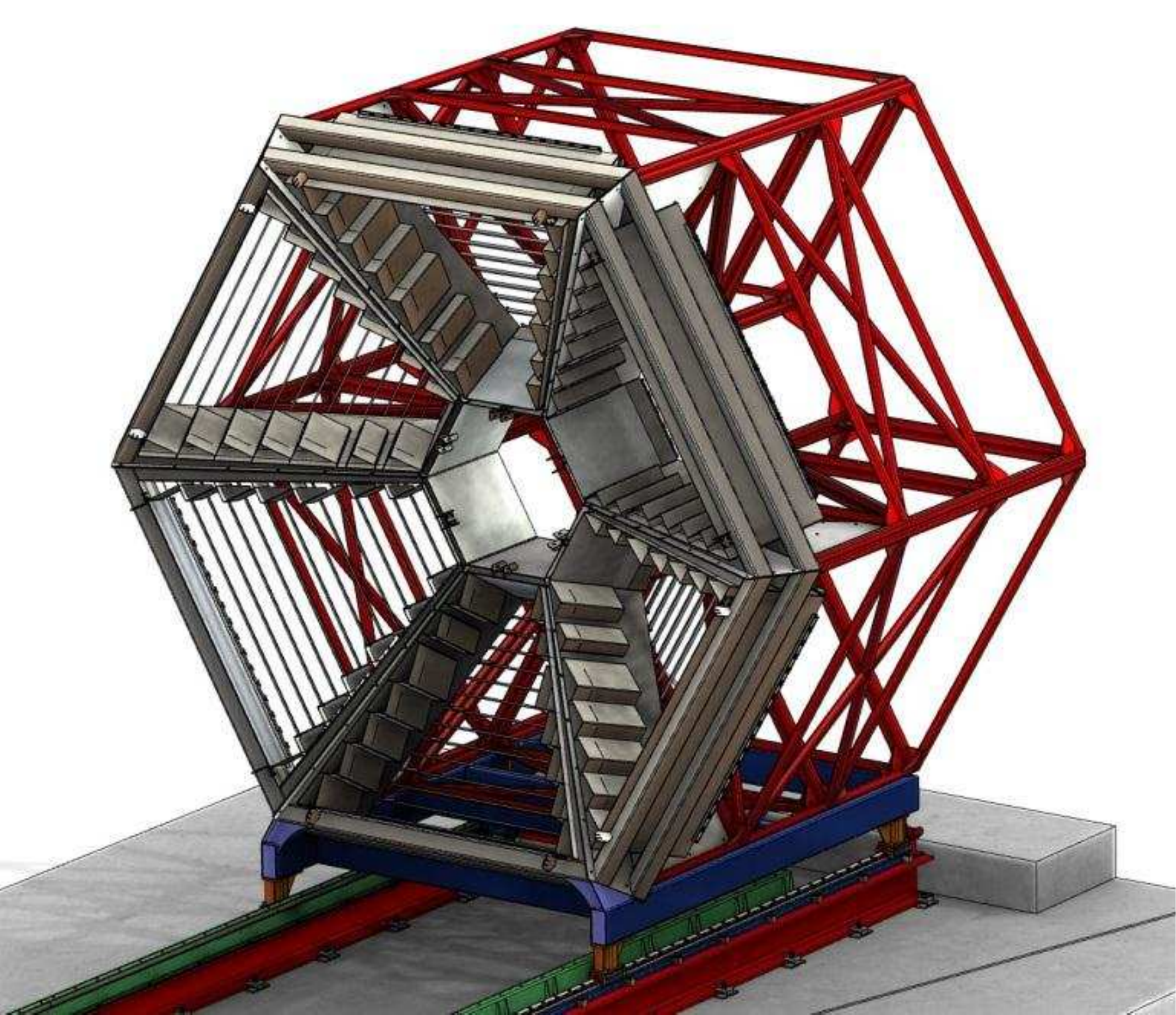}
\caption{View of the platform construction after placement of the last segment.}
\label{mech8}
\end{minipage}
\hspace{0.4cm}
\begin{minipage}[b]{0.45\linewidth}
\centering
\includegraphics[scale=0.3]{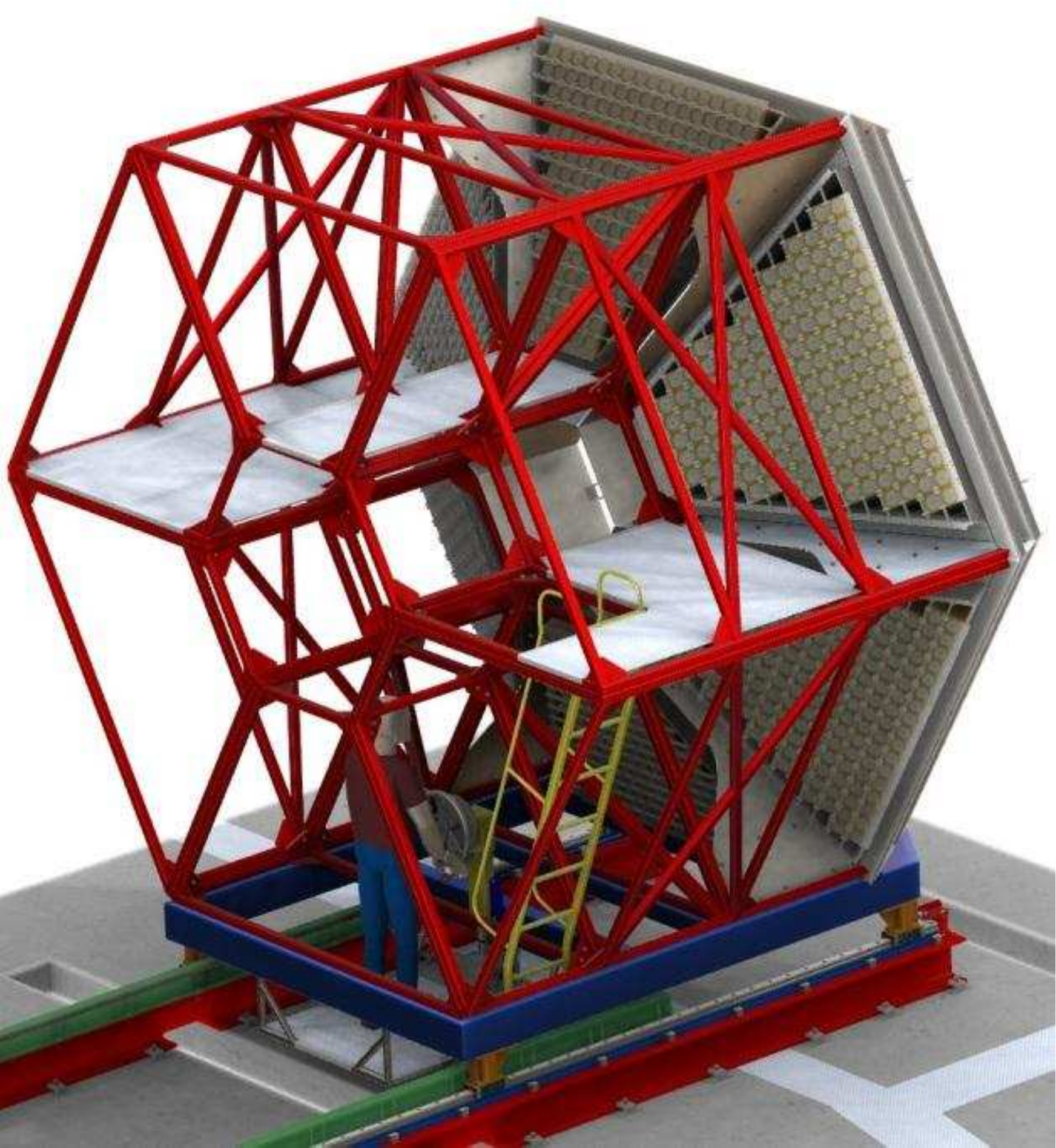}
\caption{Final assembly of the drive system and detector modules.}
\label{mech9}
\end{minipage}
\end{figure}

\begin{figure}[ht]
\begin{minipage}[b]{0.42\linewidth}
\centering
\includegraphics[scale=0.24]{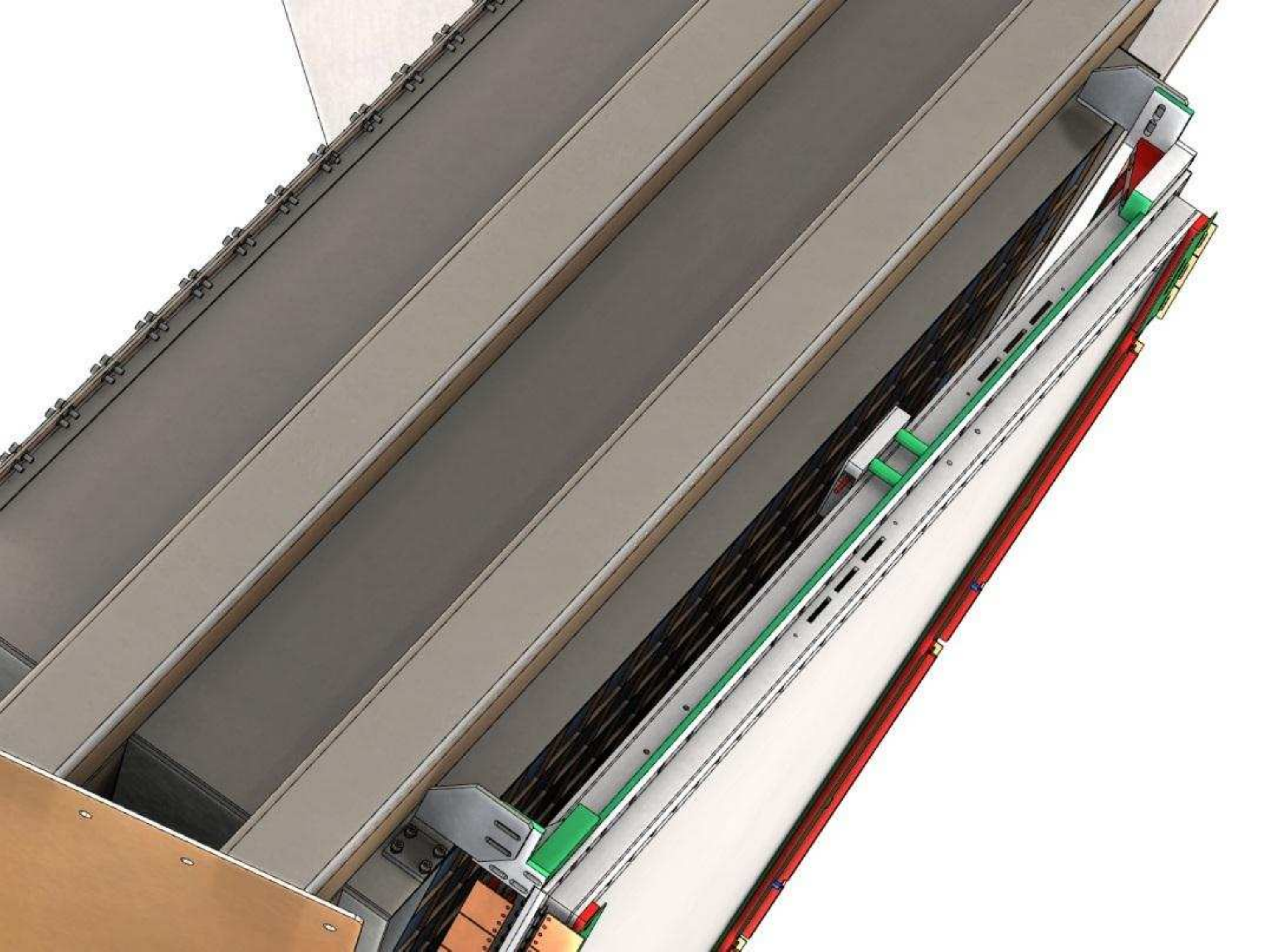}
\caption{Detail of the top mounting of the RPC.}
\label{mech11}
\end{minipage}
\hspace{0.3cm}
\begin{minipage}[b]{0.45\linewidth}
\centering
\includegraphics[scale=0.25]{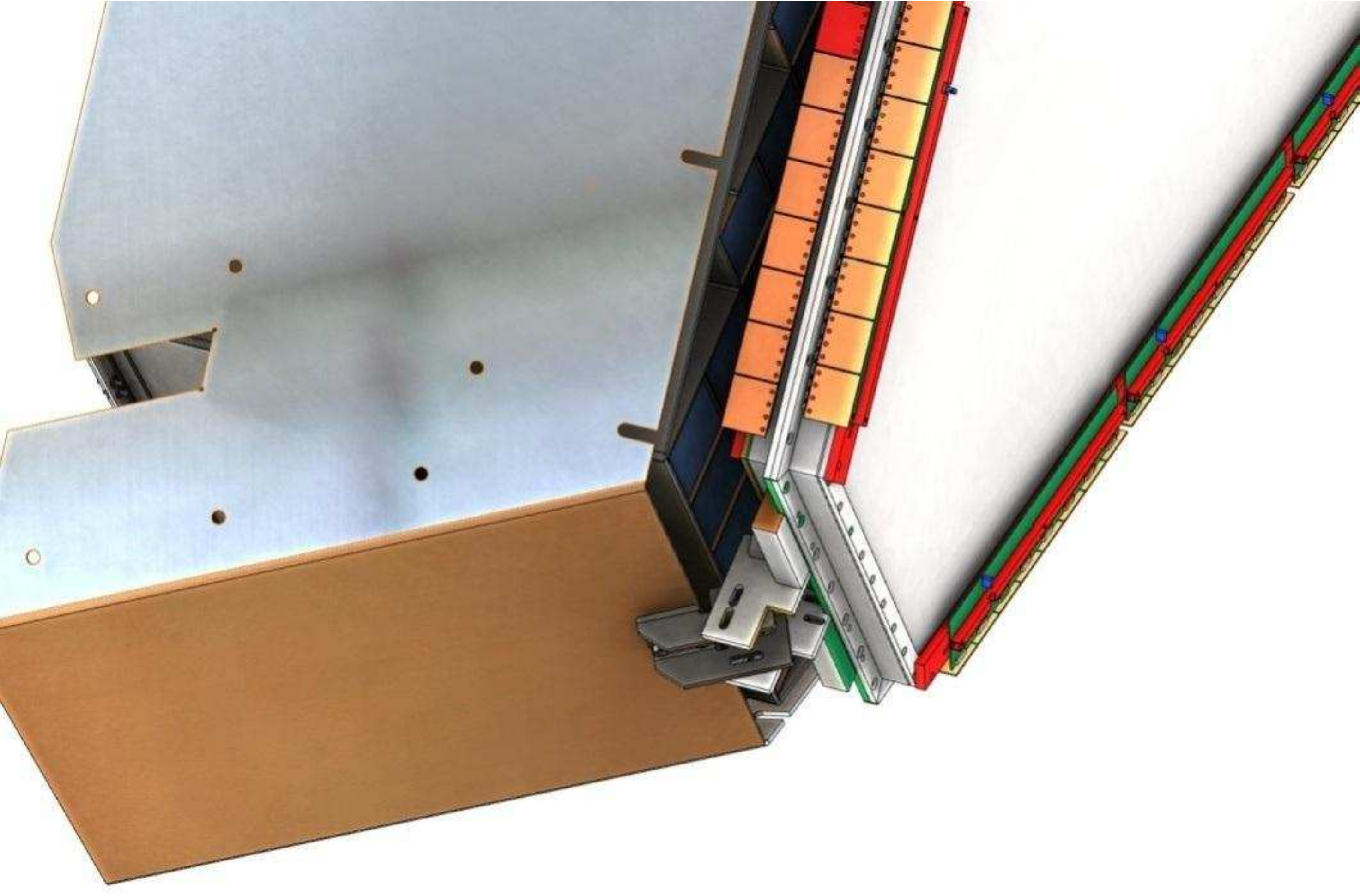}
\caption{Detail of the bottom mounting of the RPC.}
\label{mech12}
\end{minipage}
\end{figure}

\begin{figure}[!htb]
\begin{center}
\includegraphics[angle=0,scale=0.51]{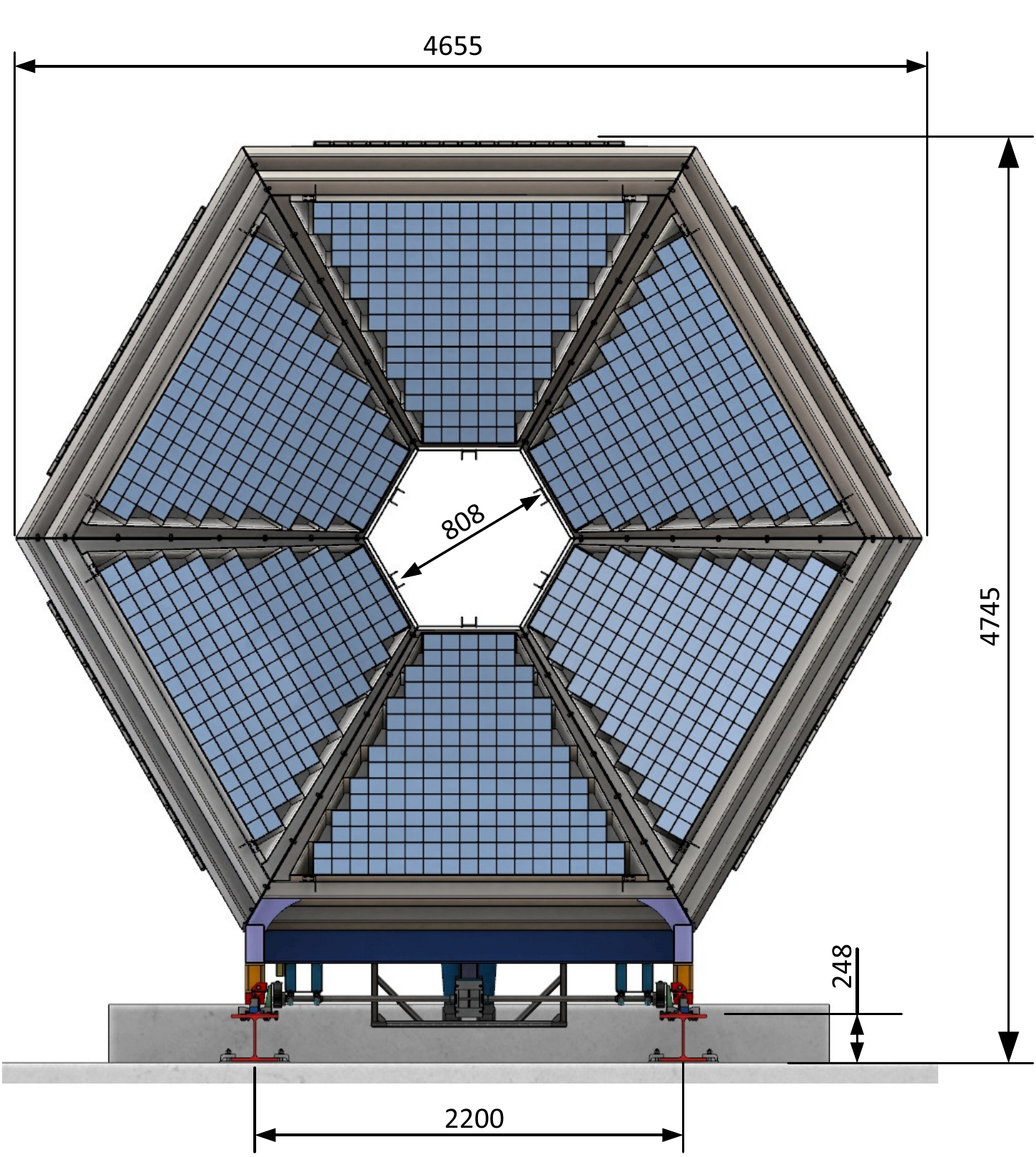}
\caption[]{Dimensions of the HADES electromagnetic calorimeter, front view.
\label{mech15} }
\end{center}
\end{figure}

\begin{figure}[!htb]
\begin{center}
\includegraphics[angle=0,scale=0.61]{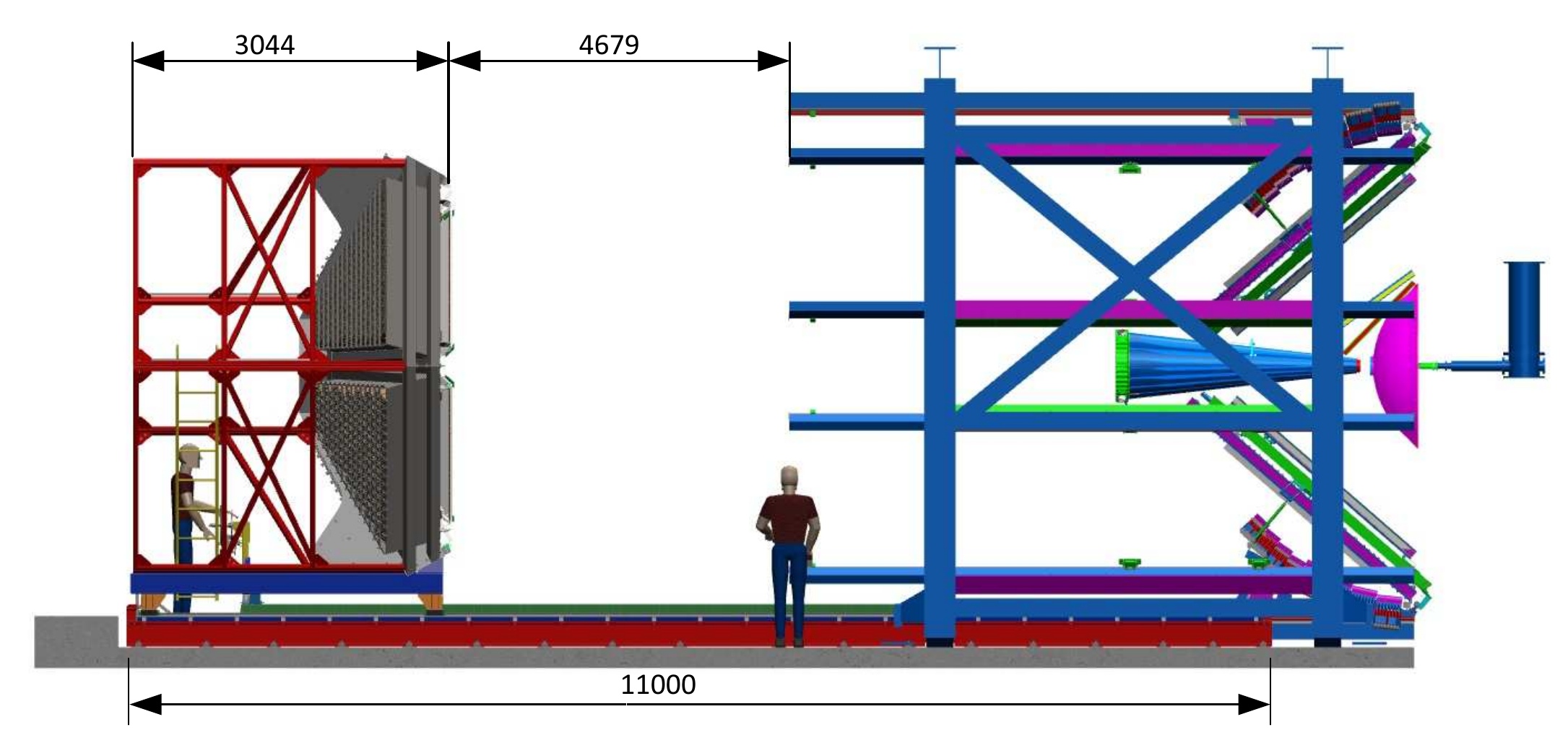}
\caption[]{Dimensions of the HADES electromagnetic calorimeter, side view.
\label{mech16} }
\end{center}
\end{figure}

\begin{figure}[!htb]
\begin{center}
\includegraphics[angle=0,scale=0.41]{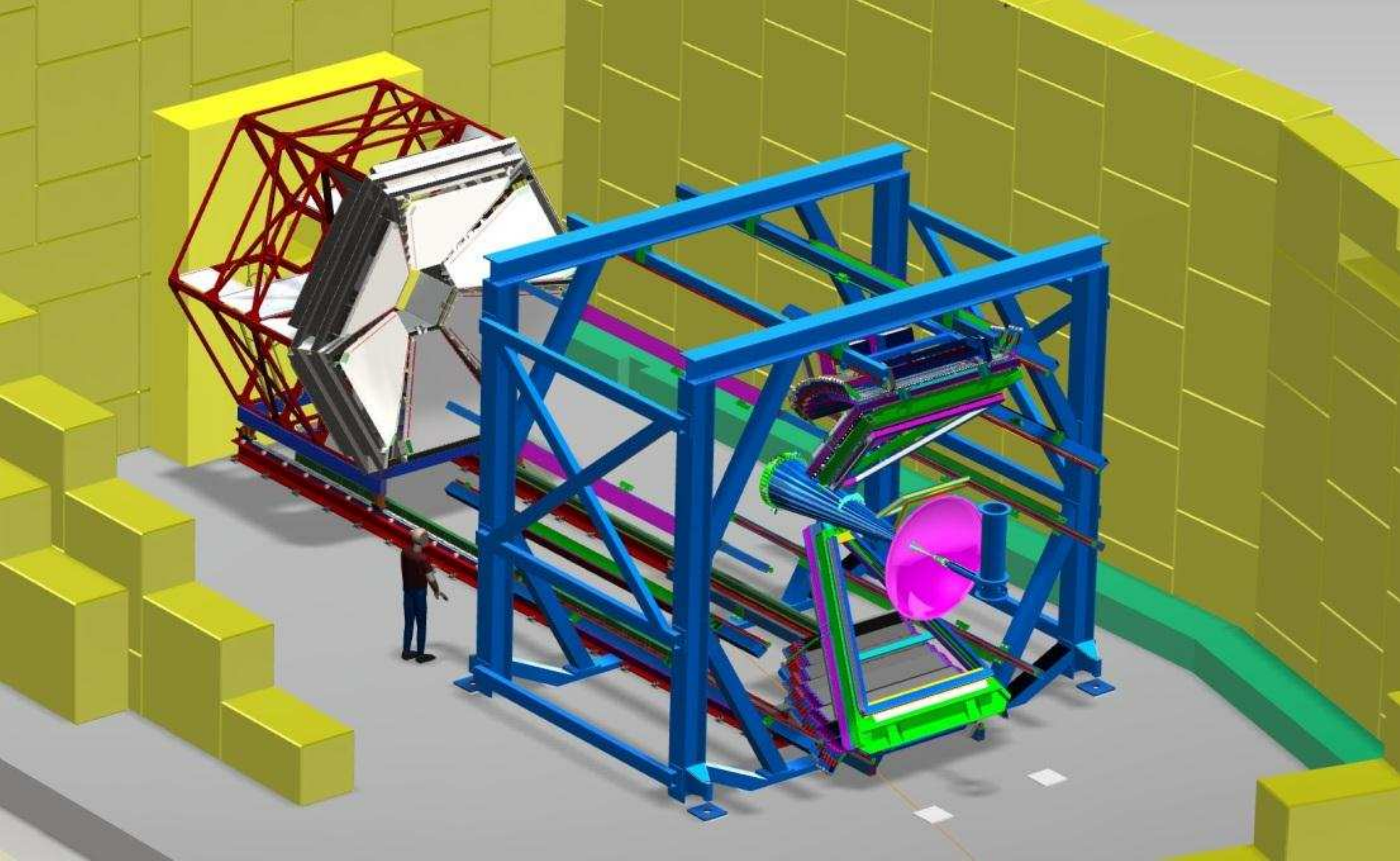}
\caption[]{3D model of the construction -- calorimeter maximally pulled out of HADES in the HADES Cave.
\label{mech14} }
\end{center}
\end{figure}

The main aim of the mechanical construction is to support the modules of calorimeter and RPC detectors, keeping them in the proper position and locating them along the axis of the beam for precise positioning. The structure was determined by positioning the 6 modules array such to cover the polar angle from 12$^{\circ}$ to 45$^{\circ}$, as shown in Fig.~\ref{mech1}. The spatial arrangement of the glass modules is shown in Fig.~\ref{mech2}. The calorimeter is composed of 978 modules, arranged in the 6 trapezoidal sectors. Each sector include 15 layers with modules. The mass of each module is approximately 15 kg that leads to an overall mass of about 15 tones.
To mount the modules a mobile platform is required. Easy movement and positioning will be provided by using linear bearing which have low motion resistance, high precision and repeatability of position. Thanks to this technology the platform can be man operated even when highly loaded. A view of the platform with description of elements is shown in Fig.~\ref{mech3}.
Each of the six sectors is hosted by a container made of stainless steel 1.4301 (X5CrNi18-10) acc. EN 10088-3 (Fig.~\ref{mech4}). The container is mounted to the steel truss made of steel S355J2, acc. EN-10025-2 as shown in Fig.~\ref{mech5}. This operation is repeated for all six sectors.
Due to large dimensions and weight of the construction, this will be transported in segments and assembled on the experimental site. The construction is dismountable in case of moving to another site. Installation of the modules will be available after assembling the construction. The final assembly of mechanical construction is shown in Fig.~\ref{mech8} and \ref{mech9}. 
Once the 6 segments are in place, the RPC detector has to be mounted. The mounting of the RPC is conducted in the following way: there is a gap of about 20 mm between the front plane of the glass modules and the back plane of the RPC for allowing airflow. Both top and bottom fixture points enable adjustment of position. The top and bottom mounting points of the RPC to the calorimeter segment for one sector are shown in Fig.\ref{mech11} and \ref{mech12}.
The final dimensions of the HADES calorimeter mechanical construction are presented in Fig.\ref{mech15} and \ref{mech16} in a front view and side view, respectively.
The maximum height and width of the construction are 4745 and 4655 mm, respectively. The construction is placed on the platform, which moves along the rails located at a height of 248 mm, where 240 mm  and 8 mm are the height of double-tee bar profile and of the pad of the profile, respectively. In the central part of the calorimeter there is a hexagonal hole which allows the insertion of a carbon pipe. The distance of the innermost sides of the hexagon is 900 mm. However, the clearance is reduced to 808 mm, because of mounting of the RPC. The length of rails mounted to move the Glass Calorimeter is 11 000 mm, which after pulling completely out  the calorimeter from the HADES frame provides a free space of 4679 mm in-between (Fig~\ref{mech16}).
A mating study has been carried out to avoid collision points between HADES and the glass calorimeter.
 At the design stage, an assembly of glass calorimeter with the HADES frame was made. Potential collisions were detected and removed. The final assembly of glass calorimeter in the HADES frame together with the HADES Cave is presented in Fig.~\ref{mech14}. This picture corresponds to the situation when the glass calorimeter is maximally pulled out. As it is shown in this figure, the space of HADES Cave has been fully utilized.
\subsection{Stress and deflection}
The aim of finite element (FE) analysis was to check the level of stresses and deflections of the construction under load. 
The material of the supporting structure is
stainless steel 1.4301 acc. EN 10088-3 and has the following properties:
\begin{itemize}
\item 	Yield stress, Rp0.2 = 190 MPa, 
\item 	Tensile stress, Rm = 500-700 MPa, 
\item 	Elongation, Amin = 22 \%, 
\item 	Modulus of Elasticity, E = 193 000 MPa 
\end{itemize}
Material of other components: 
steel S355J2 acc. EN-10025-2, with the following properties:
\begin{itemize}
\item   Yield stress, Rp0.2 = 355 MPa
\item   Tensile stress, Rm = 470-630 MPa, 
\item   Elongation, Amin = 45 \%, 
\item   Modulus of elasticity, E = 220 000 MPa
\end{itemize}
\begin{figure}[!htb]
\begin{center}
\includegraphics[angle=0,scale=0.51]{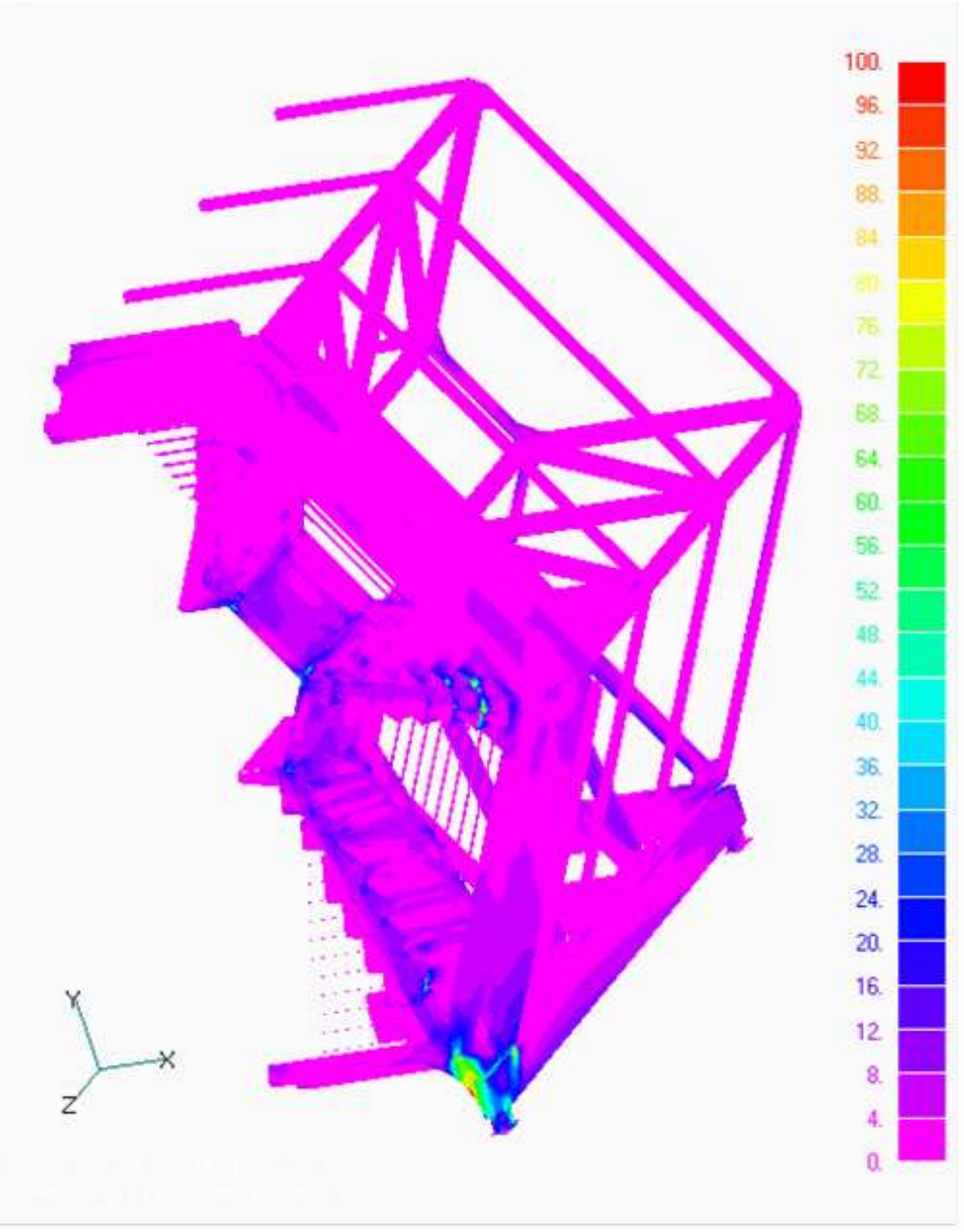}
\caption[]{Maximal von Mises stress in MPa.
\label{mech20}}
\end{center}
\end{figure} 
\begin{figure}[!htb]
\begin{center}
\includegraphics[angle=0,scale=0.41]{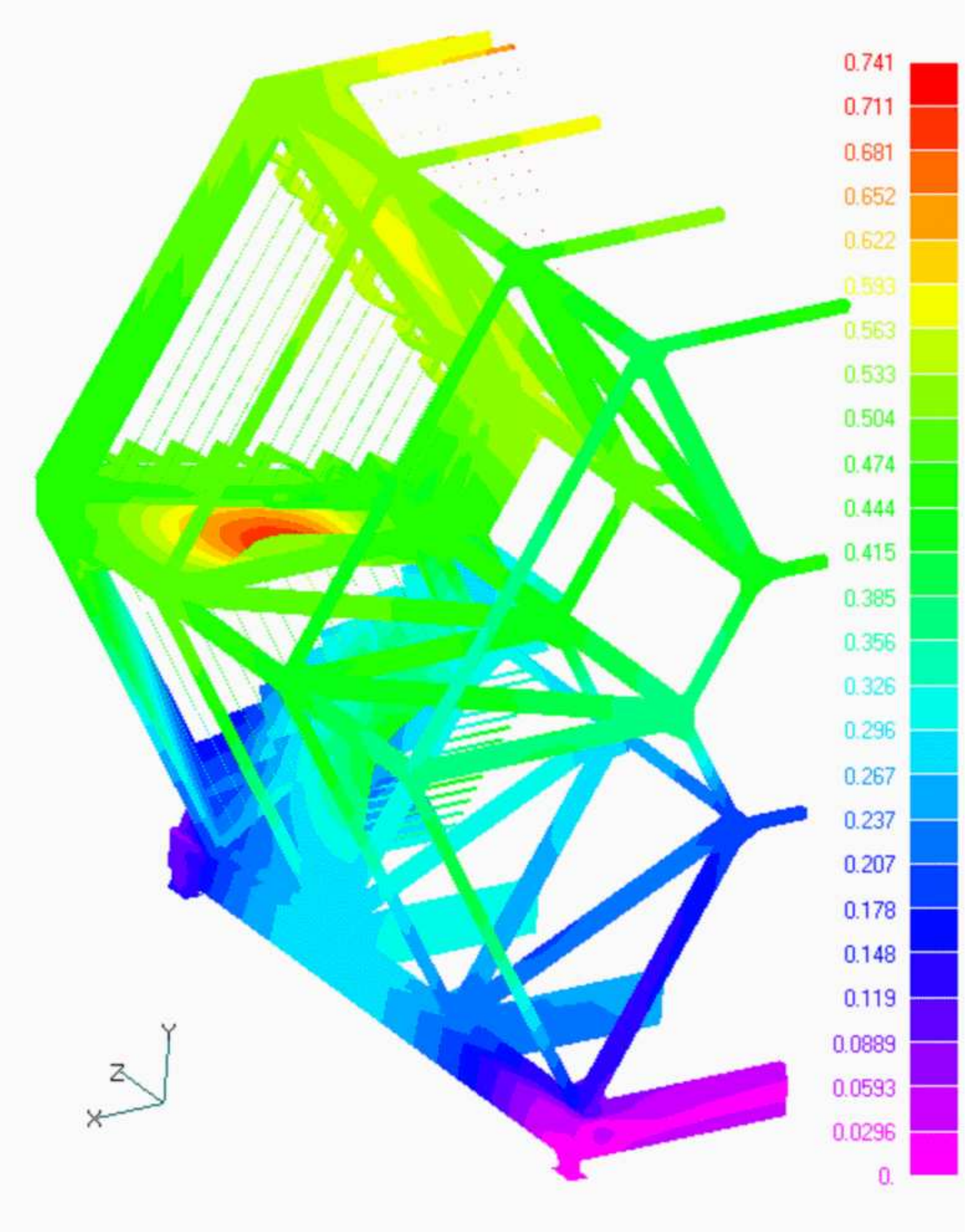}
\caption[]{Total translation im mm, rear view.
\label{mech21} }
\end{center}
\end{figure} 
To conduct FE calculations, a FE model was prepared with the MSC/Nastran program for Windows package. The model was created with the use of plate elements, which were generated on the  mid-surfaces of the structure components.
The boundary conditions were defined in the following way. Note that the cart is the carriage operating on the rail:	
\begin{enumerate}
\item	 No translations along X axis in carts area. 
\item  No translations along Y axis (vertical direction) in carts area.
\item  The interaction between segments for supporting glass was included.
\item	 The weight of the whole structure was taken into account
\item  Only half of the model was used in FE calculations.
\end{enumerate}
The stress values and the displacement have been obtained by the FE calculations. The outcoming 'von Mises' stress and displacements along particular axis are shown in Fig.~\ref{mech20} und ~\ref{mech21}. A maximal 'von Mises' stress value of 118 MPa  is predicted in correspondence of the area where the cart is supported on the rail (see Fig.~\ref{mech20}).The total translation in the three directions X-Y-Z is presented in Fig.~\ref{mech21}.
\subsection{Summary}
The designed mechanical support of the calorimeter allows to install 978 modules in six containers, that guarantees a coverage in the laboratory polar angle between 12¡ and 45¡. The overall mass of the construction with modules installed would be approximately 20 tons. An applied system of linear bearing guidance ensures low resistance of motion. This allows to precise move and position along the axis of the stream using the reduction gear with racks installed parallel to the tracks. The compact design and tracks on the entire length of the room allows to minimize the occupied space, which is necessary because of the mounting of other detectors or equipment.
The mass of glass modules loads mainly the front part of the construction. Therefore, maximal stresses appear in this part. However, the FE results show that permissible stress values are not exceeded and translations which appeared in the structure are less than 0.8 mm.

%% file: Crystals/Crystals.tex
\section{Module Properties}

\begin{figure}[!htb]
\begin{center}
\includegraphics[angle=0,scale=0.51]{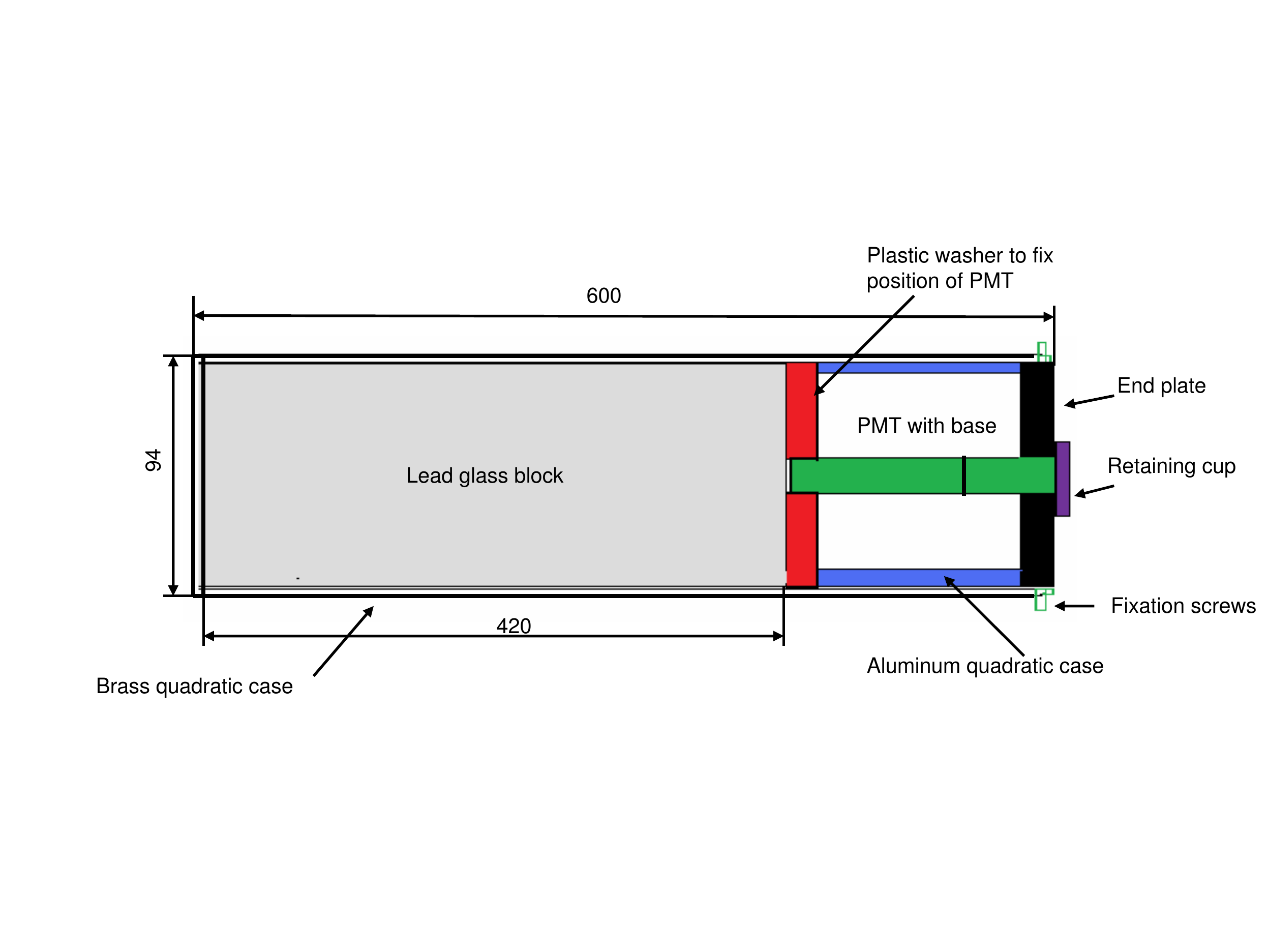}
\caption[]{Schematic view of the calorimeter module.}
\label{Kaon_mass} 
\end{center}
\end{figure}

The main element of the HADES EMC is the modified module from the OPAL end cap electromagnetic calorimeter \cite{akr90}. Lead glass CEREN 25 \cite{lev60} is used as a Cherenkov radiator in this module. The lead glass has the density 4.06 g/cm$^3$, refractive index of 1.708 (at 410 nm) and a radiation length (X$_0$)  of 2.51 cm. About 1000 OPAL modules were selected with the same length of 420 mm (16.7 radiation lengths) for the HADES EMC. The transmission coefficient for this lead glass length is about 0.96 at 400nm. Each lead glass block has transverse sizes of 92x92 mm$^2$ that are comparable to the transverse size of the electromagnetic showers.
A schematic view of the module is shown in Fig. \ref{Kaon_mass}. The lead glass block is housed in a brass can. The wall thickness of the can is 0.45 mm. All surfaces of the lead glass blocks are mirror polished. All sides of the block (excluding the end to which the PMT is attached) are wrapped with white paper (tyvec). The intended study confirmed that this reflector in combination with the optical grease (RHODORSIL Pate No. 7) between PMT and lead glass improves the energy resolution for about 10\% in comparison with aluminum or mylar reflector materials.  The position of the PMT on the rear side of the lead glass block was fixed by the plastic plate with the hole at the center. The PMT together with HV divider is fixed by retaining cup on the end plate.  The end plate is connected to the aluminum quadratic ensuring that the mechanical structure of the module is more rigid. 
The can provides the mechanical support of the assembly via the end plate of the can that is secured by a rigid back plate. This support allows the individual assembly to be withdrawn from the array without  removing or displacing its neighbors. The lead glass wrapping, the brass can and the space between the counters add additionally 2 mm in the transverse direction. Therefore the counters are mounted on a 94 mm pitch.

There are several options for the PMTs. At the present the 1.5'' EMI 9903KB 
PMTs are used to read-out the prototypes. About 600 such PMTs, tested
with a small size Na(Tl) scintillator and a radioactive source, are available 
for the calorimeter. The diameter is rather small to collect light from the 
rear side  with size of 92x92 mm$^2$. However, tests with larger size 
Hamamatsu H1949 PMT  (2'' diameter) did not show significantly better energy
resolution, as shown in section \ref{testRes}. Also resolution $> 5\%$
at 1 GeV was achieved in the OPAL barrell detector using the 
lead-glass modules equipped with  3'' phototubes \cite{opal_barrell}. 
It seems that other factors like fluctuations in the shower propagation 
limit the energy resolution.  
The tests will be continued to select the optimal 
PMT type. At present 35 pices of 3'' Hamamatsu R6091 PMTs have beed 
ordered to test the performance, hence still 373 are missing (when 
counting 30 pieces as spares).

As for the HV power supply, we propose to use the CAEN SY1527LC HV system
based on the 24 channels A1535 modules providing voltage up to 3.5 kV and
current up to 3 mA. At the present, more than 300 channels are available for
the project.

%% file: Electronics/FE_Electronics.tex
\section{Electronics and Read-Out}
The lead glass calorimeter read-out electronics is based on separate time and energy measurements. The input signal is split on the analog front-end board into fast and slow, shaped paths. The fast signal is delivered to a TDC (Time to Digit Converter) on a new development of the TRB \cite{Fro08}. This TRB3 will have up to 256 channels TDCs with 10 ps time resolution RMS between two channels, 10ns dead time, and  hit buffer for hit storage. The amplitude of the shaped signal is measured by an ADC (Analog to Digital Converter) on the add on card attached on the TRB3. The add on card designed for the Shower detector is considered.
The following parameters have been assumed for the calorimeter PMT signals:
\begin{itemize}
\item  Rise time: 3 ns
\item Falling time: 50 ns
\item S/N ratio:	$>$12.5
\item Pulse amplitude (20 MeV):		50 mV
\item Pulse amplitude (600 MeV):		1.5 V
\end{itemize}
The required read out parameters are:
\begin{itemize}
\item Time resolution: $<$ 500 ps
\item Dynamic range of signal amplitude (energy): 5 mV - 5 V
\item Accuracy needed for the energy measurement: 5 mV
\end{itemize}
Each calorimeter sector will deliver 163 independent signals for a total of 978 signals. The expected hit rate amounts to 10 kHz.
\subsection{Time and energy measurement}
\paragraph{Front-end board}
The front-end board comprises twenty four identical channels. The description of one channel follows.
Since the calorimeter signal may have high amplitude, it has to be attenuated before further processing. The attenuated signal is then split into the fast path and the slow, integrating path. A fast discriminator is used to get the timing signal for the TDC. The ADCMP604 from Analog Devices is considered. The discriminator delay is 1.6 ns. The input signal walk will be compensated off-line making use of the measured amplitude level. The discriminator has a differential output working with the LVDS standard. In the slow path the signal is integrated, converted to differential and sent to the ADC for the amplitude measurement. The peaking time of the integrator is 400 ns. The AD8099 and AD8139 amplifiers from Analog Devices are proposed.
The discriminator threshold is set by a 10-bit DAC. The output voltage step is 3.2 mV. The DAC is set externally via slow control lines entering the front-end board.\\
A small programmable digital integrated circuit is needed as a glue logic. Two 80-way cables are foreseen to deliver time and energy signals to the digitizing board. The front-end board is supplied with $\pm$ 5 V.

\paragraph{The digitizing board}
The TRB3 recently developed for HADES, PANDA (Barrel-DIRC and Straws) and CBM (MVD readout) is foreseen as the digitizing board for the Calorimeter. TRB3 will be used for a time measurement. Four identical mezzanine cards equipped with 10-bit ADCs will be used for the amplitude measurement. The TRB3 card with four mezzanine cards will have 192 (4*48) time and amplitude measurement channels. The digitizing board controls the parameters set on the front-end boards by separate slow control lines.

\paragraph{DAQ System}
After digitization of the signals by the TRB3 and the ADC card the data is combined to HADES-sub-events and sent to the HADES DAQ-system physically via optical links.  The TrbNet protocol will be used for LVL1 trigger and slow control. Gb/s Ethernet will be used for the transport of the data to the event-builders. This HADES development is finished and can be reused for the calorimeter.

%% file: Aging/Aging.tex
\section{Aging Studies}

Effects of radiation damage and recovery of CEREN 25 lead glass.

It is well known that the lead glass is damaged by ionizing radiation that creates the color centers \cite{lev60} and causes the strong absorption of light in the near ultraviolet, visible and near infrared spectra.  Usually UV light and temperature cycling was used to anneal the radiation damage in lead glass \cite{gol73}.
Detailed investigation of the radiation damage effects in CEREN 25 glass was done by OPAL collaboration \cite{akr90}. Two full-size blocks of CEREN 25 with the length 520 mm were irradiated with the total absorbed dose of 10 Gy. The transmission of each block was studied with a blue light-emitting-diode, LED and compared with that of third block that was not irradiated. The relative transmission, T is the ratio of the optical transmission after and before the irradiation. Immediately after the irradiation, the relative transmissions of two blocks were measured to be 29\% in both cases. The first block was stored at $0^\circ$ C for 4 days and after that the relative transmission increased to 42\%. Then during 24 days the block stayed at the room temperature ($25^\circ$ C) and T rose to 60\%. Following further 16 hours at $120^\circ$ C the relative transmission recovered to 98\%.
The transmission of the second block was gradually increased up to 50\% at room temperature during 50 hours after the irradiation. Then the block was illuminated with a 125 W
Hanova fluorescence lamp and T was recovered to 91\% in 20 days. More details about this study can be found in \cite{akr90}.
Taking into account the lead glass radiation damage it would be necessary to estimate the full absorbed dose for the modules in the HADES environment at SIS 100.

%% file: Simulation/Simulation.tex
\section{Simulation of the lead-glass calorimeter}
In order to assess the performance of the proposed electromagnetic calorimeter (EMC) design a dedicated simulation and reconstruction software was developed.

\subsection{Implementation in HGeant}

A calorimeter geometry was implemented in the HGeant code (GEANT3-based simulation framework used by the HADES collaboration) in accordance with the engineering drawing. EMC is planned to be installed approximately on a place of the SHOWER detector, directly behind the RPC planes. 142 individual calorimeter blocks are bound in one sector. Compared to the considered technical design one row of modules at high polar angles is missing.
Each individual block consists of a lead-glass crystal (92x92x420 mm) wrapped in a mylar foil and encased by a brass shielding. So far no support structures were implemented in the current version of the calorimeter geometry.

\subsection{Tracing of Cherenkov photons}

The simulation of the lead-glass calorimeter response can be split into two independent parts: 1) shower development in a module (production of secondary particles and Cherenkov photons) and 2) transport of Cherenkov photons. Note, that since individual calorimeter blocks are optically isolated, a Cherenkov photon is confined to the calorimeter block in which it was produced.

The transport of Cherenkov photons is a complex procedure that involves a set of processes: 
absorption of the photons inside the crystal volume and on the mylar wrapping, specular and diffuse reflection and refraction.

The input parameters for this part of the simulation are the following:
\begin{enumerate}

\item Refraction index of the lead-glass CEREN 25 (SF-5 in other nomenclature), $n = 1.708$;

\item Light attenuation length in a lead-glass crystal as a function of a Cherenkov photon energy;

\item Reflectivity of the mylar wrapping (specular and diffuse); 

\item Refraction index of the optical grease, connecting the crystal and the PMT, $n = 1.5$;

\item Diameter ($d = 3.6$~cm) and quantum efficiency of a PMT (EMI 9903B). 

\end{enumerate}
 
During  the testing of the simulation code, it was found out that GEANT3 fails to reproduce the expected energy resolution of the individual calorimeter block and its dependence to the energy of the incident particle ($\approx 5\%/\sqrt{E/\mathrm{GeV}}$). The reason of this malfunction was traced back to the Cherenkov photon transport procedure.
To solve this problem a stand-alone Cherenkov photon transport code, developed by M.~Prokudin, was employed \cite{Arefev:2008zz,Alekseev:2008zza}.
This code was demonstrated to describe the Cherenkov photon propagation correctly even for the very complicated calorimeter geometry. 

Thus, the development of an electromagnetic shower \emph{and} the Cherenkov photon production is performed by GEANT3, whereas the Cherenkov photon transport is delegated to a stand-alone program.

The tuning and testing of the simulation code was based on two experimentally known benchmarks: 1) the energy resolution of $\approx 5 \%/\sqrt{E/\mathrm{GeV}}$) and 2) as was shown by tests with cosmics (see Section \ref{testRes}), the response of an individual block (photoelectron yield) to a normally incident cosmic muon is equivalent to the response to a photon with an energy of 580 MeV. 

Due to the large Cherenkov photon multiplicity---around a thousand produced photons per 1 GeV deposited energy---and the complexity of the processes involved, the photon transport demands a lot of CPU time. In order to make the simulation faster a look-up table approach was developed. A look-up table contains the averaged probabilities of the photoelectron production for the Cherenkov photon with a given set of parameters. In our case a 4-dimensional table was prepared, taking as relevant parameters the $z$-coordinate and transverse coordinate $t=\sqrt{x^{2}+y^{2}}$ of the photon production point relative to the center of the crystal, the photon emission angle $\theta$ and the photon energy $\epsilon$. 

To prepare the look-up table $3 \cdot 10^{9}$ trial photons were generated and traced. Their distribution was optimized to provide more statistics in those regions of the chosen parameter space ($z, t, \theta, \epsilon$) that are mostly populated by the Cherenkov photons produced in a typical electromagnetic shower.

\begin{figure}
  \centering
  \begin{minipage}[b]{8 cm}
  \includegraphics[width=8.4cm]{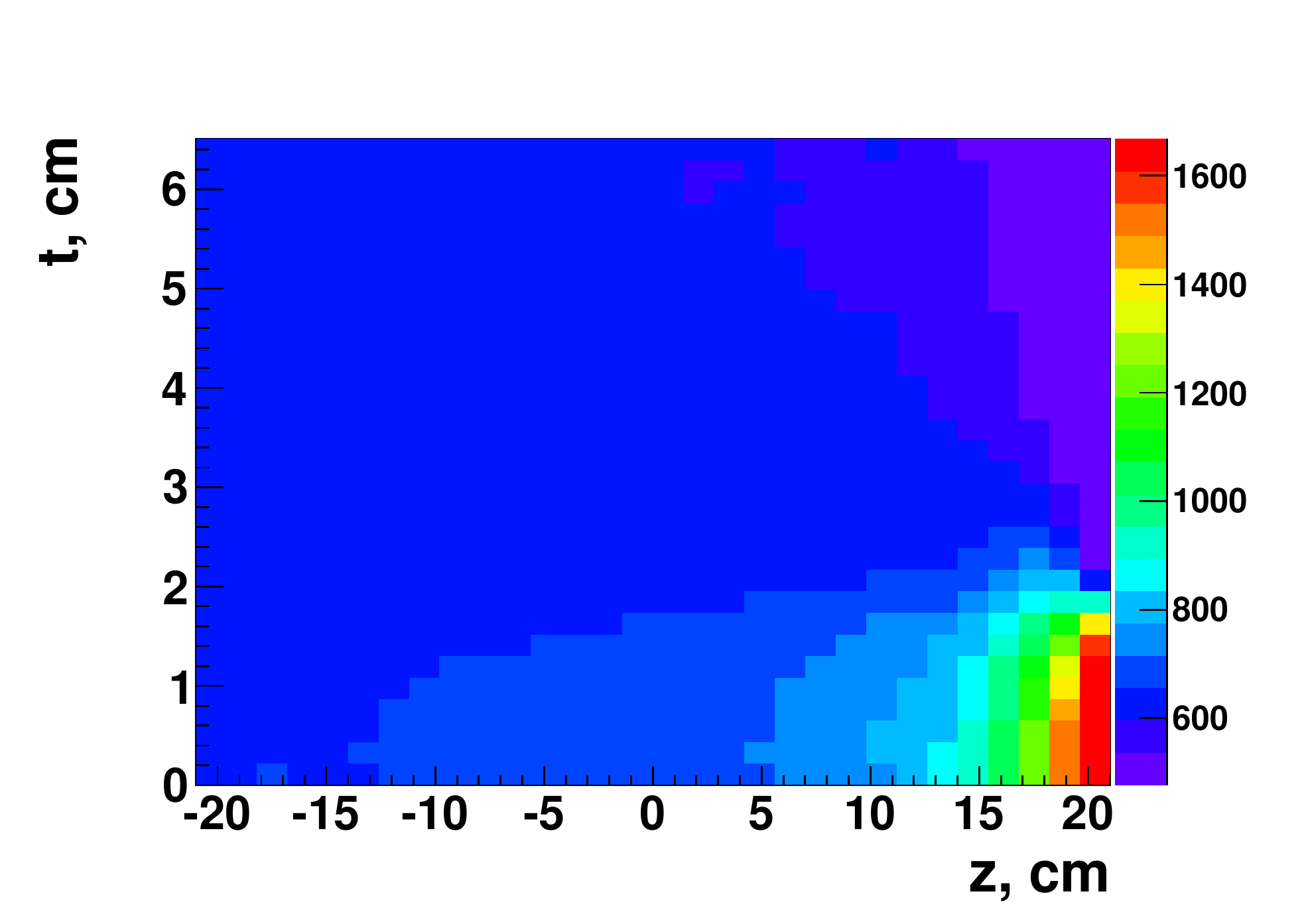} 
  \end{minipage}
  \begin{minipage}[b]{8 cm}
  \includegraphics[width=8.4cm]{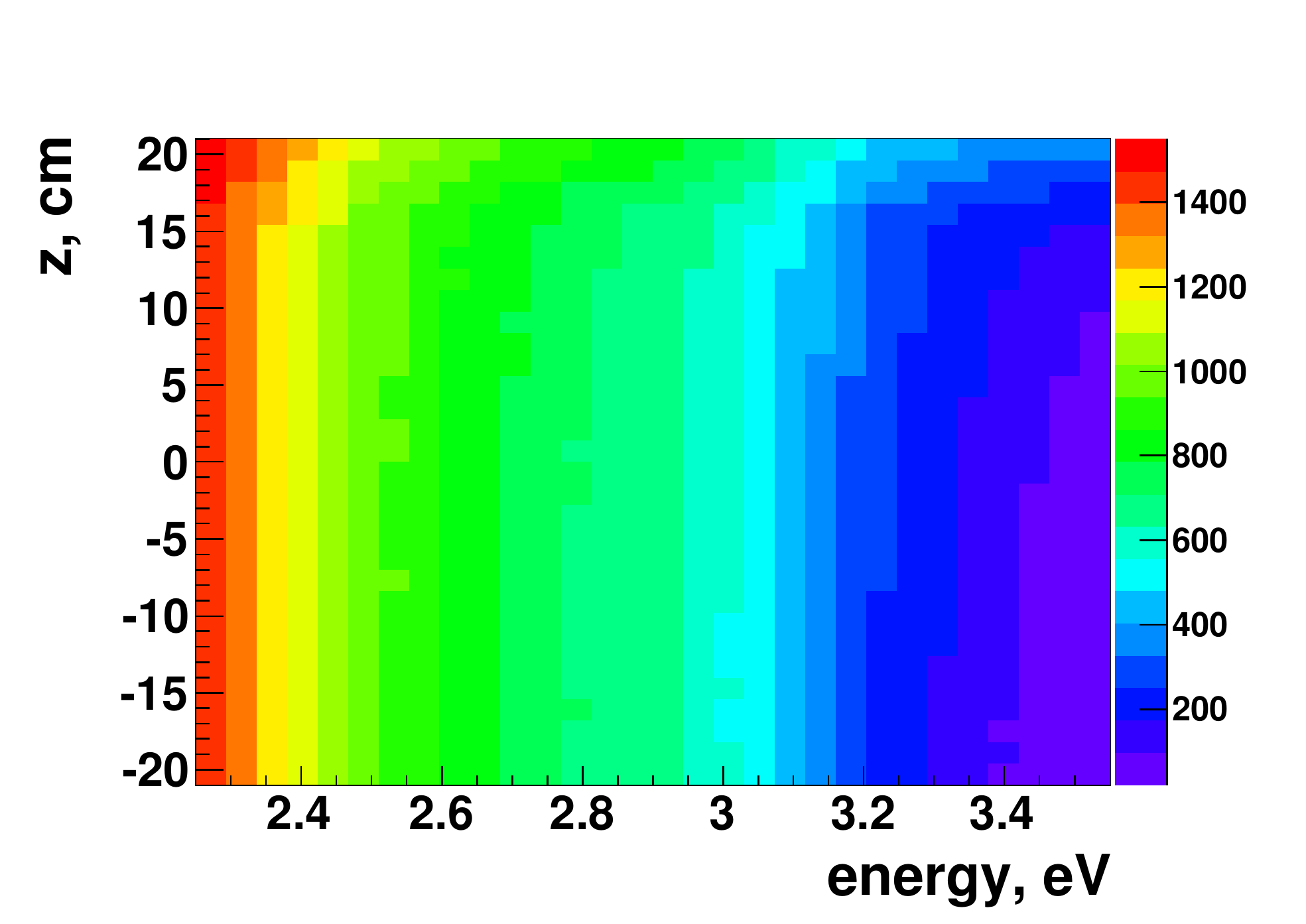} 
  \end{minipage}
  \caption{Projections of the 4-dimensional look-up table onto $z-t$ (left) and $\epsilon-z$ (right) axes. The z-axis shows the cell content, that is proportional to the photoelectron production probability.}
  \label{fig:proj_lookup}
\end{figure}

Two-dimensional projections of the obtained 4D look-up table are presented on Fig.~\ref{fig:proj_lookup}. The cell content (indicated by the color scale on the z axis) is proportional to the photoelectron production probability. One can clearly see that, as expected, the probability to produce a photoelectron is highest for the Cherenkov photons produced close to the PMT entrance window (large $z$ and small $t$). The right part of the Fig.~\ref{fig:proj_lookup} reveals the dependence of the light attenuation length on the photon energy.

As a final step, it was checked, that the 4D look-up table approach is able to reproduce the results obtained with the full transport of Cherenkov photons both for photons and hadrons bombarding the calorimeter block at different angles.
\subsection{Occupancy considerations}

The occupancy of the calorimeter modules was studied with simulations taking as an input $^{56}${Ni}+$^{56}$Ni collisions at 8~AGeV and impact parameter range $b<1$~fm generated with the UrQMD transport code. The choice of the collision system and the impact parameter range ensures the highest possible multiplicity conditions to be studied with the HADES set-up.  

\begin{figure}[htbp] 
   \centering
   \includegraphics[width=15cm]{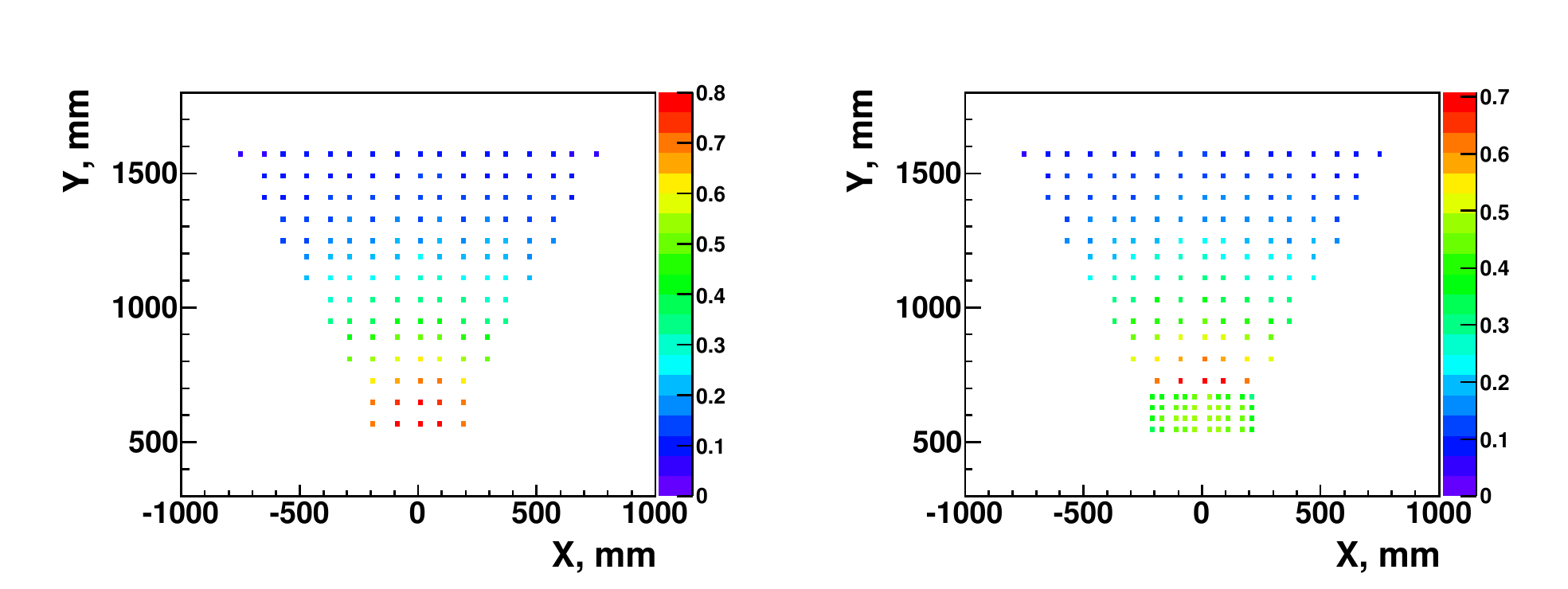}
   \caption{Average module occupancy in Ni~+~Ni collisions at 8~AGeV beam kinetic energy and impact parameter $b < 1$~fm for the standard (left) and increased granularity (right) calorimeter setup}
   \label{fig:occupancy_15}
\end{figure}

Fig.~\ref{fig:occupancy_15} (left) illustrates the event-averaged calorimeter cells occupancy. A cut on the minimal energy deposition of 15 MeV was applied. Squares correspond to the centers of the calorimeter modules. In the investigated conditions the highest occupancy of 0.8 is observed for the modules in the two innermost rows. 

In order to cope with the high occupancy at low polar angles, a high-granularity version of the calorimeter geometry was considered in which the lead-glass crystals in the two innermost rows have lateral dimension that is one fourth of the usual crystal size. Right part of the Fig.~\ref{fig:occupancy_15} illustrates the occupancy for the high-granularity version. In this case it drops to the value of 0.4 for the small modules.

In the following, results for the electron-pion separation and reconstruction of neutral mesons will be presented, as obtained in the standard version of the calorimeter geometry.

\subsection{Electron-pion separation}
An important function of the calorimeter is to improve the electron(positron)-pion separation at high values of particle momenta ($p > 500$~MeV$/c$). To achieve this, a cut on the $E_{EMC}/p$ ratio is employed, where $E_{EMC}$ is the energy, deposited by a particle in the calorimeter and $p$ --- the particle momentum, measured by the HADES tracking system. 

As an input for the simulation events generated with the UrQMD program were used: $^{12}$C+$^{12}$C and $^{56}${Ni}+$^{56}$Ni collisions at a kinetic beam energy of 8~AGeV.


As a first step, the coordinate of a track intersection with the calorimeter plane is calculated. The track is parameterized as a straight line crossing MDC planes III and IV. The finite spatial and momentum resolution of the tracking system is taken into account.


The $E_{EMC}$ is then calculated as an energy deposition in a 2 x 2 crystal matrix formed around a track-calorimeter intersection point.

\begin{figure}
   \centering
   \includegraphics[width=14.5cm]{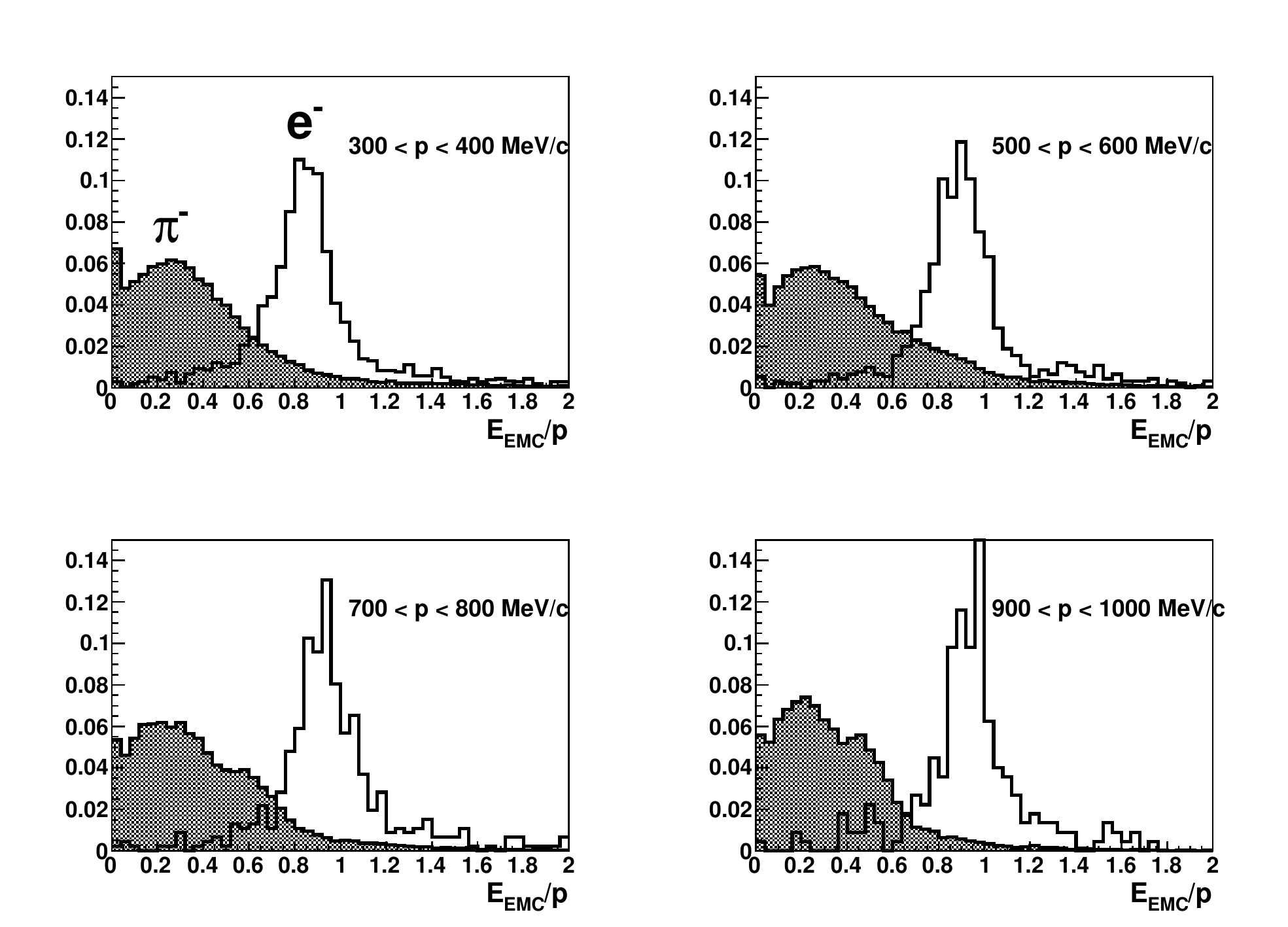} 
   \caption{$E_{EMC}/p$ distributions for electrons (unfilled histograms) and $\pi^{-}$ (shaded histograms) obtained in Ni+Ni collisions at impact parameter $b<2.5$~fm.}
   \label{fig:em_pim_NiNi}
\end{figure}

As an example, the $E_{EMC}/p$ distributions obtained with described procedure for Ni+Ni collisions at an impact parameter $b<2.5$~fm for different regions of the particle momentum are shown on Fig.~\ref{fig:em_pim_NiNi}.

\begin{figure}
  \centering
  \begin{minipage}[b]{8 cm}
   \includegraphics[width=8.2cm]{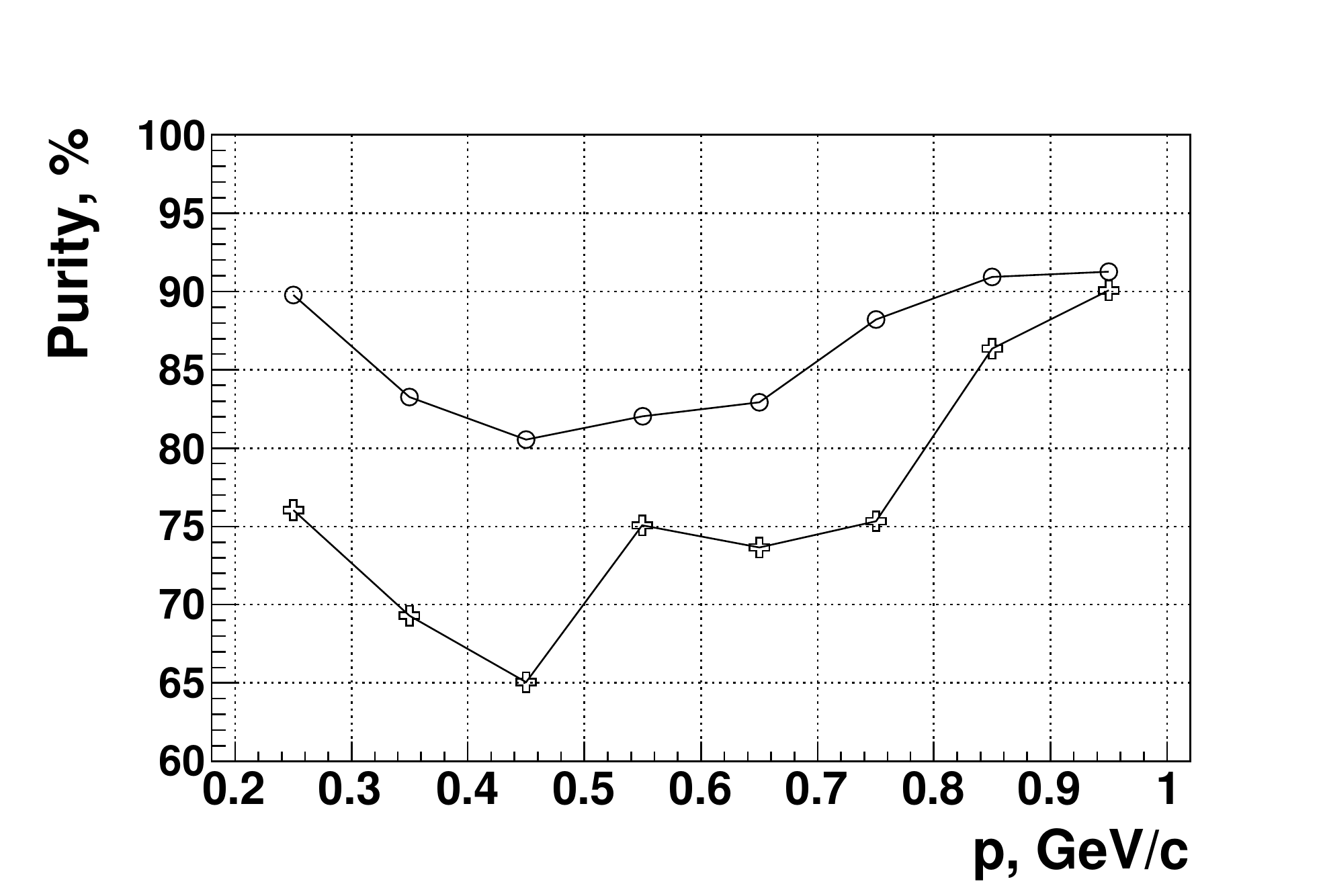} 
  \end{minipage}
  \begin{minipage}[b]{8 cm}
   \includegraphics[width=8.2cm]{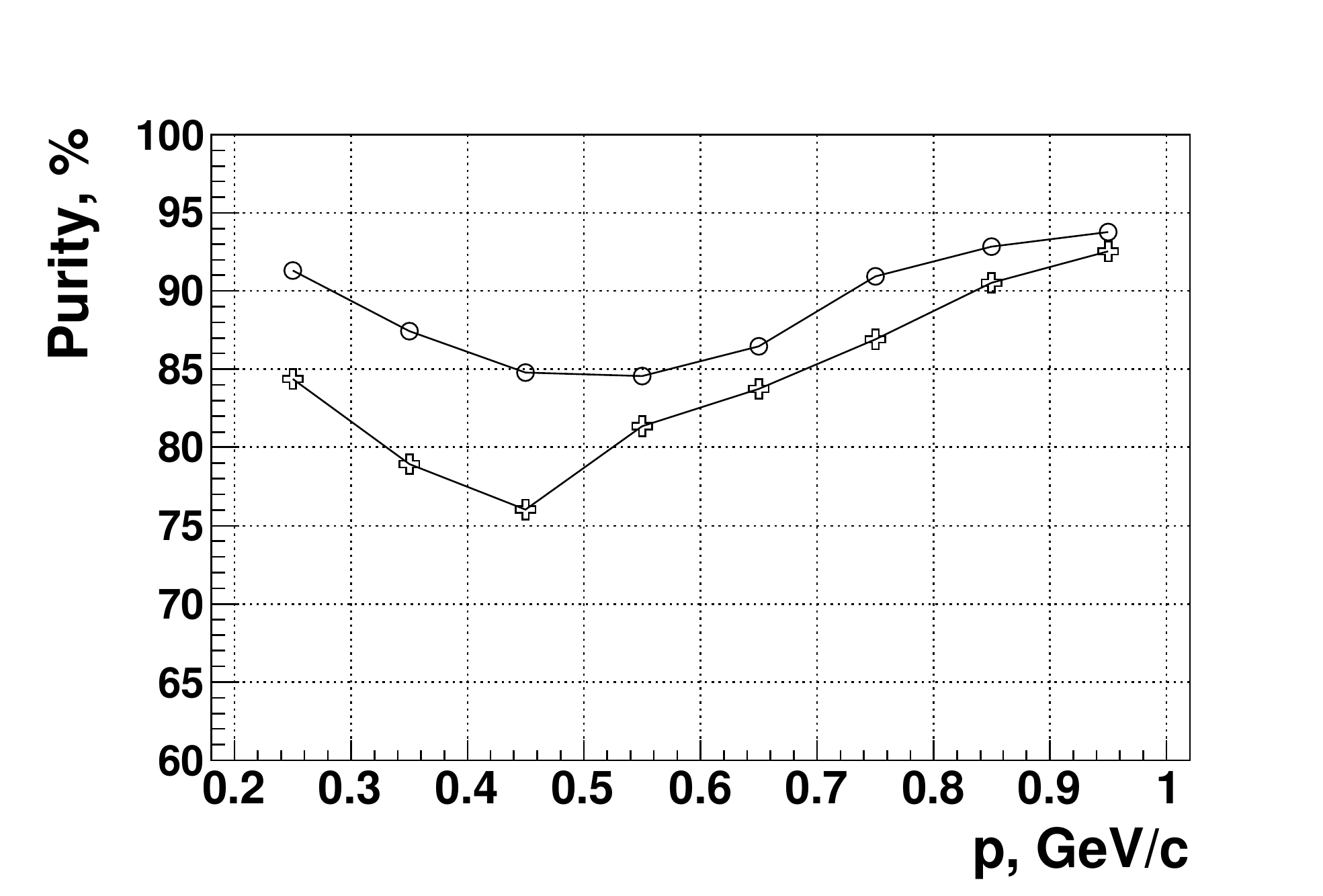} 
  \end{minipage}
  \caption{Purity of the electron (upper curve) and positron (lower curve) identification with $E_{EMC}/p$ ratio cut in Ni+Ni collisions at an impact parameter $b < 2.5$~fm at the cut efficiencies of 85\% (left) and 75\% (right).}
  \label{fig:purity_NiNi_8_imp2.5_eff75}
\end{figure}

\begin{figure}
  \centering
  \begin{minipage}[b]{8 cm}
   \includegraphics[width=8.5cm]{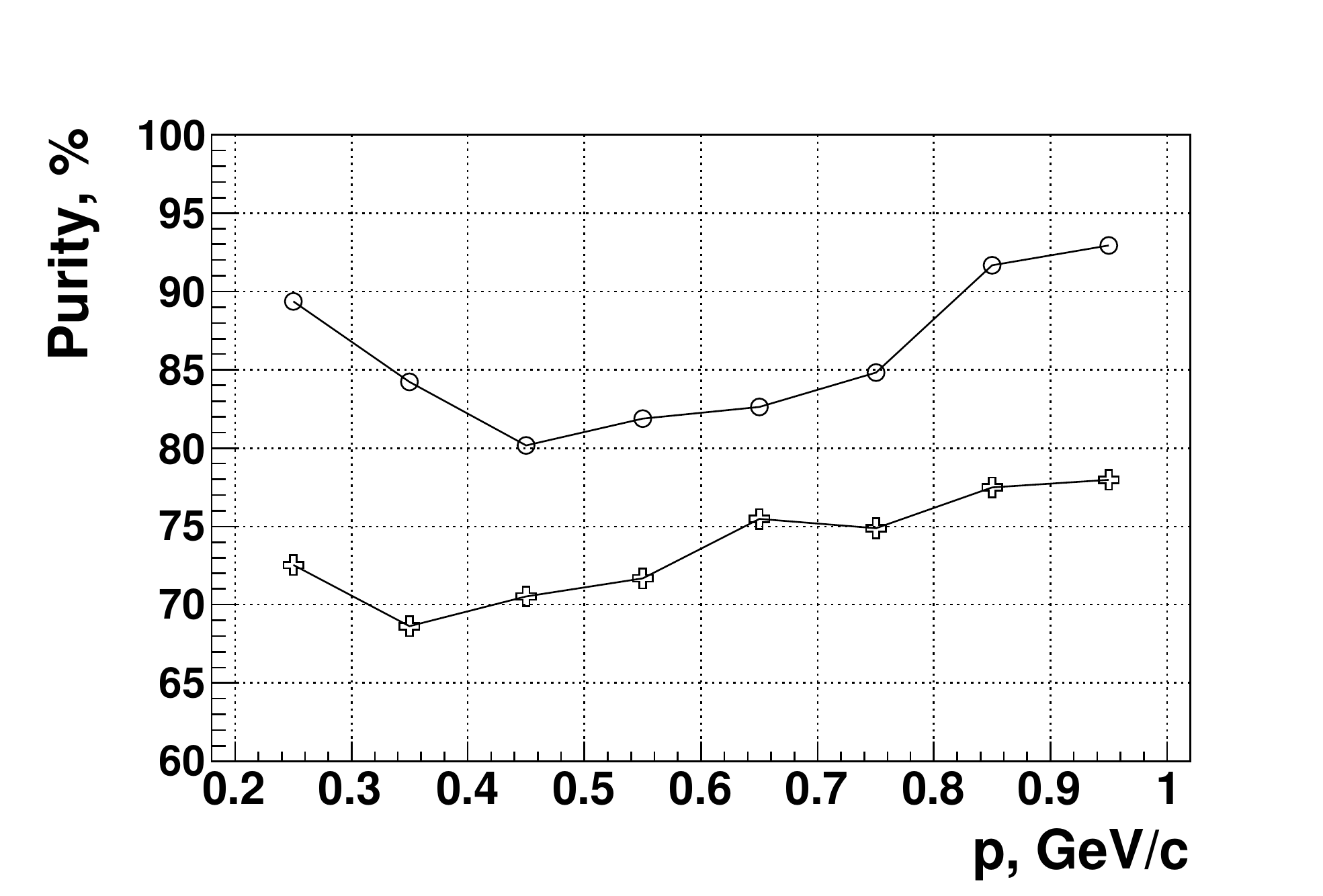} 
  \end{minipage}
  \begin{minipage}[b]{8 cm}
   \includegraphics[width=8.5cm]{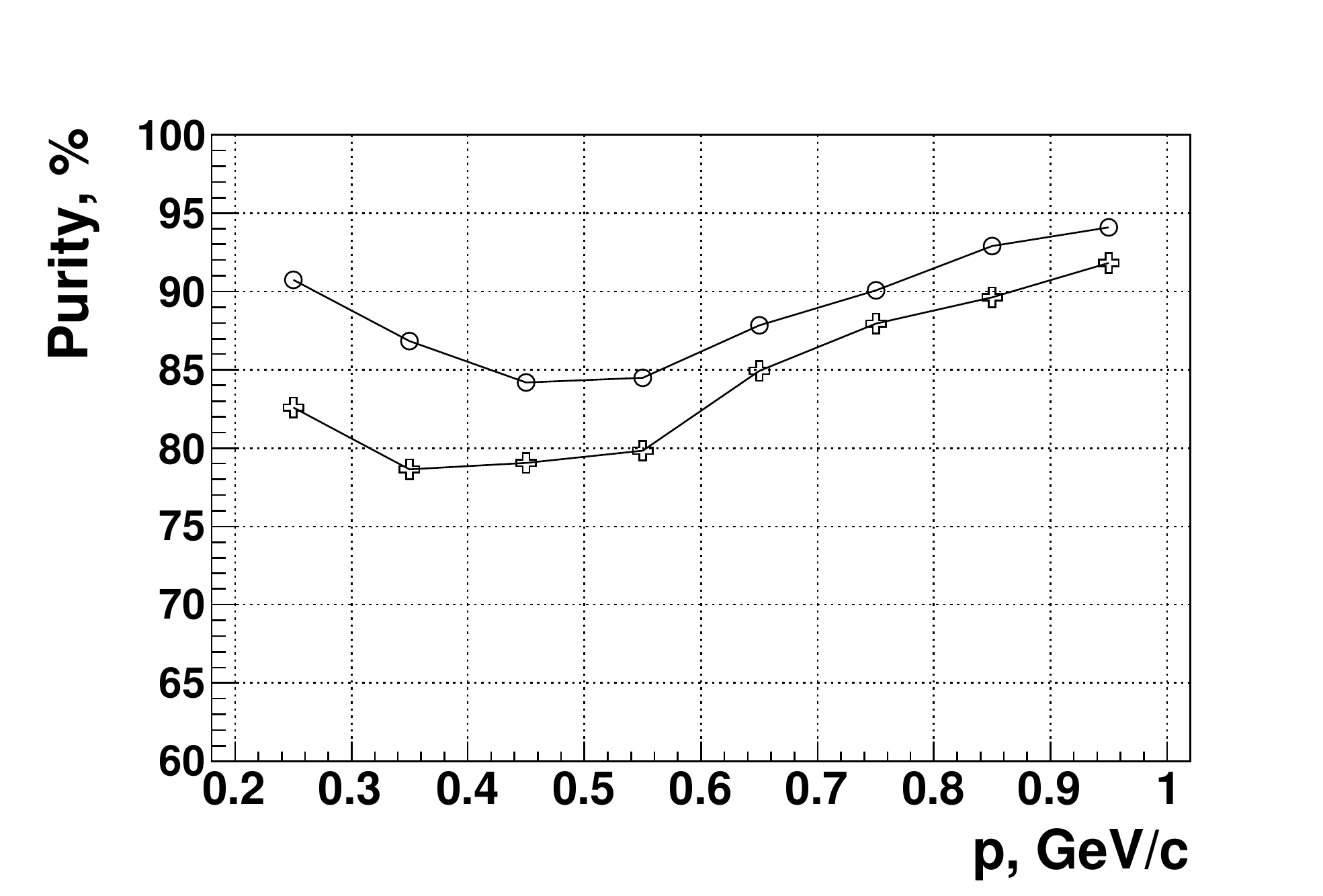} 
  \end{minipage}
  \caption{Purity of the electron (upper curve) and positron (lower curve) identification with the $E_{EMC}/p$ ratio cut in Ni+Ni collisions at an impact parameter $b < 1.0$~fm at the cut efficiencies of 85\% (left) and 75\% (right).}
\label{fig:purity_NiNi_8_imp1.0_eff75}
\end{figure}

\begin{figure}
  \centering
  \begin{minipage}[b]{8 cm}
   \includegraphics[width=8.5cm]{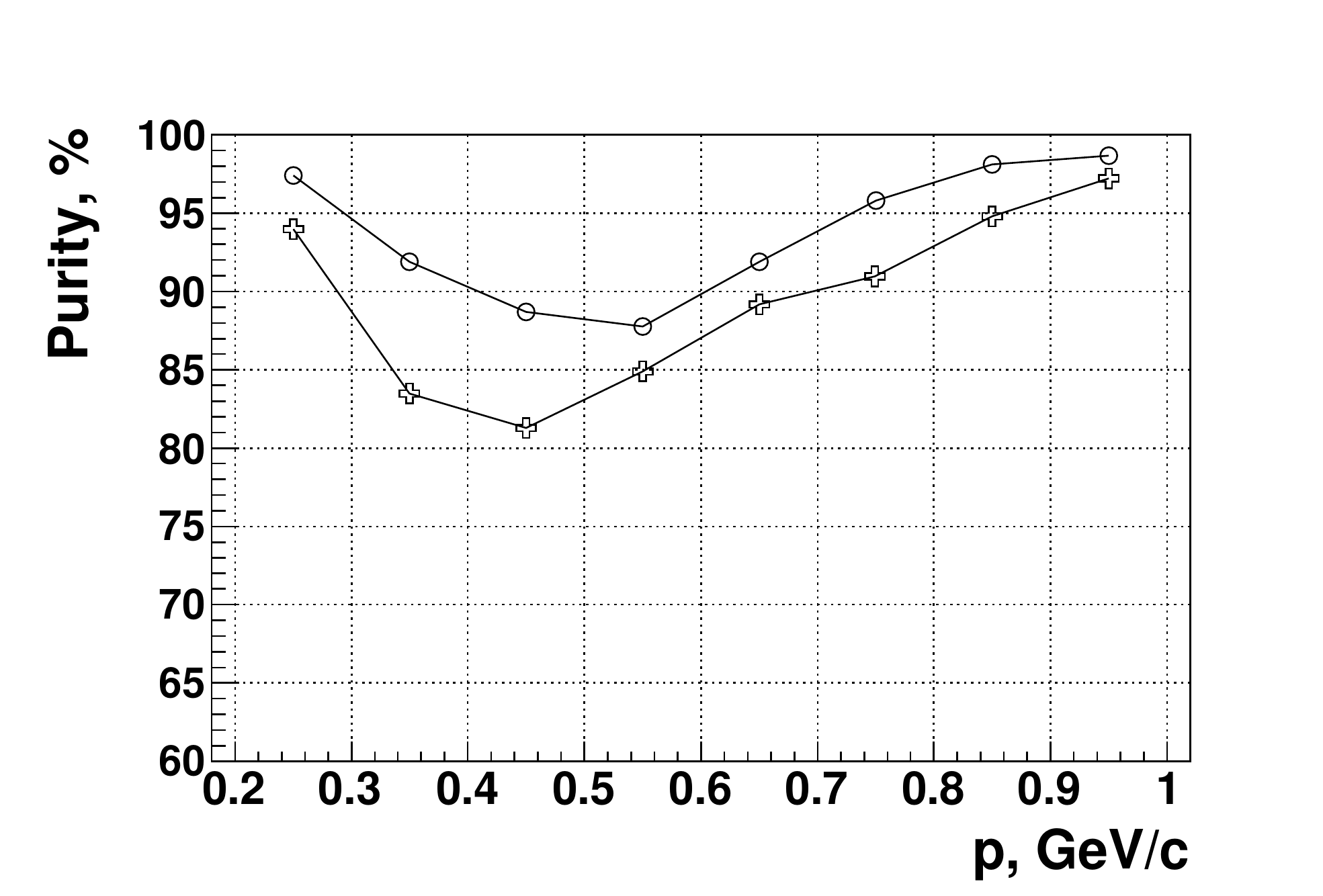} 
  \end{minipage}
  \begin{minipage}[b]{8 cm}
   \includegraphics[width=8.5cm]{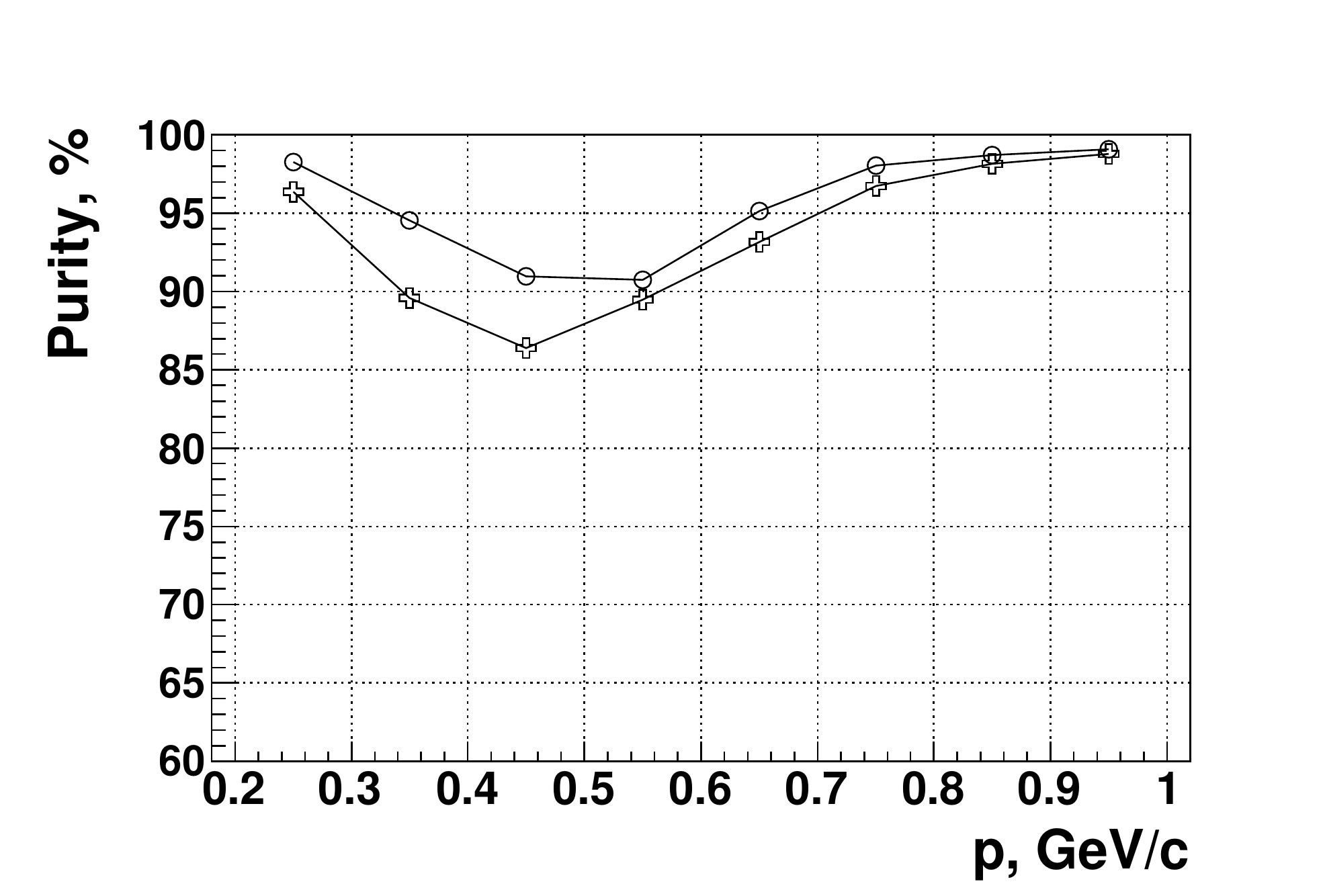} 
  \end{minipage}
  \caption{Purity of the electron (upper curve) and positron (lower curve) identification with the $E_{EMC}/p$ ratio cut in C+C collisions at an impact parameter $b < 1.0$~fm the cut efficiencies of 85\% (left) and 75\% (right).}
  \label{fig:purity_CC_8_imp1.0_eff75}
\end{figure}

Fig.~\ref{fig:purity_NiNi_8_imp2.5_eff75}, \ref{fig:purity_NiNi_8_imp1.0_eff75}, \ref{fig:purity_CC_8_imp1.0_eff75} illustrate the purity of the electron identification (defined as the fraction of the leptons in the sample selected by applying a cut to the initial sample that contained the equal amount of electrons and pions) with help of the cut on the $E_{EMC}/p$ ratio at cut efficiencies of 85\% and 75\% (a cut efficiency defined as the fraction of the initial lepton subsample that satisfied the cut) as a function of the momentum for different colliding systems and different ranges of the impact parameter. In most cases, the calorimeter provides a sufficient degree of the pion rejection. The purity of the positron identification is worse as compared to electrons, due to the concentration of positrons in the inner part of the calorimeter, which has a higher charged particle occupancy. 

\begin{figure}
   \centering
   \includegraphics[width=10cm]{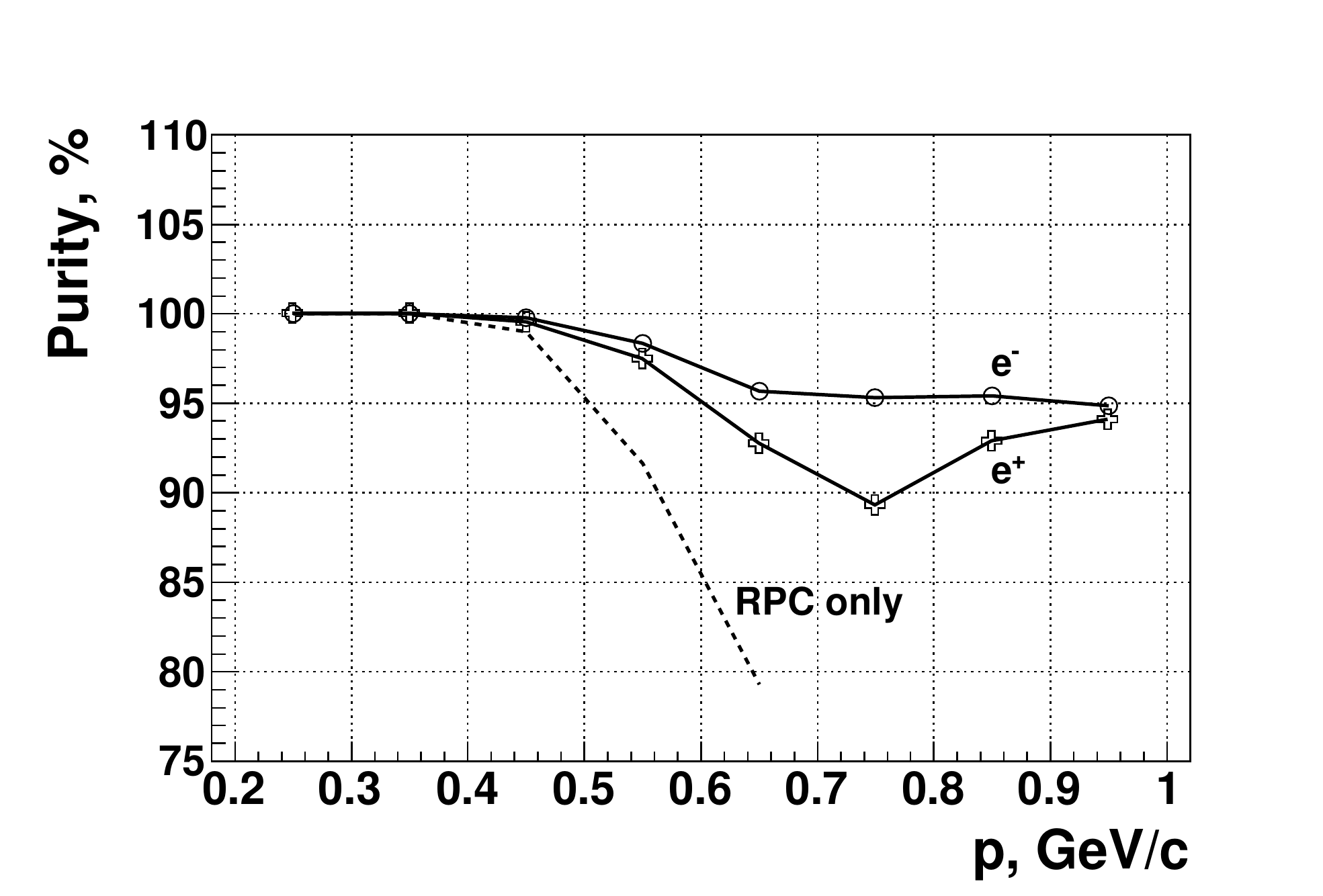} 
   \caption{Purity of the electron (upper solid curve) and positron (lower solid curve) identification ($E_{EMC}/p$ + time-of-flight cuts) in Ni+Ni collisions at impact parameter $b < 2.5$~fm in comparison with the lepton purity achieved using the RPC time-of-flight cut only (dashed curve).}
   \label{fig:purity_NiNi_8_imp2.5_eff85_tof_cut_rpc_only}
\end{figure}

It is clear, that the lepton identification will be based not only on the information coming from the calorimeter, but will exploit data from other detectors of the HADES ensemble as well. As an example of such cooperation, Fig.~\ref{fig:purity_NiNi_8_imp2.5_eff85_tof_cut_rpc_only} shows the purity of the lepton sample that can be achieved applying a cut on $E_{EMC}/p$ ratio and a cut on the time-of-flight measured with RPC (assuming time resolution of $\sigma_{t} = 100$~ps) simultaneously. Compared to the single EMC cut (cf. Fig.~\ref{fig:purity_NiNi_8_imp2.5_eff75}, left) a much higher purity is achieved (above 90\% for all values of the particle momenta). On the other hand, the purity delivered by the time-of-flight cut only (shown on Fig.~\ref{fig:purity_NiNi_8_imp2.5_eff85_tof_cut_rpc_only} by a dashed curve) rapidly degrades for momenta values $p > 550$~MeV$/c$ and becomes unacceptable. 

\subsection{$\pi^{0}$- and $\eta$-reconstruction}

Besides improving the electron-pion separation, the calorimeter allows to measure photons. This opens a possibility to identify light neutral pseudoscalar mesons ($\pi^{0}$ and $\eta$) via their decay into two photons (a diphoton): $\pi^{0}, \eta \to \gamma\gamma$. In this section we briefly discuss the developed photon identification algorithm and present simulated diphoton invariant mass spectra.

\subsection{Photon identification}
Within the current approach the photon identification is performed in several consecutive steps:

\begin{enumerate}
\item In order to suppress the contribution from charged particles, the crystals, spatially correlated with the RPC hits, are removed from the further analysis.

\item A clustering is performed. A cluster is defined as a set of adjacent crystals with an energy deposition above a certain threshold ($T_{1} = 60$~MeV).

\item Local maxima are identified within each cluster. Those are cells with energy deposition exceeding a threshold ($T_{2} = 150$~MeV) that are surrounded by the cells with lower energy deposition.

\item The total energy of the cluster is distributed among \emph{bumps} originating from independent showers. A bump is a set of cluster cells with assigned weights corresponding to the fraction of the cell energy in the bump. The weights are computed taking into account the distance between a cell and a local maximum and the energy of the latter.
\begin{figure}
   \centering
   \includegraphics[width=10cm]{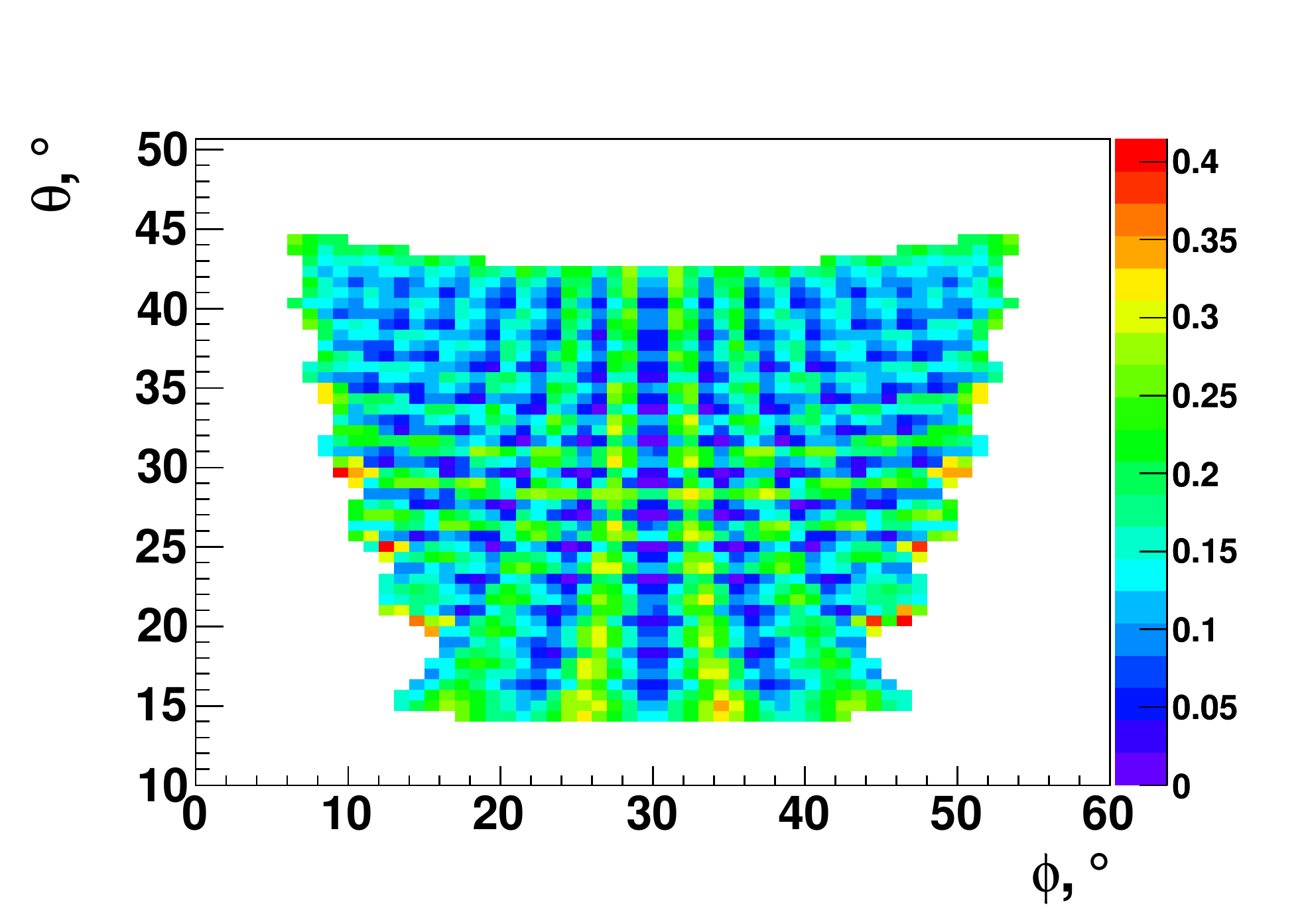}
   \caption{Deviation of the reconstructed photon energy from its real value (in per cents) in $\phi-\theta$ coordinates.}
   \label{fig:phi_theta_endiff}
\end{figure}
\begin{figure}
   \centering
   \includegraphics[width=10cm]{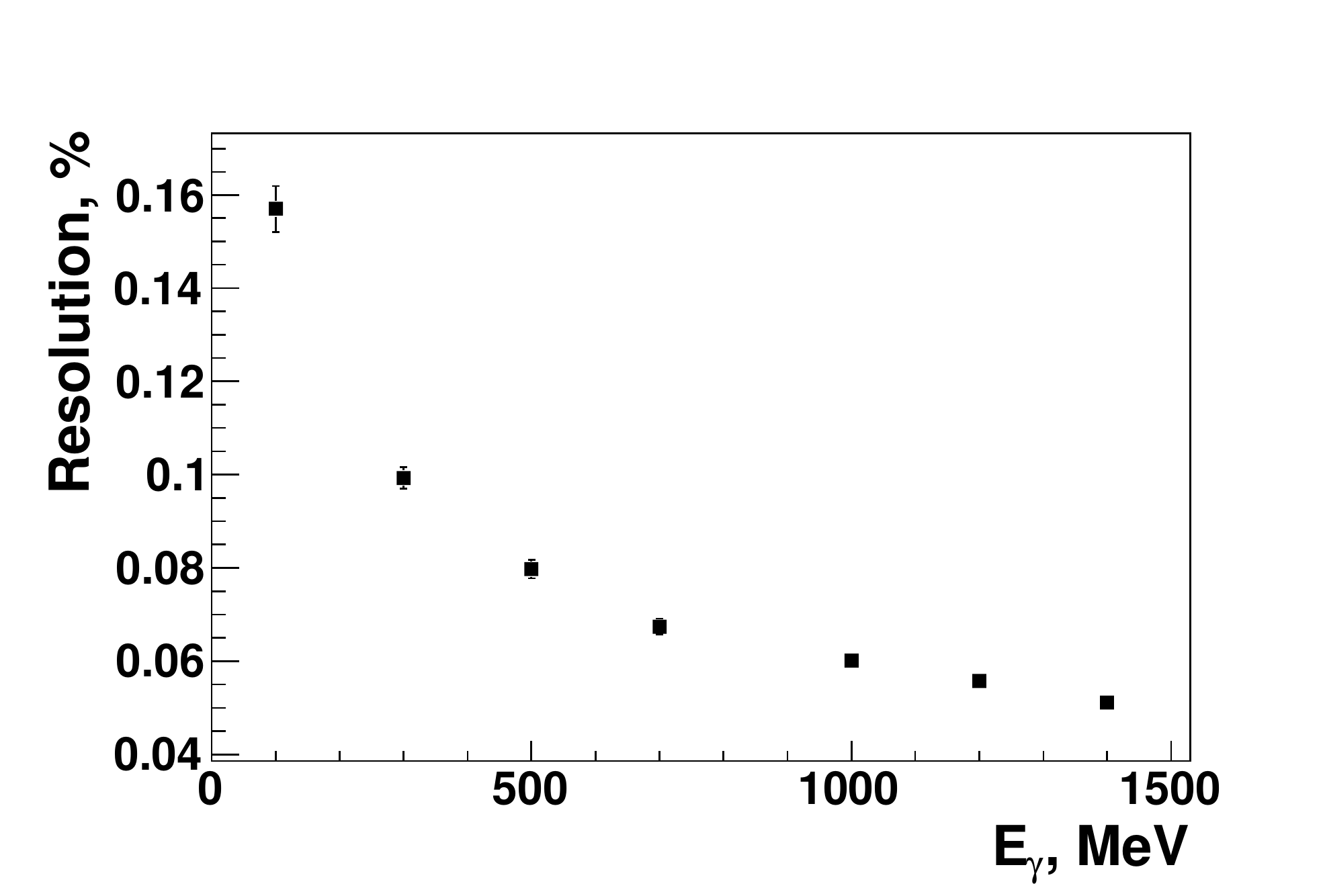}
   \caption{Energy resolution for single photons as a function of the photon energy.}
   \label{fig:res_energy_sim}
\end{figure}
\item An identified bump corresponds to the shower coming from the photon candidate. The energy of the bump and it geometrical center (calculated with logarithmic-weights method) provide information on a photon candidate 4-vector. Besides energy and momentum, various auxiliary data are computed for the photon candidate: the fraction of the shower energy contained in the local maxima, the second moment of the shower, a number of the Zernike moments etc.

\item The reconstructed energy and position of the photon are (optionally) adjusted with help of the correction tables, obtained via dedicated simulations that used as input a sample of $10^7$ photons with uniform energy (0--3~GeV) and angular distributions. Fig.~\ref{fig:phi_theta_endiff} shows the deviation of the reconstructed energy from its real value as a function of the photon polar ($\phi$) and azimuthal ($\theta$) angles. Reconstructed coordinates are corrected in a similar way.
Fig.~\ref{fig:res_energy_sim} shows the energy resolution for single photons as a function of the photon energy obtained with the aforementioned simulations.
\end{enumerate}

\subsection{Diphoton spectra}

To investigate feasibility of the $\pi^{0}$ and $\eta$ reconstruction, events coming from the Pluto generator were used to provide enough statistics ($\sim10^{7}$ events) in a reasonable time. A simulated particle cocktail for C+C and Ni+Ni collisions is presented in table~\ref{table:cocktail}. 

A full simulation of the EMC detector and all other detectors of the HADES set-up was made and the simulation output was passed to the reconstruction algorithms. Diphotons were formed taking into account all possible combinations of the identified photons. A  combinatorial background was constructed with help of the mixed event technique: uncorrelated pairs were formed taking the photons produced in different events. Afterwards, the background was normalized to the right tail of the invariant mass distribution ($M > 620$~MeV$/c^{2}$).

\begin{figure}
  \centering
  \begin{minipage}[b]{8 cm}
   \includegraphics[width=8.5cm]{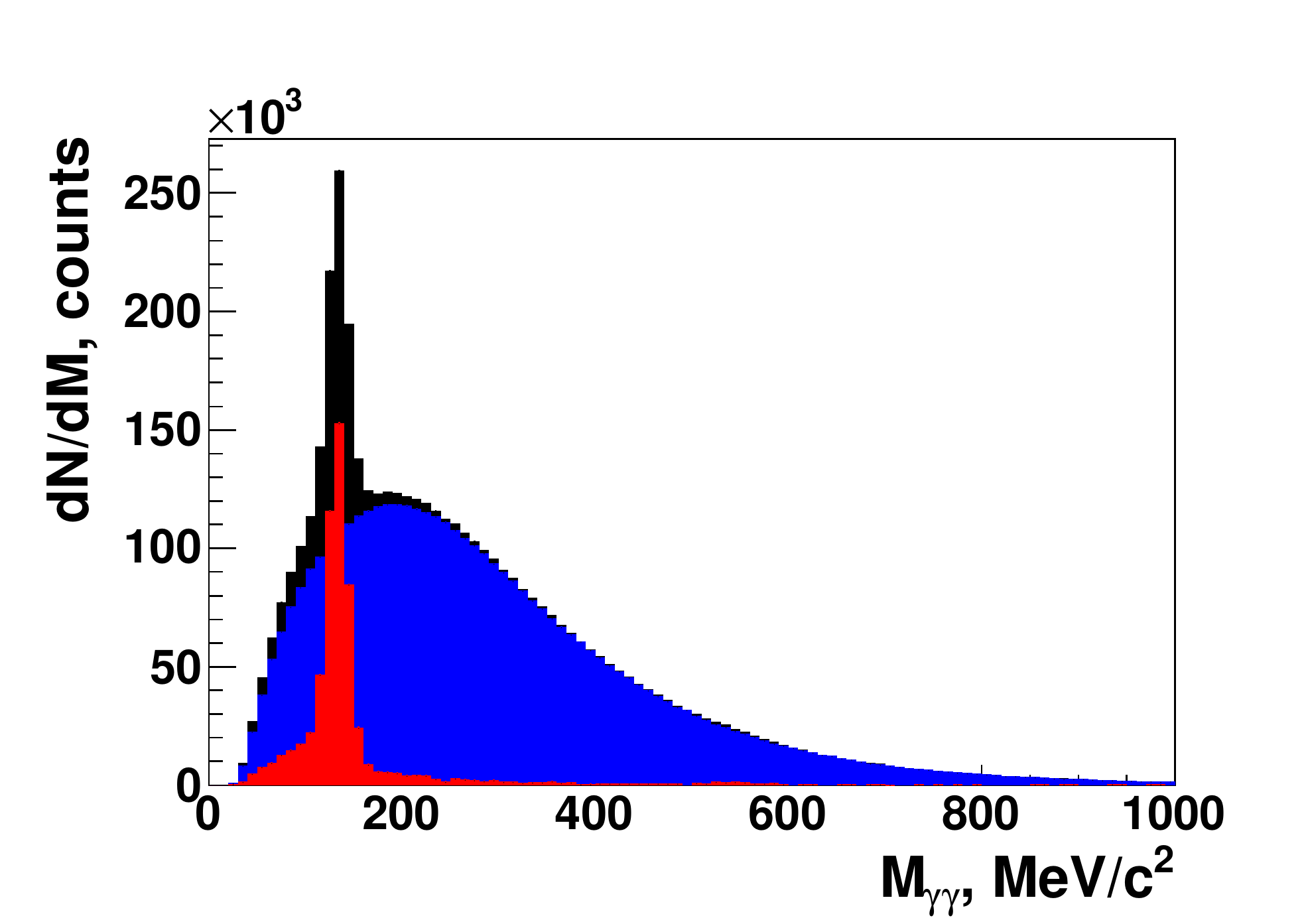} 
  \end{minipage}
  \begin{minipage}[b]{8 cm}
  \includegraphics[width=8.5cm]{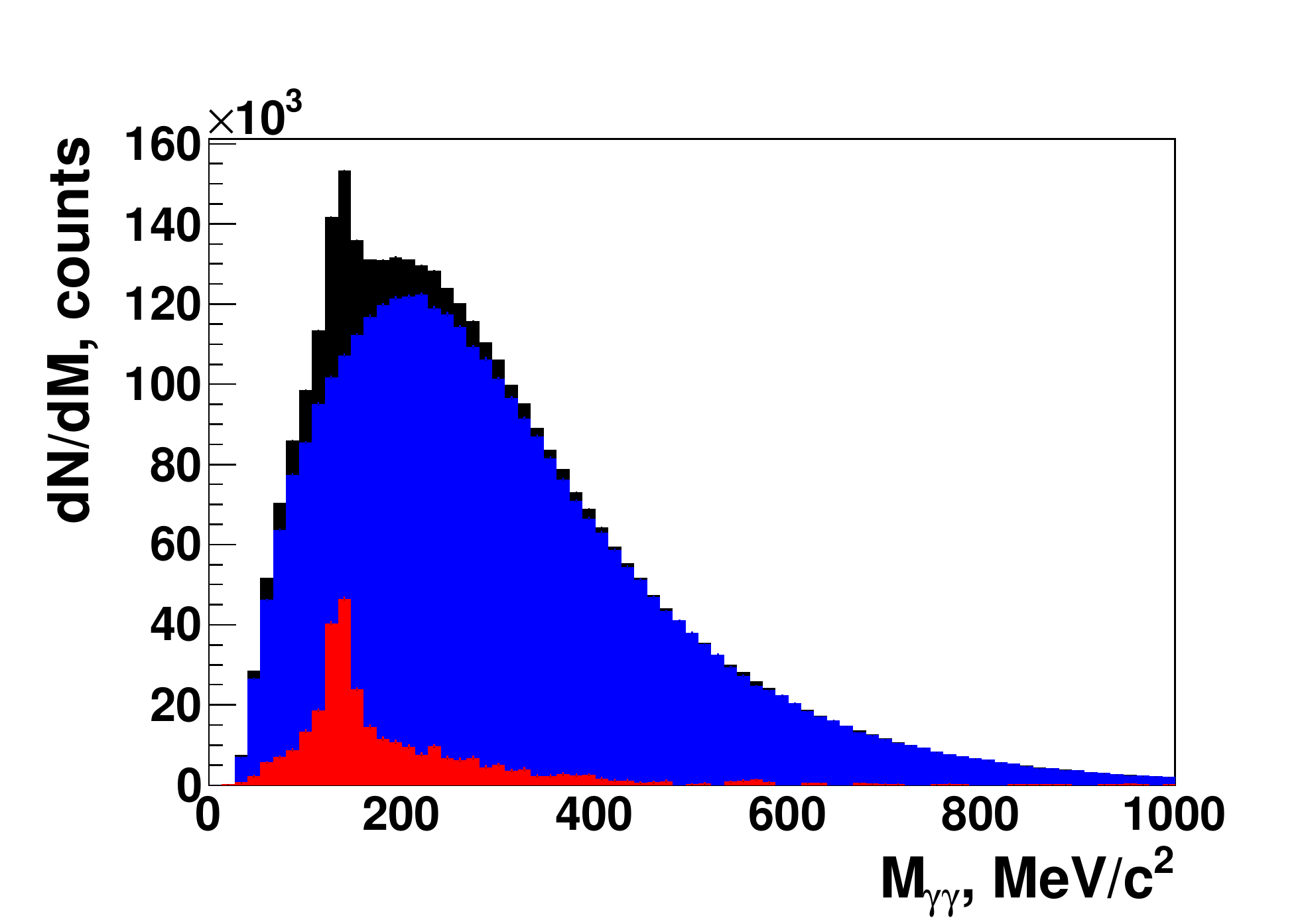} 
  \end{minipage}
  \caption{Diphoton invariant mass spectra reconstructed in C + C (left) and Ni + Ni (right) collisions at 8~AGeV beam kinetic energy (black histograms), combinatorial background (blue histograms) and signal after background subtraction (red histograms)}
  \label{fig:CC_NiNi_8_color}
\end{figure}

\begin{figure}
  \centering
  \begin{minipage}[b]{8 cm}
   \includegraphics[width=8.35cm]{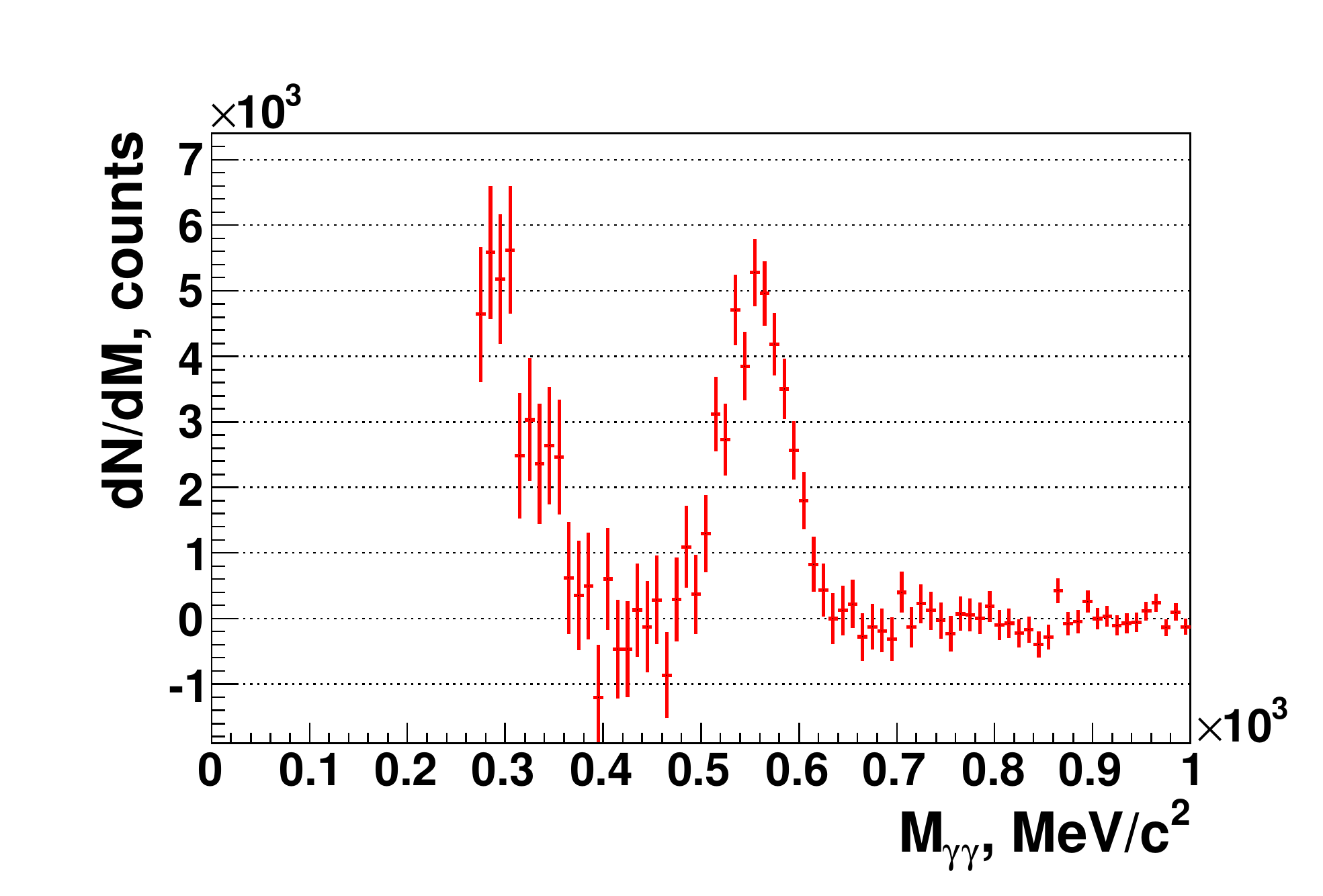} 
  \end{minipage}
  \begin{minipage}[b]{8 cm}
   \includegraphics[width=8.35cm]{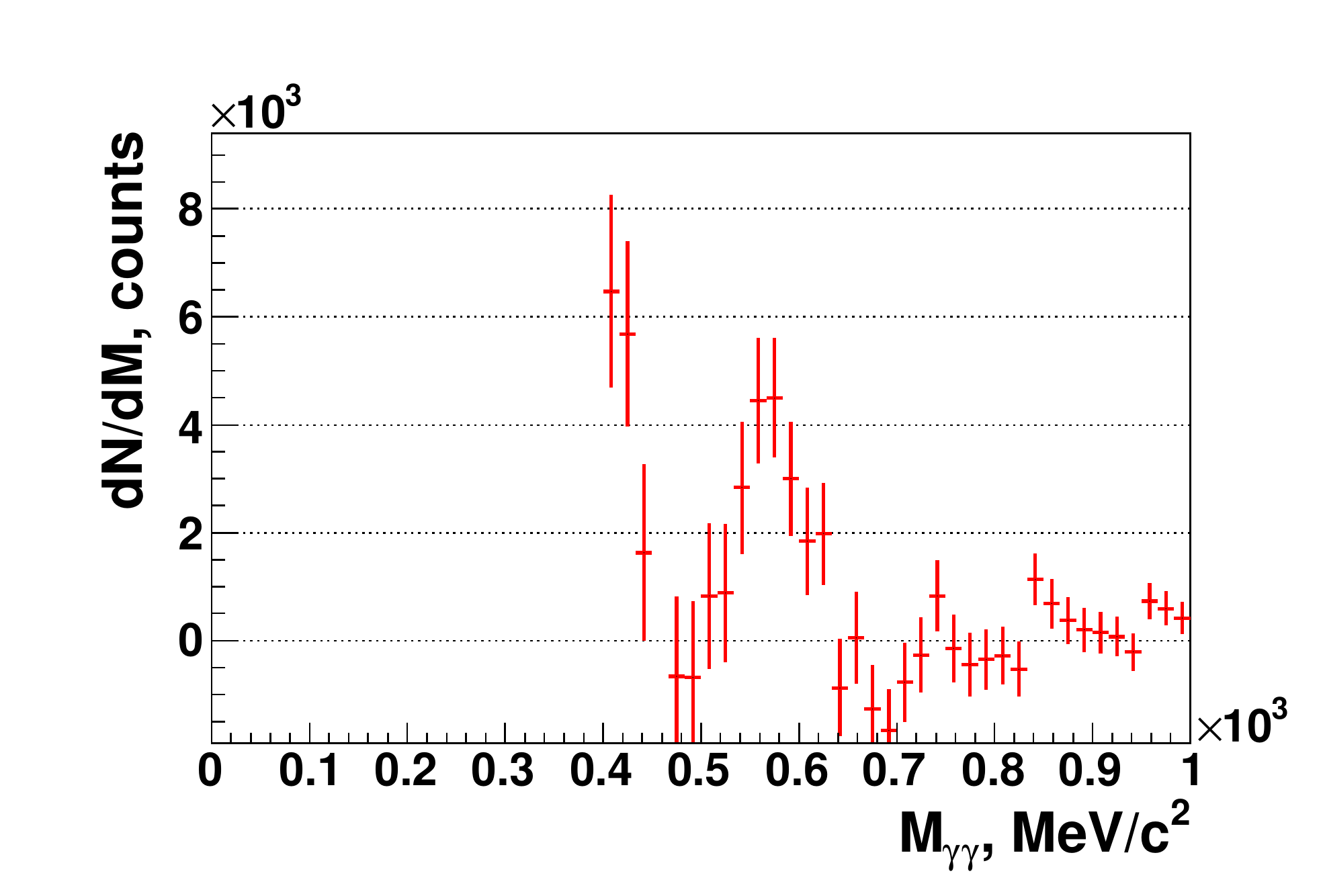} 
  \end{minipage}
  \caption{$\eta$-meson signal in the diphoton invariant mass spectrum obtained in C + C (left) and Ni + Ni (right) collisions at 8~AGeV beam kinetic energy.}
  \label{fig:eta_signal}
\end{figure}

Fig.~\ref{fig:CC_NiNi_8_color} shows diphoton invariant mass distributions, obtained for \mbox{C + C} (left) and \mbox{Ni + Ni} (right) collisions at 8 AGeV, together with the combinatorial background and the signal. For both systems a clear $\pi^{0}$-signal is visible on top of the combinatorial background. The $\eta$-meson signal appears after background subtraction (Fig.~\ref{fig:eta_signal}) only.

Table~\ref{table:meson_quality} summarizes the quality of the $\pi^{0}$ and $\eta$ reconstruction in C+C and Ni+Ni collisions.

\begin{table}
  \begin{tabular}{  c  c  c  c }
    \hline
    ~Reaction~ & ~$M_{p, n}$~ & ~$M_{\pi^{+},\pi^{-},\pi^{0}}$~& ~$M_{\eta}$~ \\ \hline \hline
    C~+~C & 2.7 & 1.86 & 0.093 \\
    Ni~+~Ni & 13.2 & 8.7 & 0.45 \\
  \end{tabular}  
  \caption{Hadronic cocktail used for the investigation of the $\eta$ reconstruction in C+C and Ni+Ni collisions. $M_{x}$ stands for the multiplicity of the particle $x$.}
  \label{table:cocktail}  
\end{table}

\begin{figure}
  \centering
  \begin{minipage}[b]{8 cm}
   \includegraphics[width=8.35cm]{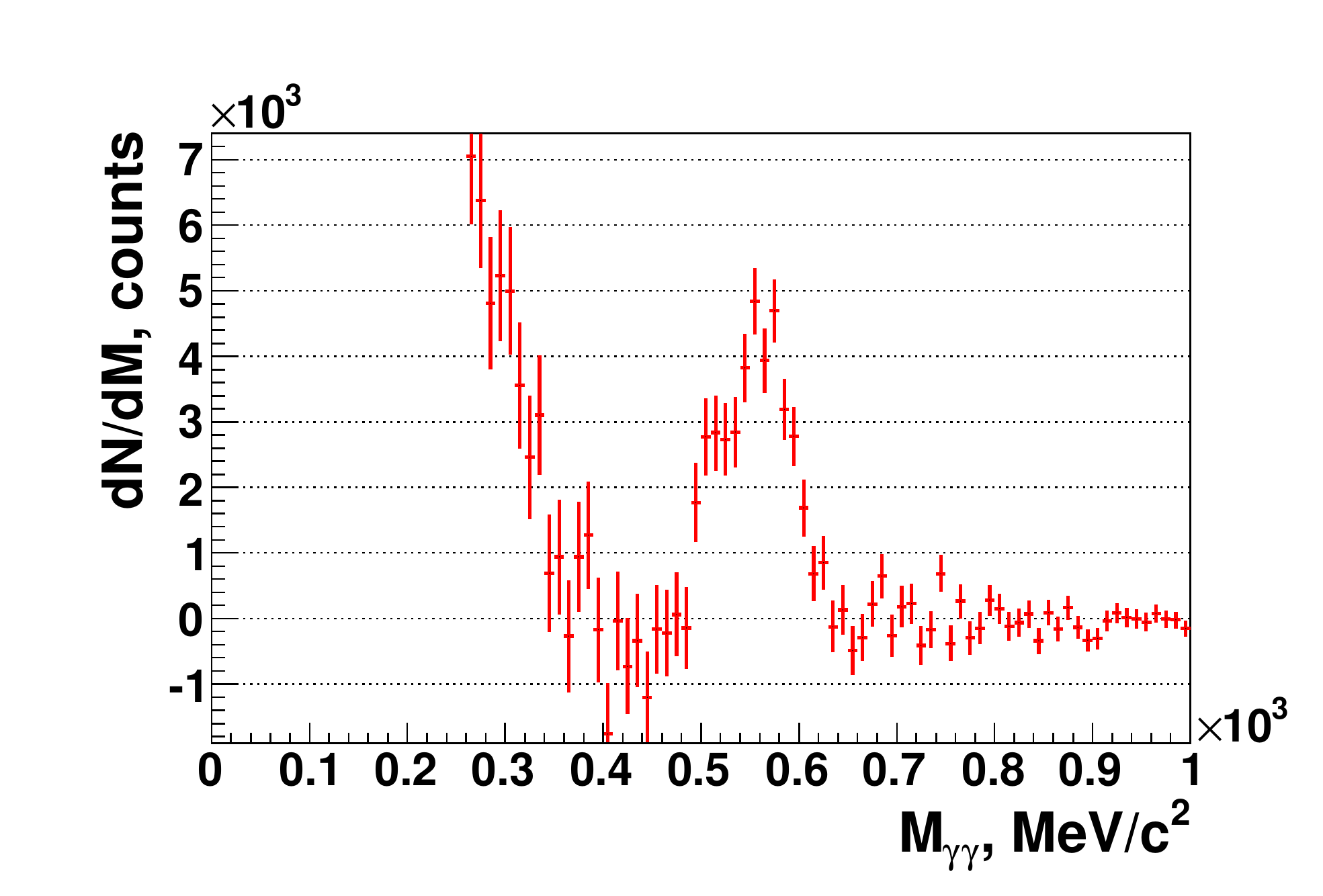} 
  \end{minipage}
  \begin{minipage}[b]{8 cm}
   \includegraphics[width=8.35cm]{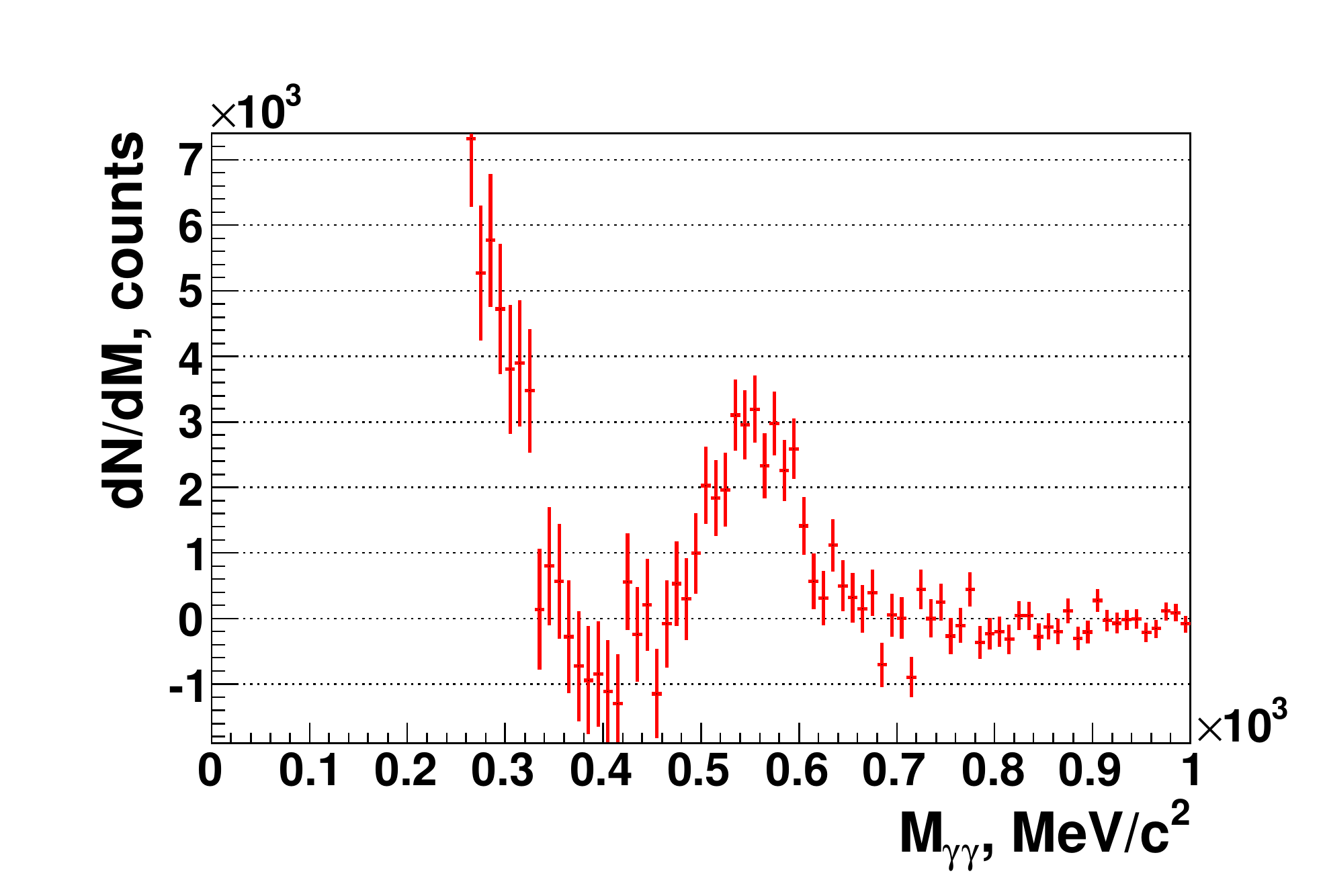} 
  \end{minipage}
  \caption{$\eta$-meson signal in the diphoton invariant mass spectrum obtained in C + C collisions at 8~AGeV employing energy resolution of $6 \%/\sqrt{E/\mathrm{GeV}}$ (left) and $9 \%/\sqrt{E/\mathrm{GeV}}$ (right)}
  \label{fig:eta_signal_resolution}
\end{figure}

In order to study effects of energy resolution, simulations of C+C collisions employing energy resolution of $6 \%/\sqrt{E/\mathrm{GeV}}$ and $9 \%/\sqrt{E/\mathrm{GeV}}$ were performed. Resulting invariant mass distributions demonstrating the $\eta$-meson peak are shown in Fig.~\ref{fig:eta_signal_resolution}. A systematic broadening of the signal is clearly visible. However, even with these values for energy resolution, the $\eta$-meson can be reconstructed.

\subsection{Conclusions}
Extensive simulations of the lead-glass calorimeter have shown that its performance satisfies the imposed requirements: the calorimeter provides necessary charged pion rejection power and allows reconstruction of the neutral pseudoscalar mesons in the diphoton invariant mass spectra. Further development and tuning of the reconstruction algorithms will make possible to improve the results that can be achieved with the considered calorimeter design.

As the occupancy of the innermost calorimeter modules in 
$^{56}${Ni}+$^{56}$Ni collisions at 8~AGeV
is very high and exceeds 0.7 in the most demanding case of the central
reactions, using of modules with higher granularity in this region
is under consideration.
It should be stressed that high occupancy of the cells is caused mainly 
by the particles with low energy deposition (hadrons) and does not 
necessarily obstruct realization of the main objectives---electron/pion 
separation at high values of particle momenta and neutral meson 
reconstruction. 
For the most demanding case of the $\eta$-meson reconstruction
energy resolution plays major decisive role and is determined by lead glass
properties. 
On the other hand for electron identification energy resolution is not so
important (momentum determination is obtained from high resolution
tracking) and multi-hadron hits can be suppressed by an energy cut on single 
cell energy deposition, a time-of-flight cut and track matching with the high
resolution RPC detector. 
For example, taking into account cells with minimal energy deposition of
100 MeV highest occupancy (in the standard geometry) drops already to 0.45.

The simulation will be continued to evaluate the effect of the detector
configuration on the meson reconstruction.

\begin{table}
   \begin{tabular}{  c  c  c  c c }
    \hline
    ~Reaction~ & ~Meson~ & ~S/B, \%~&~$S/\sqrt{S+B}$~ &~$\sigma_{M}/M, \%$~ \\ \hline \hline
    C~+~C & $\pi^{0}$ & 92.8  & 1193.6 & 9 \\
     & $\eta$ & 2.2 & 25.0 & 6  \\ \hline
    Ni~+~Ni & $\pi^{0}$ & 30.7 & 998.7 & 15 \\
    & $\eta$ & 0.4 & 7.6 & 5 \\
  \end{tabular}  
  \caption{The quality of the $\pi^{0}$ and $\eta$ reconstruction using the electromagnetic calorimeter.}
  \label{table:meson_quality}  
\end{table}

%% file: TestResults/TestResults.tex
\section{Test Results}
\label{testRes}
\subsection{Measurements of energy resolution at the MAMI accelerator (Mainz)}
A dedicated test exploiting a photon beam with an energy
between 0-1500 MeV was carried out to measure the energy resolution of 
detector modules.
\subsubsection{Test conditions and setup}

The test took place in the A2 hall of the MAMI-C facility in Mainz \cite{MAMI}. 
Detectors were positioned in the secondary gamma beam produced via
Bremsstrahlung from a primary electron beam. 
The obtained gamma energy varies from 0 to the beam energy of the primary electron, with an intensity
exponentially falling with increasing energy.
Exception made for a study of the influence of the particle hit position on the 
resolution, the detectors were hit in the centre of their front side,
and the beam proceeded along their longitudinal axis.
The beam diameter at the detector position was about 6 mm.
The gamma energy was determined by measuring the scattered electron angle 
(``tagged electron'' technique) \cite{MAMI_tag}. As a trigger, the signals from
8 selected scintillators in electron tagger has been used; in this way events with 8 known gamma energies in all range of energies up to the energy of the electron beam have been selected.
Exception made for the energy response of the studied modules, the time signals from
8 taggers were included via a CAMAC based DAQ system to data stream, so that for
each written event the gamma energy is known. The energy spread for the gammas
for the tagged events was below 1\%. This did not influence the obtained energy resolution of the tested lead glass modules.
\begin{figure}[htb]
\centering
\includegraphics*[width=140mm]{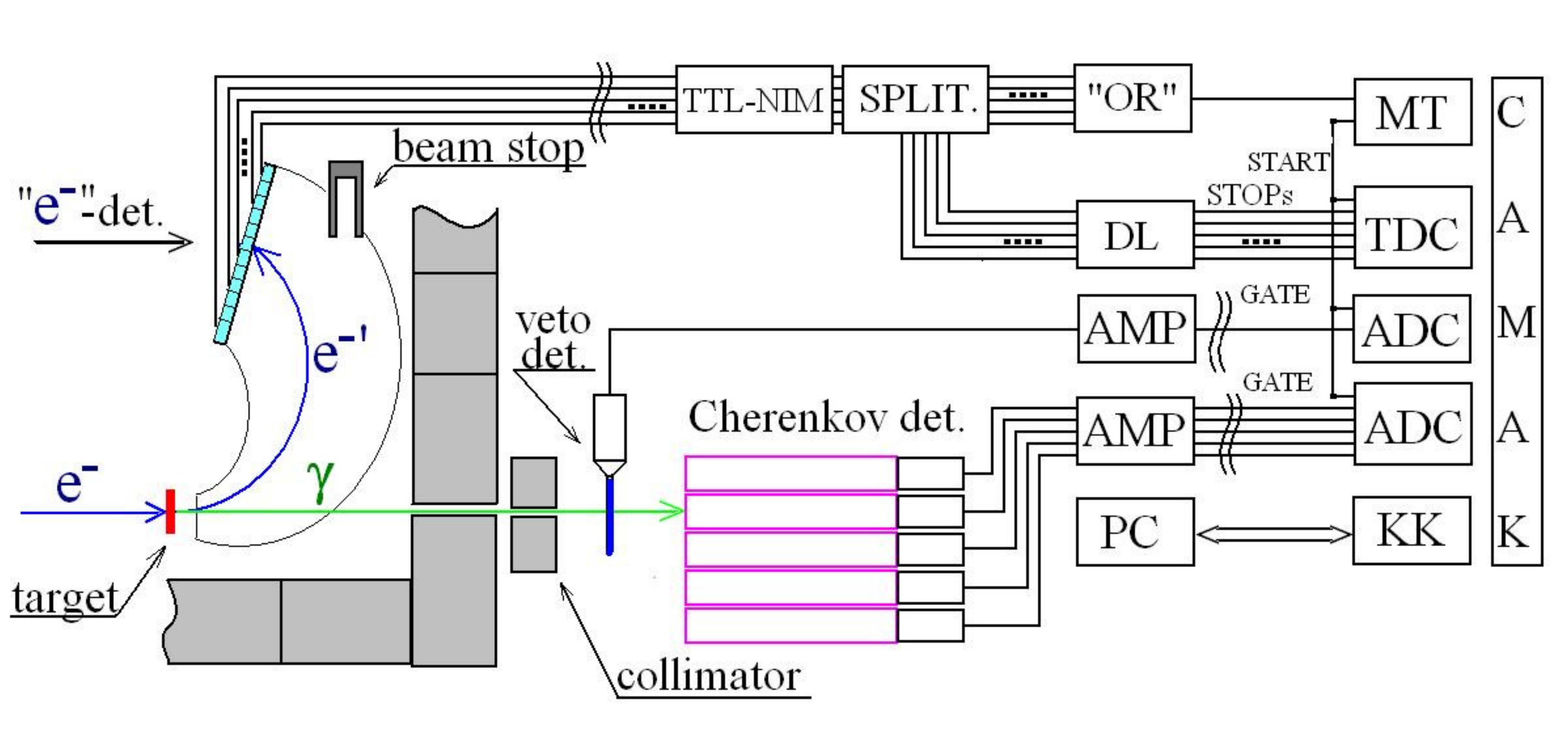}
\caption{Setup of the test at MAMI Mainz}
\label{setup_mami}
\end{figure}

Two electron accelerator settings were employed, one with an electron energy
of 855 MeV and the second with 1508 MeV. The gamma beam intensity was 
about 5 kHz.
A sketch of the test and DAQ setup is shown in Fig. \ref{setup_mami}.
The PMT signal was amplified and shaped with a shaping time of 0.5 $\mu$s
with a MA8000 amplifier (produced at GSI Darmstadt), and then processed with an ORTEC AD811 11 bit ADC.

5 lead-glass modules with various configurations have been used during the test, as summarized in 
table \ref{tab-det1}. Two modules were equipped with a 5 cm long light-guides 
from the same material which allowed a smooth shape transition from the detector 
back side, with dimensions of $9.2x9.2cm^2$, to the front window of the 1.5 inch 
PMT with diameter of 37 mm. On other 3 modules the PMT's were glued directly
to the center of the detector back side (optical glue TL-VT 3402 KK made by
PETERS). 
For the detector wrapping the aluminized mylar foil or white paper (made by TYVEK) were used. For comparison, one module was equipped with the 2 inch Hamamatsu H1949 PMT, as compared to a ``standard'' setup with the 1.5 inch EMI9903KB PMT. For all four other modules EMI PMTs with a similar gain (within 20\%) were selected.    

\begin{table}[h!]
\caption{
Detector configurations used during the test at MAMI.
\label{tab-det1}}
\center~
\begin{tabular}{c c c c }
\hline
module Nr. & lightguide & wrapping &  PMT type  \\ 
\hline   
1          &   yes      &  mylar   &  EMI9903KB \\
2          &   yes      &  paper   &  EMI9903KB \\
3          &   no       &  mylar   &  EMI9903KB \\
4          &   no       &  paper   &  EMI9903KB \\
5          &   no       &  mylar   &  Hamamatsu 1949 \\
\hline
\end{tabular}\\[3pt]
\end{table}

\begin{figure}[htb]
\centering
\includegraphics*[width=115mm]{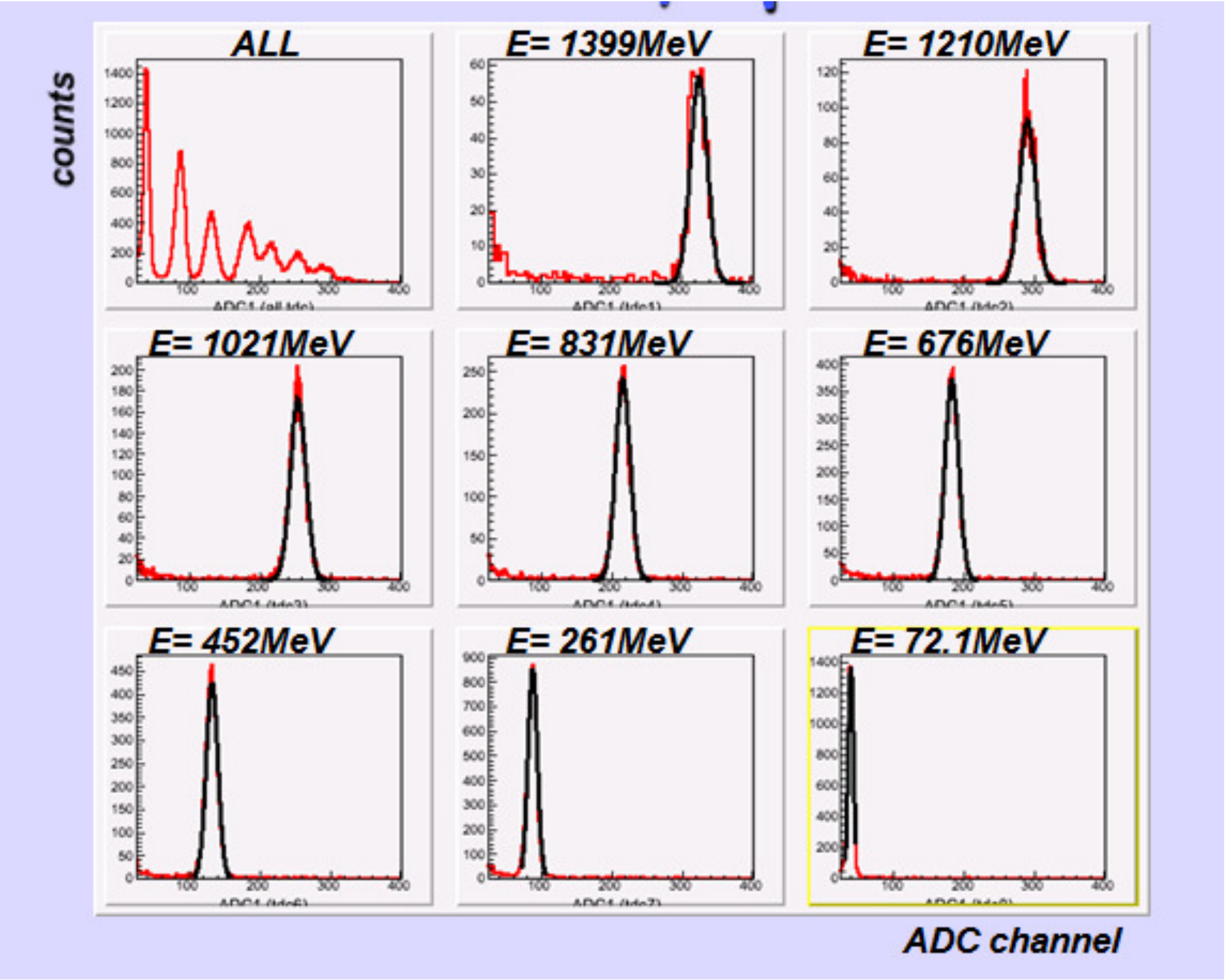}
\caption{Distribution of the photon energy (in ADC channels) deposited in the 
EMC module. Summed spectrum and 8 spectra triggered by selected tagger scintillators are shown.}
\label{fig_adc_vs_energy}
\end{figure}
\subsubsection{Results}

\begin{figure}[htb]
\centering
\includegraphics*[width=110mm]{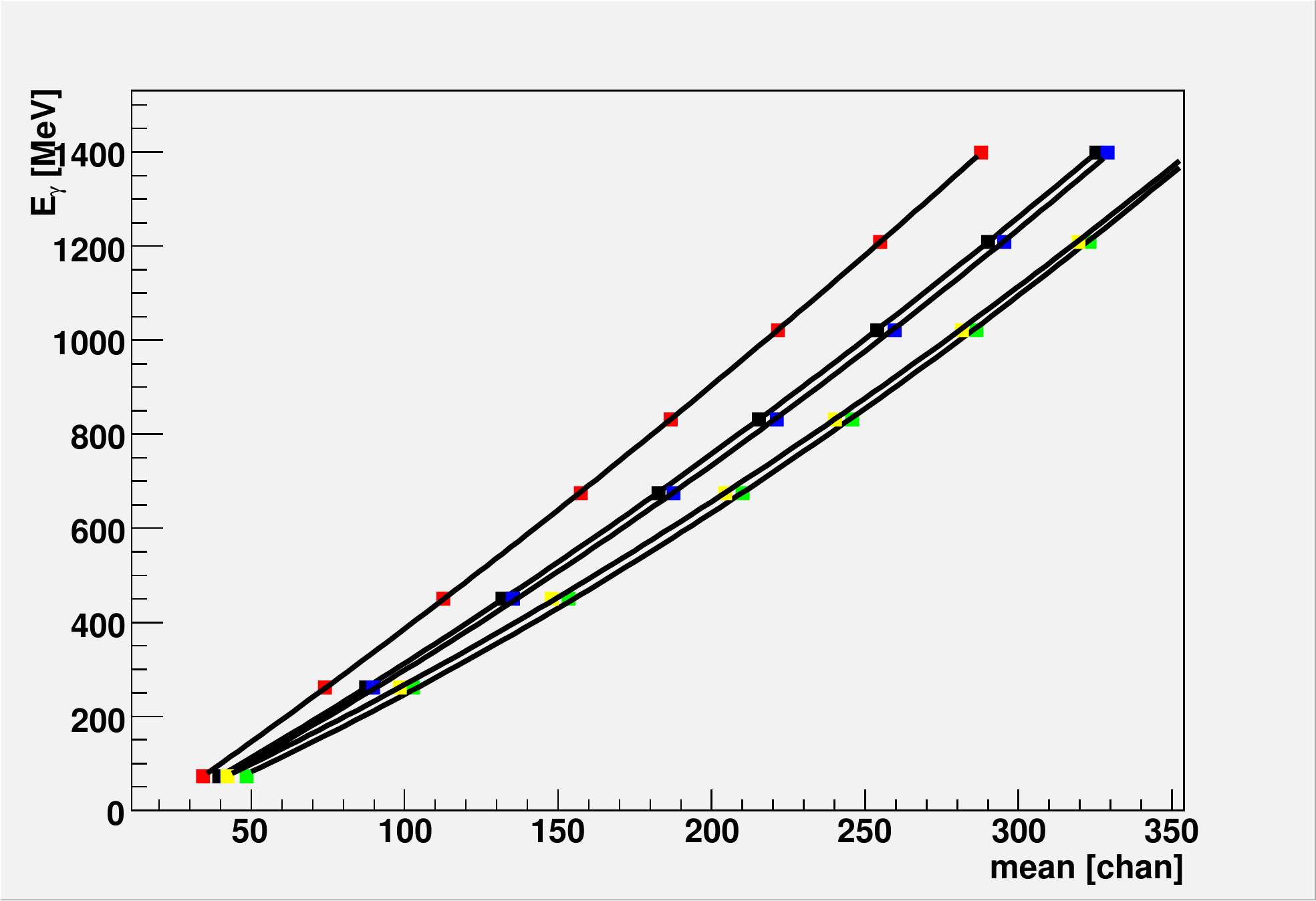}
\caption{Energy calibration of spectrometric channels from 5 studied modules for a ``high energy'' run}
\label{fig_ecal_he}
\end{figure}

The measured pulse height distributions for one lead-glass module are shown in 
Fig. \ref{fig_adc_vs_energy} for the run with higher electron energy settings.     
The distributions were then fitted by a Gaussian function. 
The results - peak positions and standard deviations - for one lead-glass 
module are presented in table \ref{tab-mr1}. One can see that the energy resolution ranges from 5 to 20 \% when varying the photon energy from 1399 to 72 MeV.
The peak positions then were used for energy 
calibration of the electronic chain including the PMT and the shaping 
amplifier (AMP). The results of the energy calibration -
energy versus ADC channel - for five spectrometric channels are shown in Fig. \ref{fig_ecal_he}. The linear behaviour of the read-out chain is well visible. 

\begin{figure}[htb]
\centering
\includegraphics*[width=115mm]{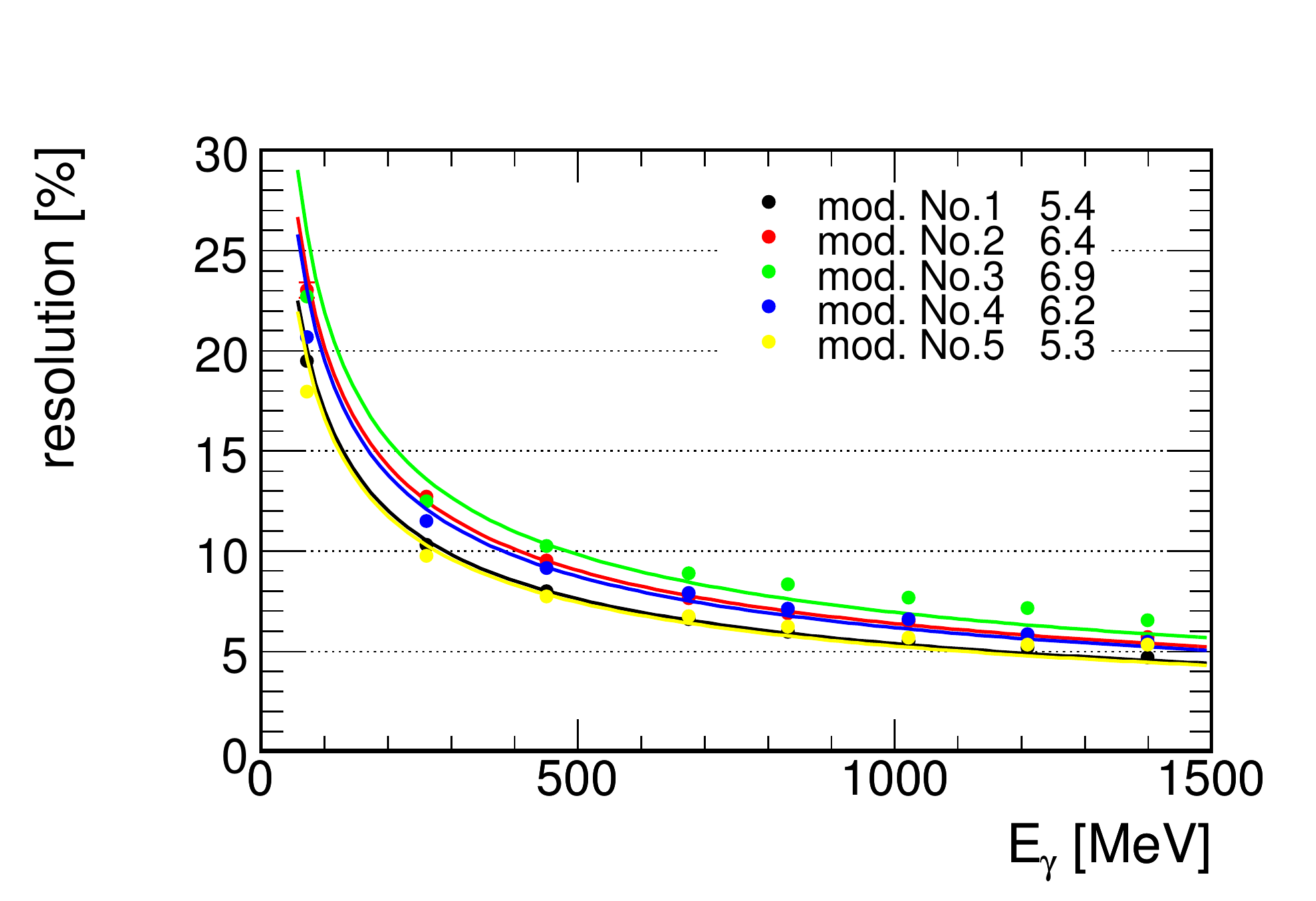}
\caption{Dependence of energy resolution upon the gamma energy for for a ``high energy'' run.}
\label{fig_res_vs_HE}
\end{figure}

\begin{table}[h!]
\caption{
Response of one lead-glass module (No.4) to photons (for ``higher energy'' run).
\label{tab-mr1}}
\center~
\begin{tabular}{c c c c c }
\hline
$E_{\gamma}$ & mean & $\sigma$ &  $\sigma_{E}$ & $\sigma_{E}/E$ \\
(MeV)        & (ADC chan.) & (ADC chan.) &  (MeV) & (\%) \\ 
\hline   
72.1   &     42.14$\pm$0.04 &   4.08$\pm$0.04 & 14.9$\pm$0.15 & 20.7 \\
261    &     89.85$\pm$0.06 &   7.60$\pm$0.05 & 30.1$\pm$0.2  & 11.5 \\
451    &    135.21$\pm$0.09 &   9.74$\pm$0.07 & 41.3$\pm$0.3  &  9.2 \\
675    &    187.63$\pm$0.10 &  11.56$\pm$0.08 & 53.4$\pm$0.4  &  7.9 \\
831    &    221.23$\pm$0.13 &  12.32$\pm$0.11 & 59.4$\pm$0.5  &  7.1 \\
1021   &    259.74$\pm$0.16 &  13.29$\pm$0.13 & 67.6$\pm$0.7  &  6.6 \\
1209   &    295.38$\pm$0.21 &  13.32$\pm$0.17 & 70.8$\pm$0.9  &  5.9 \\
1399   &    329.1$\pm$0.6   &  13.8$\pm$0.5   & 76.5$\pm$2.6  &  5.5 \\
\hline
\end{tabular}\\[3pt]
\end{table}
Figure \ref{fig_res_vs_HE} shows a dependence of the relative energy 
resolution on the gamma energy. The distribution can be well described 
by a formula 
$\Delta E = k/\sqrt(E [GeV])$, 
where the meaning of the
constant k is a relative resolution in \% at energy of 1 GeV. 
The constant k is shown for each measured lead-glass module in right 
upper corner of Fig. \ref{fig_res_vs_HE}. 

The resulted resolution for all modules is 
about 6\% at 1 GeV. The use of the light-guides does not help to increase 
the light detection, the measured pulse height is comparable with 
those from modules without the light-guide (see Fig. \ref{fig_ecal_he}).
 No significant difference in resolution is obtained either. 
We also did not observe any influence upon the wrapping material on the 
resolution. 
The module equipped with the 2 inch Hamamatsu H1949 PMT exhibits a good 
energy resolution of 5.3\%, but it is still comparable with performance 
of other modules with EMI PMTs.

At the end of the test the dependence of the resolution on position of
the beam with respect to the centre of one lead-glass module (Nr.1) was studied. 
The beam passed always
along the detector longitudinal axis, with a distance from 0 to 4 cm from
its front-end centre in a step of 1 cm. So in the last run, the beam 
passed only 0.5 cm from the edge of the detector. Together with the signal from the 
irradiated detector also the signal from the adjacent module (Nr.2) was recorded.
For beam positions up to 1 cm from the centre, no light from the adjacent module
was observed, and the energy resolution stayed constant at 5.4\%. 
For positions closer to the edge, the summed signals from both modules were considered  and 
the same energy resolution of 5.4 \% was obtained, exception made for the last position (0.5 cm from the edge), where the resulted resolution was 6.0\%.
  
\subsection{Measurement with cosmic muons}

\begin{figure}[htb]
\centering
\includegraphics*[width=115mm]{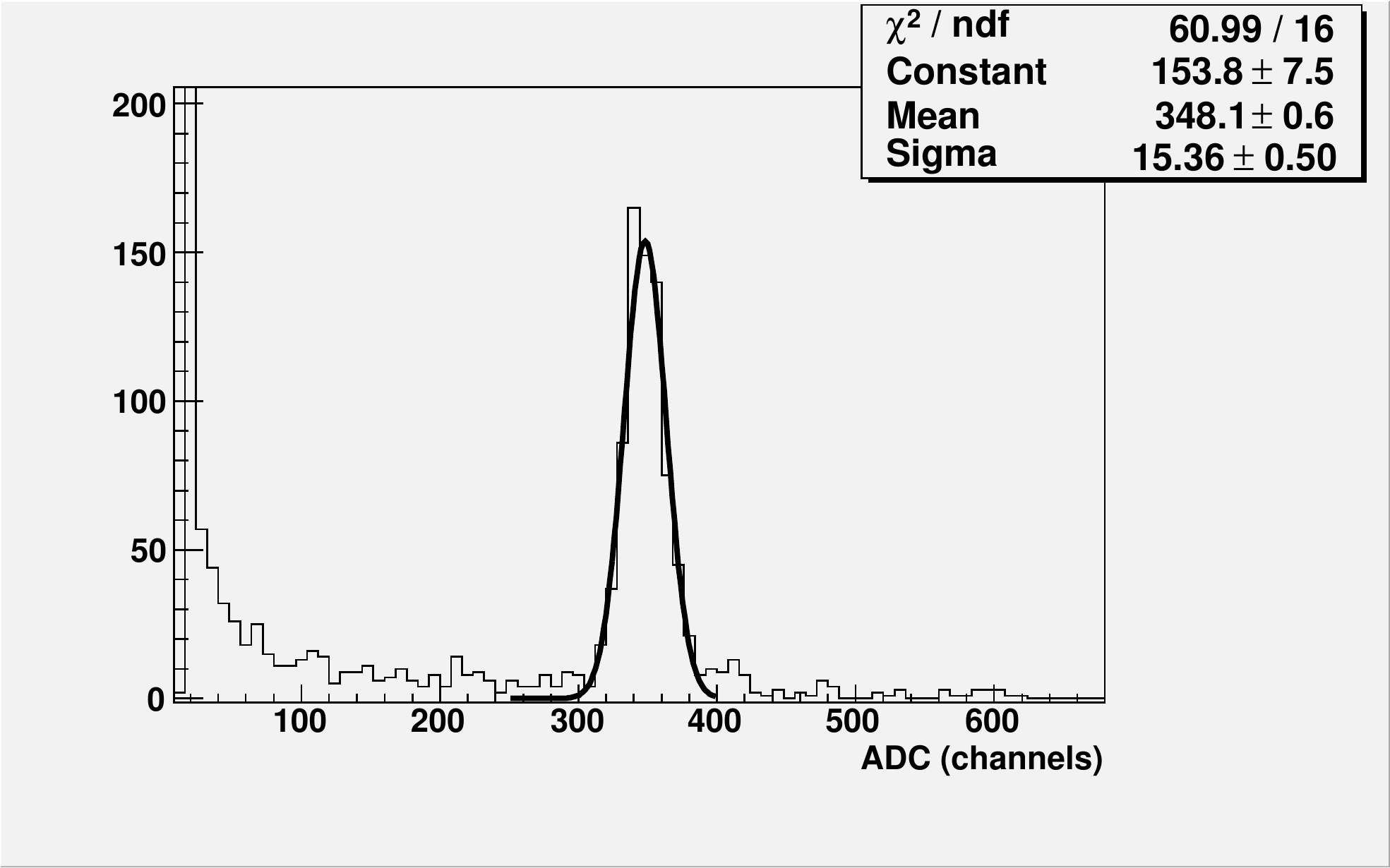}
\caption{Pulse-height response of the lead-glass module Nr.4. on cosmic muons.}
\label{cosmics4}
\end{figure}

For the same modules used in the photon beam, the response to muons from cosmic radiation was measured as well.  
Modules were placed vertically, and a coincidence signal from two 2.5 cm thick 
plastic scintillators with dimensions $8x8 cm^2$, one placed above and the other below 
the module, triggered the muons. The same electronics was used for the data
processing and the calibration obtained by the photon measurement was utilized.
The response of the detectors (amount of detected Chererenkov light) on muons 
correspond to a response on photons with energy of 577 MeV. The obtained energy resolution
for the modules was   between 5.5-7\%, so very similar to a response to photons (see Fig.
\ref{fig_res_vs_HE}). The example of measured ADC spectra for module Nr. 4 is
shown in Fig. \ref{cosmics4}.

\subsection{Measurement at CERN}

A further test experiment has been carried on in May 2010 at CERN on the T10 test beam-line of 
the PS synchrotron \cite{CERN_T10}. 
In this case, the purpose was to evaluate the electron/pion separation power of the
lead glass modules, and to measure the time resolution which is essential
for removing the hits caused by neutrons.

\begin{figure}[h!]
\centering
\includegraphics*[width=140mm]{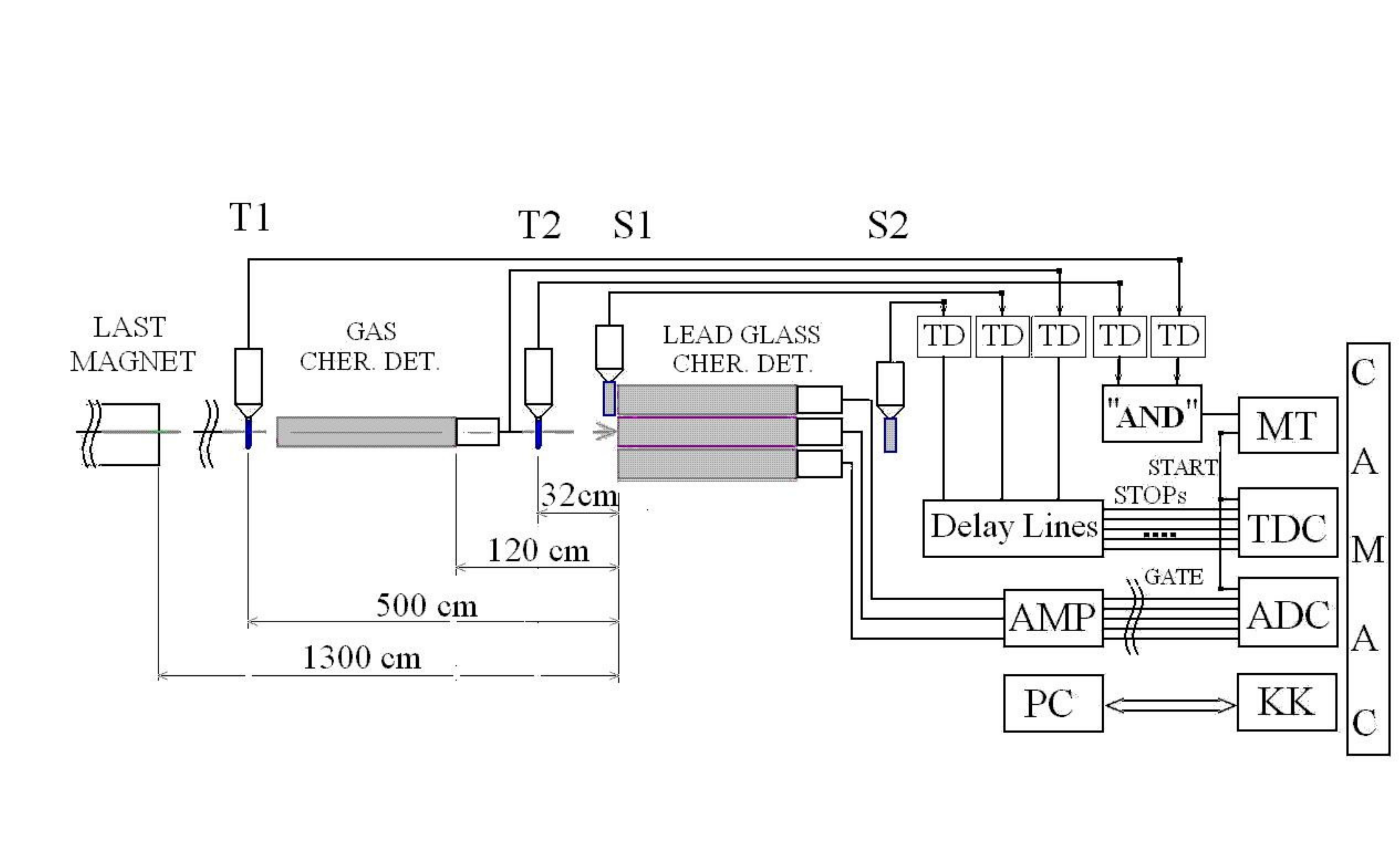}
\caption{Set-up at T10 CERN PS test beam line.}
\label{setup_cern}
\end{figure}

\begin{figure}[hb!]
\centering
\includegraphics*[width=105mm]{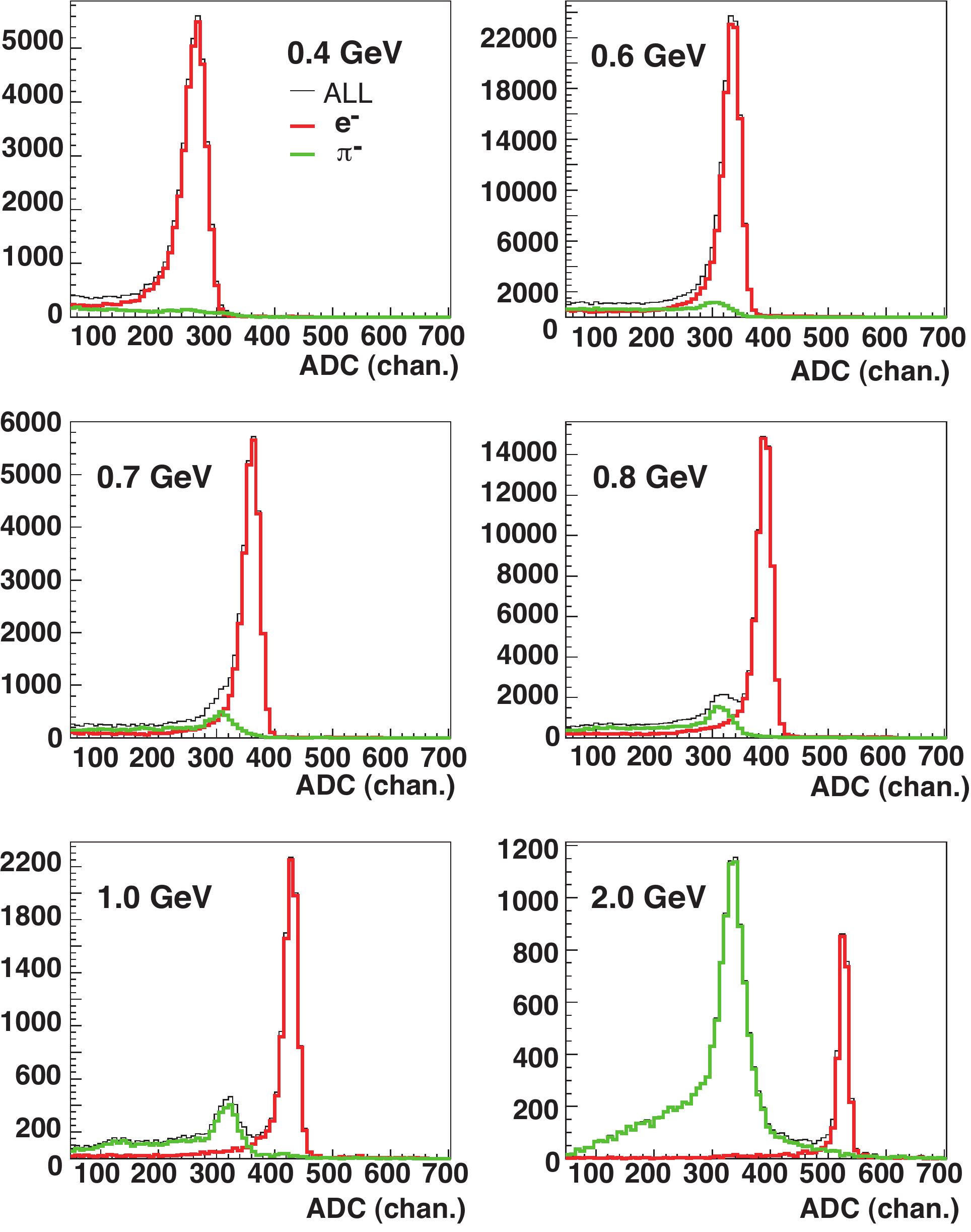}
\caption{Distribution of energy deposited in lead-glass module exposed
to the $\pi^{-}$ (green line) and $e^{-}$ (red line) beam at 8 different 
momentum setting of the T10 beam line.}
\label{fig_cern_Escan}
\end{figure}

\subsubsection{Test conditions and setup}

At this time, five identical modules were tested (glass+optical grease+PMT EMI9903KB).
The T10 test beam line of the CERN PS synchrotron provides (for the setup selecting 
negatively charged particles) a beam of secondary $\pi^-$ with a momentum range  
of 0.4 - 6 GeV/c. The electron
fraction is a few percent at  6 GeV and increases to about 50\% at the
lowest momenta). Detectors were again positioned directly in the beam, they
were hit in the centre of their front side, and the beam
proceeded along their longitudinal axis.
Beam diameter at detector position was larger than detector cross section
(9.2x9.2 cm). For our purpose, only the central region with size of 4x4 cm 
were selected, which was defined by a trigger scintillator placed in from 
of the lead glass module. 
The triggered beam intensity was 100-1000 particles/ bunch, with a
repetition rate of 45 s.
The electron identification was provided by a
 2 m long air Cherenkov 
threshold detector placed in beam about 120 cm in front of lead glass module. 
The efficiency for electrons was 98\%.
The trigger was obtained by the logic OR of the signals from the two  4x4 cm$^2$ scintillators, one placed 
directly in front of the lead glass module, the other one about 5 m upstream. 
Several runs were taken with a T10 line momentum settings from 0.4 to 2 GeV/c.

The detector and DAQ setup is shown in Fig. \ref{setup_cern}.

\begin{figure}[ht!]
\centering
\includegraphics*[width=105mm]{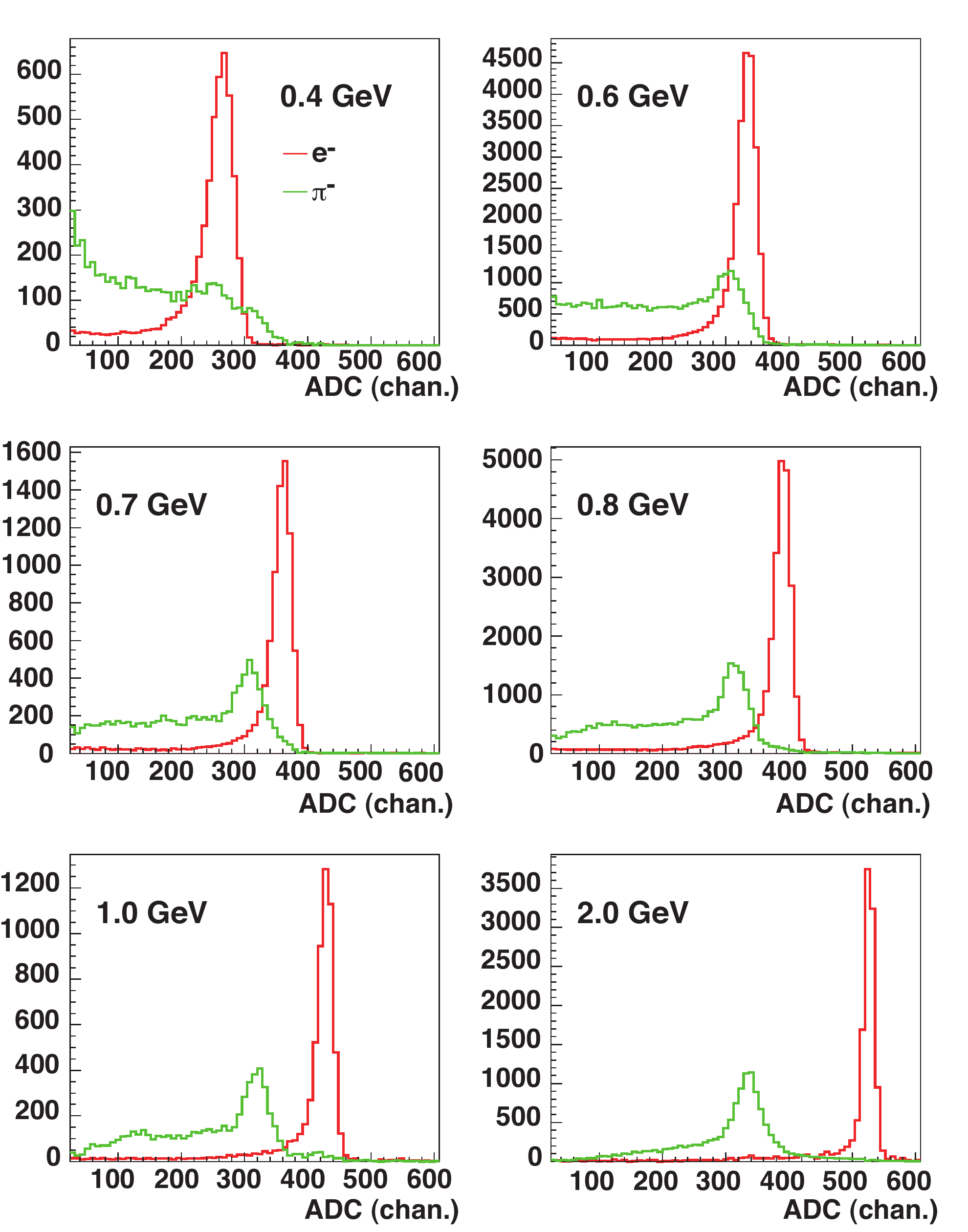}
\caption{The same as in Fig. \ref{fig_cern_Escan}, only the $\pi^-$ and e$^-$ response is normalized to the same intensity}
\label{fig_cern_Escan2}
\end{figure}
\begin{figure}[htb]
\centering
\includegraphics*[width=115mm]{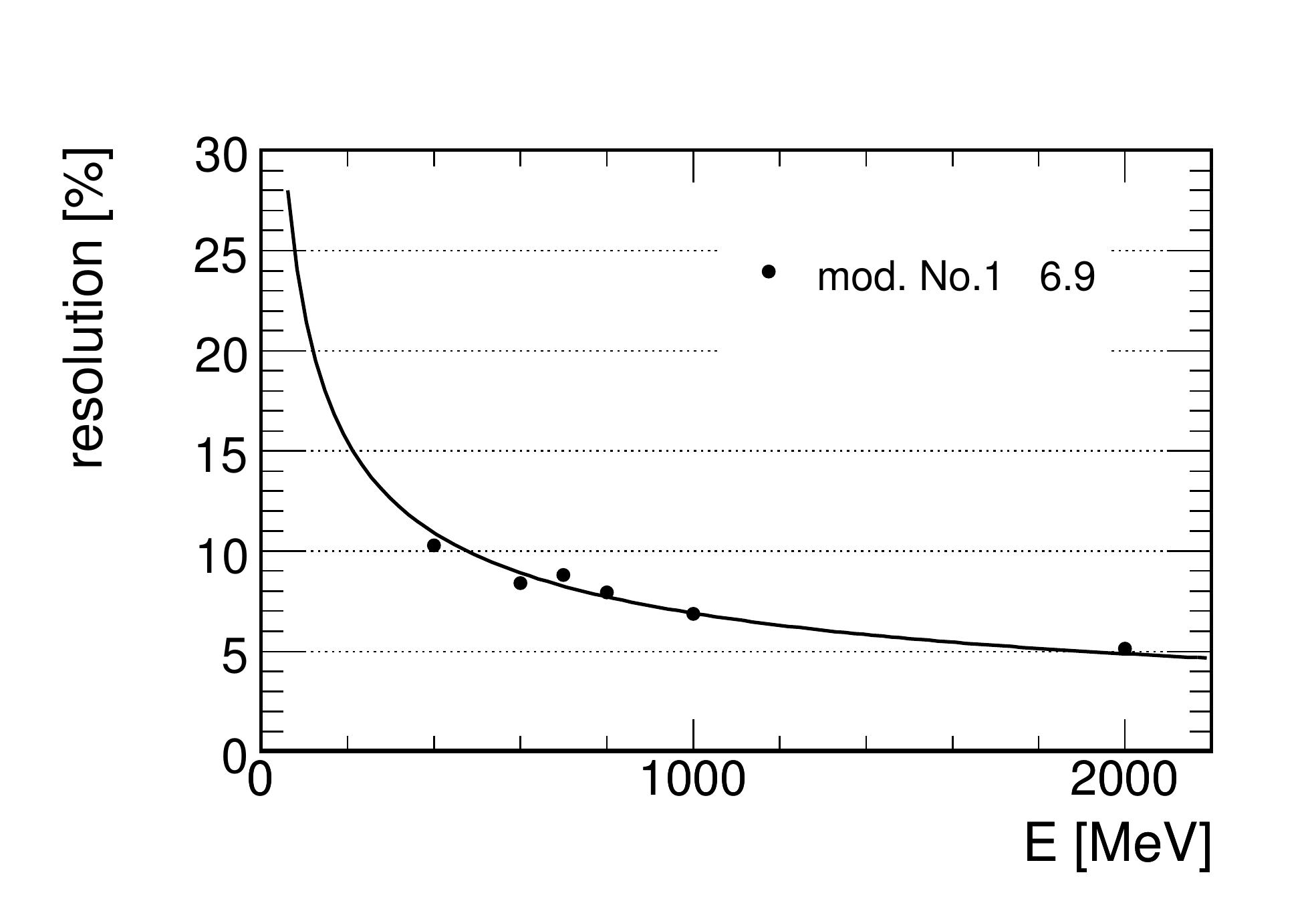}
\caption{Dependence of energy resolution on electron energy}
\label{fig_cern_e_resolution}
\end{figure}

\subsubsection{Results}

\begin{figure}[htb]
\centering
\includegraphics*[width=80mm]{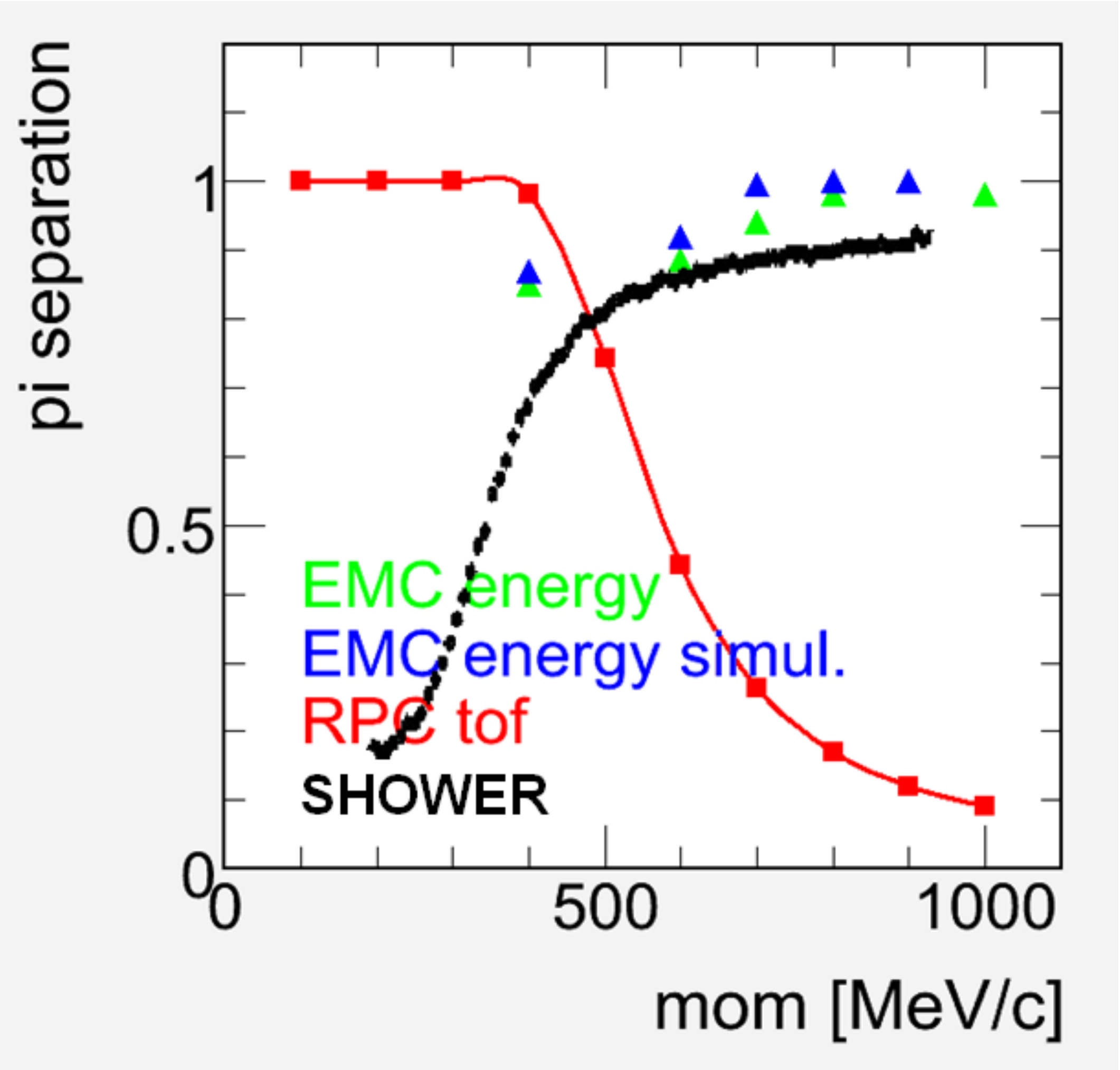}
\caption{e/$\pi$ separation factor as a function of momentum.}
\label{fig_epi_sep}
\end{figure}

The response of one lead-glass module exposed to $e^{-}$ and $\pi^{-}$ beam 
with energies from 0.4 to 2 GeV is shown in Fig. \ref{fig_cern_Escan}.
On Fig. \ref{fig_cern_Escan2} the $e^{-}$ yield is scaled to be the same as
for $\pi^{-}$ yield, and one can see more clearly difference in the response to 
different particles.
Fig. \ref{fig_cern_e_resolution} shows the relative energy resolution 
for electrons measured in this test. 
The energy calibration was done in the same way as described above for 
photon data. 
We observe an energy resolution of 6.9\% 
for electrons at 1 GeV, while on basis of the photon test a resolution 
of 6\% or better is expected. This can be  caused by the bad shape 
of the  electron peaks, with a long energy tail due to energy loss
of electrons in air ($\approx$ 15 m) and other detectors in the T10 area. 
Electron distributions measured here look worse than gamma peaks 
(cf. with Fig. \ref{fig_res_vs_HE}).
The ability of the lead-glass detectors to separate electrons from pions
was extracted from the data as shown in Fig. \ref{fig_cern_Escan2}. 
The electron/pion separation factor is defined as a number of pions outside
of the electron peak divided by the total number of pions, assuming the same
electron and pion abundancy. The result is displayed in Fig. \ref{fig_epi_sep}. 
 
\begin{figure}[htb]
\centering
\includegraphics*[width=85mm]{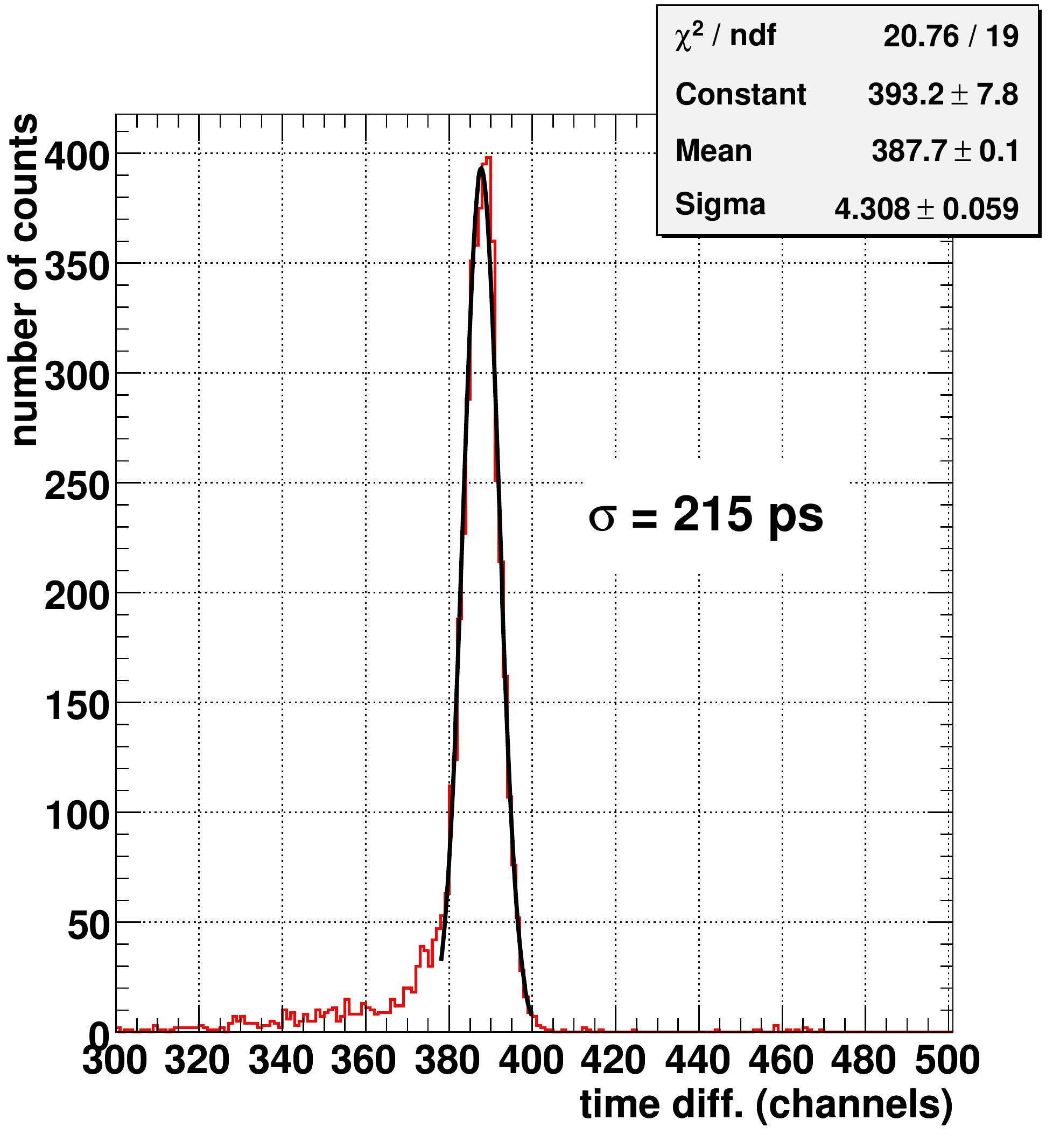}
\caption{Distribution of the time difference measured between the lead glass 
module and the beam quartz detector for electrons
with energy of 800 MeV. One TDC channel corresponds to 50 ps.}
\label{fig_cern_time_resolution}
\end{figure}

For the calculation of the separation factor of the RPC time-of-flight detector we assume 
a time resolution of 100 ps (sigma). The results show that the electromagnetic 
calorimeter will allow for an efficient $\pi$/e separation at higher momenta
where the time of flight cannot be used.
During this test, we also measured the time resolution of the lead-glass
detector, see Fig. \ref{fig_cern_time_resolution}. For the START signal 
a quartz detector with a time resolution lower than 100 ps was used.
The measured time resolution of 215 ps (sigma) will enable to efficiently
remove neutrons which are not detected by the tracking system. 

To conclude, the energy resolution of this detector is shown to be satisfactory and 
probably could be improved employing a better 
electronic. The most important result interests the $\pi$/e
separation. Indeed, results show really good
separation, much better than the actual pre-shower detector. 
These results are very promising and an
electromagnetic calorimeter with lead glass modules from 
OPAL experiment is a practical, feasible and
really efficient proposal to improve the results of HADES.

%% file: Conclusion/Conclusion.tex
\section{Summary}

We propose to build the Electromagnetic calorimeter
for the HADES spectrometer. It will enable to measure the data on 
neutral meson production, which are essential for interpretation of
dilepton data, but are unknown in the energy range of
planned experiments. The calorimeter will improve the 
electron-hadron separation, and will be used for detection of photons
from strange resonances in elementary and HI reactions.

Detailed description of the detector layout, the support structure, 
the electronic readout and its performance studied via Monte Carlo 
simulations and series of dedicated test experiments is presented.

The device will cover the total area of about 8\,m$^{2}$ at polar angles 
between $12^{\circ}$ and $45^{\circ}$ with almost full azimuthal coverage.
The photon and electron energy resolution achieved in test experiments 
amounts to $\approx$ 5-6\%/$\sqrt{E}$ which is sufficient for the $\eta$-meson 
reconstruction with S/B ratio of $\sim 0.4\%$ in 
Ni+Ni collisions at 8 AGeV. A purity of the identified leptons
after the hadron rejection, resulting from 
simulations based on the test measurements, is better than $80\%$  at  
momenta above 500~MeV$/c$, where time-of-flight cannot be used.

%% file: Timeplan/Timeplan.tex
\section{Time schedule}

\begin{figure}[htb]
\hspace*{-2.5cm}
\includegraphics*[width=210mm,height=180mm]{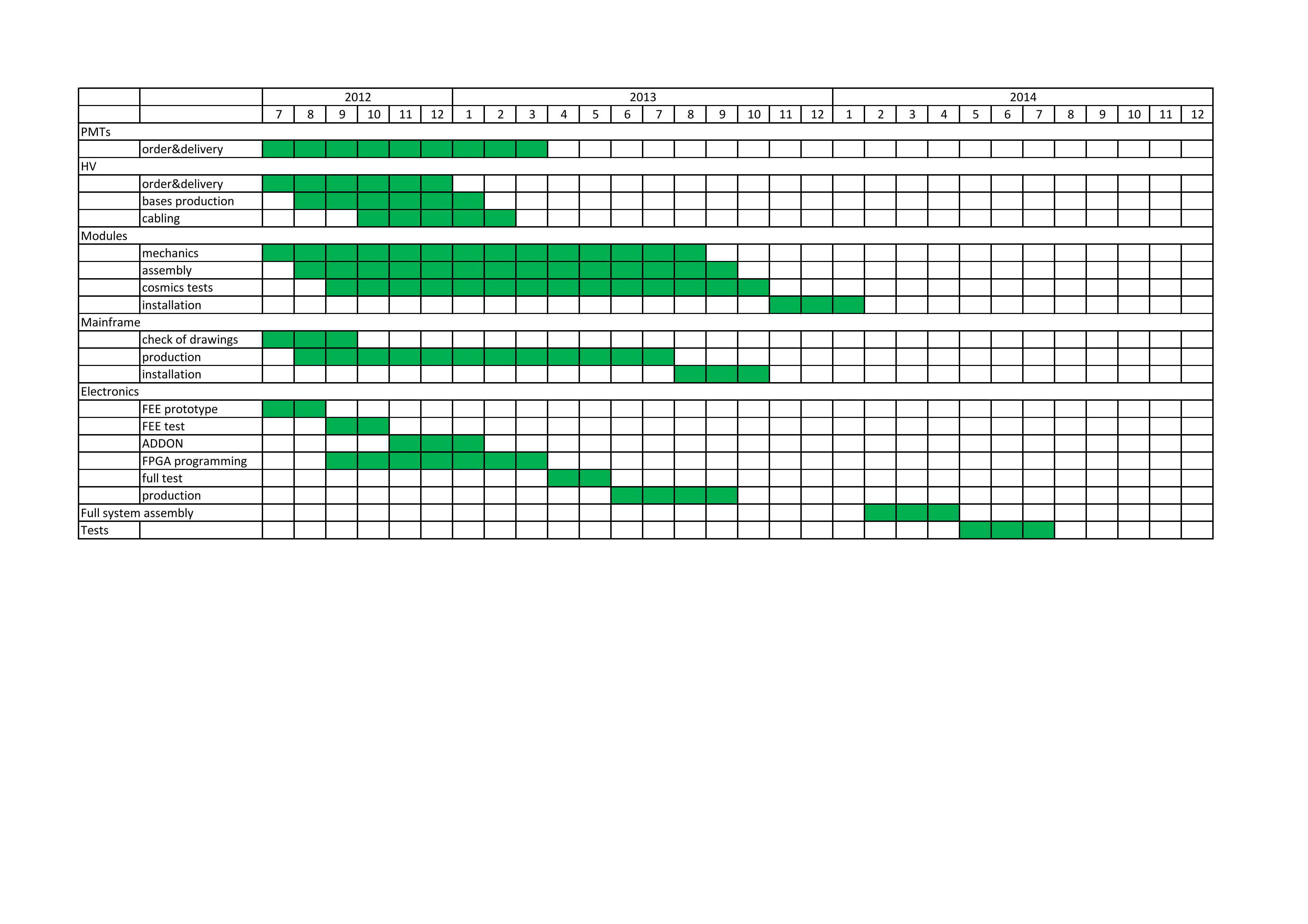}
\caption{The time schedule of the detector construction}
\label{time_plan}
\end{figure}

%% file: Ref/References.tex

%% file: TDRCalo.bbl
\begin{thebibliography}{99}


\bibitem{rapp_wambach} R. Rapp and J.~Wambach, Adv.\ Nucl.\ Phys.\  25 (2000) 1.
\bibitem{akr90} M.Akrawy et al., NIM A290 (1990) 76-94.
\bibitem{lev60} P. W. Levy, Arner, Ceramic Soc. 43 (1960) 389.
\bibitem{opal_barrell} OPAL Collaboration,  NIM A305 (1991) 275-319.
\bibitem{gol73}  M. Goldberg, P.L. Mattern, K. Lengweiler and P.W. Levy, NIM, 108 (1973) 119.
\bibitem{Fro08} I. Fr\"ohlich et al., 'A General Purpose Trigger and Readout Board for HADES and FAIR-. Experiments', IEEE Trans. Nucl. Sci. 55, 59 (2008).
\bibitem{Pal08} M. Palka et al., Nuclear Science Symposium Conference Record, 2008, NSS 08, IEEE, p. 1398-1404.
\bibitem{Arefev:2008zz} A.~V.~Arefev {\it et al.}, Instrum.\ Exp.\ Tech.\  {\bf 51} (2008) 511.
\bibitem{Alekseev:2008zza} I.~G.~Alekseev {\it et al.}, Instrum.\ Exp.\ Tech.\  {\bf 51} (2008) 491.
\bibitem{MAMI} H. J. Arendsa et al., Proceedings of The IX International Conference on Hypernuclear and Strange Particle Physics  2007, Part 1, 1-5, DOI: 10.1007-978-3-540-76367-3 1. 
\bibitem{MAMI_tag} J.C. McGeorge  et al., European Physical Journal A - Hadrons and Nuclei 37 1 129-137, DOI: 10.1140-epja-i2007-10606-0.
\bibitem{CERN_T10} CERN/PS 97-41 1997 Particle Accelerator Conference, 12-16.5.97, Vancouver, B.C., Canada.
\end{thebibliography}
